\documentclass[a4paper,12pt]{article}
\pdfoutput=1
\usepackage{subfig}
\usepackage{color}
\usepackage{graphicx}
\usepackage{amsfonts}
\usepackage{amsmath}
\usepackage{multirow}
\usepackage{mathrsfs}
\usepackage{amssymb}
\usepackage{verbatim}
\usepackage{float}
\usepackage{url,tabularx,amsfonts,slashed,amsmath,array}
\usepackage{sidecap}

\newcommand{\be}{\begin{eqnarray}}
\newcommand{\ee}{\end{eqnarray}}

\begin{document}

\begin{flushright}
preprint SHEP-12-41\\
\today
\end{flushright}
\vspace*{0.5truecm}

\begin{center}
{\large\bf Multiple $Z'\to t\bar t$ signals in a 4D Composite Higgs Model}\\
\vspace*{1.0truecm}
{\large D. Barducci$^1$, S. De Curtis$^2$, K. Mimasu$^1$ and S. Moretti$^{1}$}\\
\vspace*{0.5truecm}
{\it $^1$School of Physics \& Astronomy, University of Southampton, \\
Highfield, Southampton, SO17 1BJ, UK}\\
\vspace*{0.25truecm}
{\it $^2$INFN, Sezione di Firenze,\\
Via G. Sansone 1, 50019 Sesto Fiorentino, Italy
}\\
\end{center}
\vspace*{0.5truecm}
\begin{center}
\begin{abstract}
\noindent
We study the production of top-antitop pairs at the Large Hadron Collider as a testbed for discovering heavy $Z'$ bosons belonging to a
composite Higgs model, as, in this scenario, such new gauge interaction states are sizeably coupled to the third generation quarks of the
Standard Model. We study their possible appearance in cross section as well as (charge and spin) asymmetry distributions. 
Our calculations are performed in the minimal four-dimensional formulation of such a scenario, namely the 4-Dimensional Composite 
Higgs Model (4DCHM), which embeds five new $Z'$s. We pay particular attention to the case of nearly degenerate resonances, 
highlighting the conditions under which these are separable in the aforementioned observables.
We also discuss the impact of the intrinsic width of the new resonances onto the event rates and various distributions.
We confirm that the 14 TeV stage of the LHC will enable one to detect two such states, assuming standard detector performance and machine
luminosity.  A mapping of the discovery potential of the LHC of these new gauge bosons is given. Finally, from the latter,
several benchmarks are extracted which are amenable to experimental investigation. 

\end{abstract}
\end{center}

\section{Introduction}
\label{sect:intro}

$Z'$ states are rather ubiquitous in beyond the Standard Model (BSM) scenarios and are typically searched for at hadron colliders through a 
di-lepton signature in the neutral Drell-Yan (DY) process, i.e.,
$pp(\bar p) \to (\gamma,Z,Z') \to \ell^+\ell^-$, where $\ell=e,\mu$. In fact, the most stringent limits on $Z'$'s
at the Large Hadron Collider (LHC) come from this signature and are set at around 2.5 TeV (for a sequential $Z'$) \cite{LHC}\footnote{This is a state with
generic mass and same couplings to the SM particles as the $Z$ boson. Limits in the aforementioned models are normally obtained
by rescaling the results for a sequential $Z'$, though this implicitly assumes that the $Z'$ cannot decay into
additional extra matters present in the model spectrum.}.
Since such an experimental signature is clean and theoretical uncertainties, for sufficiently inclusive quantities, are well under control,
see, e.g., \cite{Fuks:2007gk}, including those associated to higher order effects,  both
two-loop Quantum Chromo-Dynamics 
(QCD) \cite{DYrefs} and one-loop Electro-Weak (EW) \cite{Baur:2001ze} ones, one can conceive accessing the couplings of a discovered $Z'$ 
(thereby providing a window on its high scale genesis) by studying the ensuing di-lepton observables such as 
(differential) cross sections and/or asymmetries.  

Another decay channel of $Z'$ bosons that can be phenomenologically relevant also as a search mode is the one yielding top-antitop final states
i.e., $pp(\bar p)\to (\gamma,Z,Z') \to t\bar t$. Its importance for $Z'$ searches with respect to the DY case becomes manifest in models in which 
the new neutral gauge bosons have sizeable couplings to the third generation quarks while being weakly coupled to those of the first two and, most importantly, to leptons as well. In such scenarios, the 
complications arising from a much
larger background, dominated by QCD production of top-antitop quark pairs, a more involved final state yielding six or more objects in the detector including jets as well as 
an associated poorer efficiency in reconstructing the two heavy quarks (with respect to the much simpler case of the DY process) must be overcome, if one intends
to probe the associated $Z'$ states. {{While this is an arduous challenge, it reveals  its rewards when one notices that the top 
decays before hadronising (so its spin properties are effectively transmitted to the decay products) and that the electromagnetic charge of the top can be 
tagged via a lepton and/or a $b$-jet \cite{tt-pol}}}. Also $t\bar t$ samples can be extremely useful in profiling the $Z'$ (and not only \cite{tt-pol}), 
as the aforementioned charge/spin asymmetries, particularly 
effective in pinning down the couplings of the new gauge bosons, can be defined theoretically and measured experimentally \cite{Zp-tt}. 
Furthermore, in this connection, two other key considerations, pertaining to the $t\bar t$ final state but not the $\ell^+\ell^-$ one, ought to made.
On the one hand, the multi-step decay chain $Z'\to t\bar t \to b\bar b W^+ W^-\to b\bar b X$ (with $X$ representing any possible $W^+W^-$ decay), as opposed to the simple $Z'\to \ell^+\ell^-$ one,
affords one with the possibility to define a wider variety of the aforementioned charge and/or spin asymmetries then in the DY case (albeit correlated one another). On the other hand,  
the large top mass, as opposed to a negligible one for both electrons and muons, induces non-trivial spin correlations, which are not present in the DY case 
and are also sensitive to the nature of the intervening $Z'$ state.  Guided by these considerations, 
experimental collaborations at both Tevatron \cite{Tevatron-tt} and LHC \cite{LHC-tt} have in fact been pursuing the study of $t\bar t$ data samples in BSM searches in general, and more systematically recently, 
in part driven by some anomalies that have emerged in the forward-backward asymmetry of $t\bar t$ events
at the Tevatron \cite{anomaly}\footnote{The LHC has not confirmed this, see Ref.~\cite{no-anomaly}, though the $pp$ nature of the CERN accelerator with top-antitop production dominated by gluon-gluon fusion,
as opposed to the $p\bar p$ nature of the FNAL machine with top-antitop generation driven by quark-antiquark annihilation, implies that the same signature is much harder to extract.}. 
Finally, just like in the case of DY, also for $t\bar t$ production higher-order effects from both QCD 
\cite{QCD-SM-tt} (see also \cite{earlycalc-QCD}) 
and EW \cite{EW-SM-tt} (see also \cite{earlycalc-EW}) interactions are well known, including the case of polarised (anti)tops.

The purpose of this paper is to study the sensitivity of the LHC  to the presence of $Z'$ bosons as well as
to assess the machine ability to profile them when mediating $t\bar t$ production,
in both cross sections and  (charge/spin) asymmetries, assuming as theoretical framework the 4-Dimensional Composite Higgs
Model (4DCHM) of   Ref.~\cite{DeCurtis:2011yx}. The latter is the ideal theoretical scenario to test experimentally in this context, for the following reasons.
On the one hand, the scope of DY in accessing the gauge sector of the 4DCHM is only confined to large machine energies and luminosities \cite{Barducci:2012kk}. On the other hand,
$ t\bar t$ decays are here amongst the preferred decay modes of the ensuing new heavy neutral gauge 
bosons\footnote{{{As shown in \cite{Barducci:2012kk}}}, this is no longer true when the new heavy fermion decay are channels open, as they are the preferred ones due to the large couplings involved.}. In fact, an added feature of the  4DCHM is the 
presence of multiple such resonances, i.e., five in the model spectrum, though only three are potentially accessible at the LHC.

The plan of the paper is as follows. In the next section we  recall the properties of the 4DCHM pertaining to the bosonic and fermionic sectors
affecting top (anti)quark phenomenology.
In Sect.~\ref{sec:calculation} we describe our calculation and define the observables to be studied. In Sect.~\ref{sec:results} we report and comment on our results. 
In the last one we conclude and in the Appendix we list the numerical values of the $Z't\bar t$ couplings for the benchmark points considered.

\section{Model \label{sec:model}}
We describe here the model on which we base our analysis, chiefly its neutral gauge sector and its extended fermion one, by fixing conventions and discussing its relevant features. Further, we test its parameter space against available experimental constraints.

In addition to the SM particles (the $e^-$, $\mu^-$, $\tau^-$, $\nu_{e,\mu,\tau}$ leptons, the $u,d,c,s,t,b$ quarks and the $\gamma,Z,W^{\pm}$, $g$ gauge bosons), the 4DCHM presents a Higgs boson $H$, which is a pseudo-Nambu Goldstone boson, and a large number of new particles, both in the fermionic (quark) and bosonic (gauge)  sector.
We summarise the additional particle content of the 4DCHM with respect to the SM in Tab.~\ref{table:partspec}\footnote{Hence, the $Z_{1,...5}$
states herein are our $Z'$ bosons.}. 

\begin{table}[h!]
\begin{center}
\begin{tabular}{|l|l|}
\hline
Neutral Gauge Bosons & $Z_{1,...5}$\\
Charged Gauge Bosons & $W_{1,.3}^{\pm}$\\
Charge $+2/3$ quarks    & $T_{1,...8}$ \\
Charge $-1/3$ quarks    & $B_{1,...8}$ \\
Charge $+5/3$ quarks    & $\tilde{T}_{1,2}$ \\
Charge $-4/3$ quarks    & $\tilde{B}_{1,2}$ \\
\hline
\end{tabular}
\end{center}
\caption{Extra particle content of the 4DCHM with respect to the SM. An increasing number in the label of a particle corresponds to a larger mass of 
the particle itself.}
\label{table:partspec}
\end{table}

Amongst the various new states predicted by the 4DCHM, we concentrate here on the additional neutral gauge bosons $Z_{2,3,5}$ (notice in fact,
as we shall show explicitly in the following, that
the $Z_{1,4}$ states are essentially inert for the purpose of our study, as they do not couple to first and second generation (anti)quark \cite{DeCurtis:2011yx}\footnote{Further, we can confirm
that the contribution to the process studied here induced by the subprocess $b\bar b\to \gamma, Z, Z_i\to t\bar t$ is negligible ($i=1, ... 5$), owing to the small probability
of extracting $b$-(anti)quarks from the proton sea of partons for our typical kinematic configurations, for which $x^2\approx M_{Z'}^2/s$, with
$M_{Z'}\approx 2-2.5$ TeV and $\sqrt s=7,8$ and 14 TeV ($x$ being the $b$-(anti)quark momentum fraction relative to the proton beam).})
and additional heavy quarks  $T_{1,2,...,8}$,  $B_{1,2,...,8}$, $\tilde{T}_{1,2}$ and $\tilde{B}_{1,2}$, which affect the  $Z_{2,3,5}$ widths.
We neglect the Higgs and charged gauge boson sectors, for which we instead refer the reader to \cite{DeCurtis:2011yx} in general and
\cite{Barducci:2012kk,Diboson} and \cite{HiggsPaper}  in particular, respectively, for the two contexts.

The 4DCHM can be schematised in two sectors, the elementary and the composite one, arising from an extreme deconstruction of a 5D theory.
The gauge structure of the elementary sector of the 4DCHM is associated with the $SU(2)_L\otimes U(1)_Y$ SM  gauge symmetry whereas the composite sector has a local $SO(5)\otimes U(1)_X$  symmetry
with eleven new gauge resonances.
Therefore, the spin-1 particle content of the 4DCHM is given, besides the standard $W,~Z$ bosons and the photon, by five new neutral, collectively denoted by $Z'$, and three new charged, collectively denoted by $W'$, bosons. 
The parameters for the gauge sector are: the scale $f$ of the spontaneous global symmetry breaking $SO(5)\to SO(4)$  
(typically of the order of 1 TeV) and $g_*$, the $SO(5)$ gauge coupling constant which, for simplicity, we take 
equal to the $U(1)_X$ one. The mass spectrum of the spin-1 fields is then expressed in terms of these two new parameters, 
 the gauge couplings of $SU(2)_L$ and $U(1)_Y$, namely $g_0$ and $g_{0Y}$, {{and $\langle h \rangle\sim v$}}, the Vacuum Expectation Value (VEV) of 
the Higgs boson\footnote{In the 4DCHM  the VEV of the Higgs boson is extracted by the minimum of the Coleman-Weinberg 
potential as a function of the fermion and gauge boson parameters, which, in the following analysis, will be chosen in such a way as to reproduce $\langle h \rangle\sim v=246$ GeV.  
In particular, the analytical expressions of the neutral gauge boson masses at the leading order in $\xi=v^2/f^2$,  
with $v$ 
 (see \cite{Barducci:2012kk} for details).}, are given as (here, an increasing
number in the label  indicates a particle with higher mass)
\begin{equation}\label{eq:MZ}
\begin{split}
&M^2_{\gamma}= 0,\\
&M^2_{Z}\simeq  \frac{f ^2}{4} g_*^2(s^2_\theta+\frac{s^2_\psi}{2})  \xi,\\
&M^2_{Z_1}=f ^2g_*^2,\\
&M^2_{ Z_2}\simeq\frac{f ^2g_*^2}{ c_\psi^2}(1-\frac{s^2_\psi c^4_\psi }{4 c_{2\psi}}\xi),\\
&M^2_{Z_3}\simeq \frac{f ^2g_*^2}{ c_\theta^2}(1-\frac{s^2_\theta c^4_\theta }{4 c_{2\theta}}\xi), \\
&M^2_{Z_4}=2 f ^2g_*^2, \\
&M^2_{Z_5}\simeq 2 f ^2g_*^2(1+\frac 1 {16} (\frac 1 {c_{2\theta}}+\frac 1{2 c_{2 \psi}})\xi),
\end{split}
\end{equation}
with $\tan\theta=s_\theta/c_\theta=g_0/g_*$, $\tan\psi=s_\psi/c_\psi=\sqrt{2} g_{0Y}/g_*$. The photon is massless, as it should be, and the neutral gauge bosons $Z_1$  and $Z_4$ have their masses completely determined by the composite sector. 

\begin{figure}[!h]
\begin{center}
\includegraphics[width=5cm]{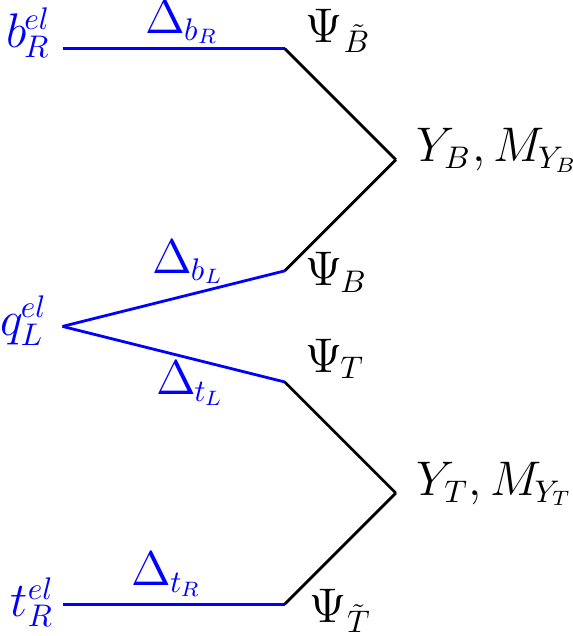}
\end{center}
\caption{Fermionic sector of the 4DCHM.  The elementary sector is on the left, the composite one is on the right. Symbolically showed are the mixing and Yukawa terms.}
\label{fig:ferm}
\end{figure}

Regarding the fermionic sector, we just recall here that the new heavy states are embedded in  fundamental representations of $SO(5)\otimes U(1)_X$ and two multiplets of resonances for each of the SM third generation quark are introduced in such a way that only top and bottom quarks
 mix with these heavy fermionic resonances in the spirit of partial compositeness. 
This choice of representation is a realistic scenario compatible with precision EW measurements and represents a discretisation to two sites of the MCHM in \cite{Contino:2006qr} (see Fig.~\ref{fig:ferm}).
As stated before, the spectrum contains four {\bf 5} representations
 indicated with $\Psi_{T,\tilde{T}/B,\tilde{B}}$ in the composite top/bottom sector, respectively.
The SM third generation quarks, both for the left-handed doublet, $q_L^{el}$, and the two right-handed singlets, $b_R^{el}$ and $t_R^{el}$, are embedded in an incomplete representation of $SO(5)\otimes U(1)_X$ in such a way that their correct quantum numbers under $SU(2)_L\otimes U(1)_X$ are reproduced via the relation $Y=T^{3R}+X$. 
The fermionic Lagrangian of the 4DCHM considered in \cite{DeCurtis:2011yx} is (for simplicity we take $m_T=m_{\tilde{T}}=m_B=m_{\tilde{B}}=m_*$):
\begin{equation}
\label{eq:lagferm}
\begin{split}
\mathcal{L}_{fermions}&=\mathcal{L}_{fermions}^{el}+ (\Delta_{t_L}\bar{q}^{el}_L\Omega_1\Psi_T+\Delta_{t_R}\bar{t}^{el}_R\Omega_1\Psi_{\tilde{T}}+h.c.)\\
&+\bar{\Psi}_T(i\hat{D}-m_*)\Psi_T+\bar{\Psi}_{\tilde{T}}(i\hat{D}-m_*)\Psi_{\tilde{T}}\\
&-(Y_T\bar{\Psi}_{T,L}\Phi_2^T\Phi_2\Psi_{\tilde{T},R}+M_{Y_T}\bar{\Psi}_{T,L}\Psi_{\tilde{T},R}+h.c.)\\
&+(T\rightarrow B),
\end{split}
\end{equation}
where $D$ indicates the covariant derivative related to the composite gauge fields,
$\Delta_{t_L,t_R,b_L,b_R}$ are the mixing parameters relating the elementary and the composite sector, whilst $Y_{T,B}$ and $M_{Y_{T,B}}$ are the Yukawas of the composite sector. In eq.~(\ref{eq:lagferm}) the fields $\Omega_{1}$, and $\Phi_2$  trigger the symmetry breaking
and are expressed in terms of the   Goldstone bosons (see \cite{{DeCurtis:2011yx}}  for details).

The top and bottom quark masses are proportional to the EWSB parameter and to the elementary-composite sector mixings (shown in Fig.~\ref{fig:ferm}) as suggested by the partial compositeness hypothesis. Due to the bottom-top mass hierarchy, we will require $\Delta_{b_L,b_R}\sim 10^{-1} \Delta_{t_L,t_R}$.
The fermionic mass spectrum, at the leading order in $\xi$ for the top and bottom quark, and for $\xi=0$  for the lightest new fermions, is given by:
\begin{equation}
\label{fermionmass}
\begin{split}
m_b^2 &\simeq\xi  \frac{m_*^2} 2   \frac{ \tilde \Delta_{b_L}^2 \tilde \Delta_{b_R}^2 \tilde Y_B^2}{(1+F_L) }, \\
m_t^2 &\simeq \xi \frac{m_*^2} 2   \frac{ \tilde \Delta_{t_L} ^2\tilde \Delta_{t_R} ^2\tilde Y_T^2}{(1+F_L)(1+F_R) },\\
m^2_{T_1} &\simeq\frac{m_*^2}{2} \left(2 +\tilde{M}_{Y_T}^2-\tilde{M}_{Y_T}\sqrt{4 +\tilde{M}^2_{Y_T}}\right)=m^2_{\tilde T_1} ,\quad |M_{Y_T}|>|M_{Y_B}|\\
m^2_{B_1} &\simeq\frac{m_*^2}{2} \left(2 +\tilde{M}_{Y_B}^2-\tilde{M}_{Y_B}\sqrt{4 +\tilde{M}^2_{Y_B}}\right)=m^2_{\tilde B_1},
\end{split}
\end{equation}
where we have defined $\tilde \Delta_{t_L,t_R,b_L,b_R}=\Delta_{t_L,t_R,b_L,b_R}/m_*$, $\tilde Y_{T,B}= Y_{T,B}/m_*$,  $\tilde M_{Y_{T,B}}=M_{Y_{T,B}}/m_*$, 
\begin{equation}
\label{FLR}
F_L=\tilde\Delta_{t_L}^2(1+\tilde M^2_{Y_T}),\quad\quad F_R=\tilde\Delta_{t_R}^2(1+(\tilde M_{Y_T}+\tilde Y_T)^2)
\end{equation}
and, for simplicity, we have taken $ \Delta_{b_L}=\Delta_{b_R}=0$ except in the  bottom mass expression.

For the process of interest here, $pp(q\bar q,gg)\to t\bar t$ (gluon-gluon fusion clearly inducing only background events),
 we need 
the couplings of the  $Z_{1,...,5}$  to the first two generations of (anti)quarks, which live in the elementary sector and those to the third generation (anti)quarks, which interact with the composite fermionic sector. {{While the former comes only from the mixing of the $Z_{1,...,5}$ with the elementary gauge bosons in the latter also the mixing of the third generation (anti)quarks with the new heavy fermions has to be taken in account.}}

In order to give an idea of the order of magnitude of this effect, we provide here the analytical expression for the neutral current interaction Lagrangian
of the 4DCHM at the leading order in $\xi$,  limited to the case of the first two generations of (anti)quarks, as for the case of the third generation it is sufficient to show the $\xi=0$ terms since they are already sizeable due to the elementary-composite sector mixing terms.
(In both instances though, in all the forthcoming calculations of cross sections and asymmetries, we have used  the corresponding full numerical expressions without any approximations.)  

Let us first consider the neutral-current  Lagrangian of the 4DCHM for leptons and the first two generation quarks. Starting from the elementary sector, where the neutral gauge fields of $SU(2)_L\times U(1)_Y$ are coupled with the fermion currents, we get, after taking into account the mixing among the fields the following expression:
\begin{equation}\label{LNC}
{\cal L}_{NC}^{l,u,d,c,s}=\sum_f\big[ e\bar\psi^f \gamma_\mu Q^f \psi^f A^\mu+ \sum_{i=0}^5   (\bar\psi^f_L  g_{Z_i}^L(f) \gamma_\mu  \psi^f_L+\bar\psi^f_R  g_{Z_i}^R(f) \gamma_\mu  \psi^f_R ) Z_i^\mu \big], 
\end{equation}
where $\psi_{L,R}=[(1\pm\gamma_5)/2]\psi$ and we have identified $Z_0$ with the neutral SM gauge boson $Z$. The photon field, $A_\mu$, 
is coupled to the electromagnetic current in the standard way, namely with the electric charge, which in the 4DCHM is defined as:
\begin{equation}\label{e}
e=\frac{g_L g_Y}{\sqrt{g_L^2+g_Y^2}},\quad\quad g_L=g_0 c_\theta,\quad \quad g_Y=g_{0Y} c_\psi ,
\end{equation}
while the couplings of the $Z_i$'s have the following expressions,
\begin{equation}
g_{Zi}^L(f)= A_{Z_i}T^3_L(f)+ B_{Z_i} Q^f, \quad\quad
g_{Zi}^R(f)=  B_{Z_i}Q^f,
\end{equation}
where $A_{Z_i}=(g_0 \alpha_i - g_{0Y} \beta_i) $, $B_{Z_i}=g_{0Y} \beta_i$  and
$\alpha_i$ and $\beta_i$ the diagonalization matrix elements, namely, 
\begin{equation}
W_3=\sum_{i=0}^5 \alpha_i Z_i, \quad\quad  Y=\sum_{i=0}^5 \beta_i Z_i,
\end{equation}
with $W_3$ and  $Y$ the elementary gauge field associated to $SU(2)_L$ and $U(1)_Y$, respectively.
As intimated, as a result, the $Z_1$ and $Z_4$ bosons are not coupled to leptons and to the first two quark generations, so they are completely inert for the process we are here considering.

At the leading order in $\xi$ we get:
\begin{eqnarray}
A_{Z_0}= \frac{e}{s_\omega c_\omega}\big[1+(c^2_\omega a_Z+ s^2_\omega b_Z)\xi\big],\quad\quad
&&B_{Z_0}= - \frac{e s_{\omega}}{c_{\omega}}(1+b_Z \xi), \\
A_{Z_2}=  -\frac{e}{c_\omega} \frac{s_\psi}{c_\psi} \Big[1+(\frac{c_\omega}{s_\omega} a_{Z_2}-b_{Z_2})\xi\Big], \quad\quad  &&B_{Z_2}= \frac{e}{c_\omega} \frac{s_\psi}{c_\psi} \Big[1-b_{Z_2}\xi\Big],\\
A_{Z_3}=  -\frac{e}{s_\omega}\frac{s_\theta}{c_\theta}\big[1+(a_{Z_3}+\frac{s_\omega}{c_\omega} b_{Z_3})\xi\big], \quad \quad   &&B_{Z_3}=   \frac e{c_\omega} \frac{s_\theta}{c_\theta} b_{Z_3}\xi,\\
A_{Z_5}=\frac e {s_\omega} ( a_{Z_5}-\frac{s_\omega}{c_\omega} b_{Z_5})\sqrt{\xi},\quad\quad   &&B_{Z_5}=  \frac e {c_\omega} b_{Z_5}\sqrt{\xi},
\end{eqnarray}
with
\begin{equation}
\tan\omega=\frac{g_Y}{g_L},\quad\quad e=g_L s_\omega=g_Y c_\omega,\quad\quad  \frac e{s_\omega c_\omega}= \sqrt{g_L^2+g_Y^2},
\end{equation}
and
\begin{equation}
\begin{split}
a_{Z_0}=  (2 s_\theta^2+s_\psi^2)(4 cˆ_\theta^2-1)/32,\quad \quad
&b_{Z_0}=  (2 s_\theta^2+s_\psi^2)(4 cˆ_\psi^2-1)/32, \\
a_{Z_2}= \frac{\sqrt{2} s_\theta s_\psi c_\psi^6}{4(c_\psi^2-c_\theta^2)(2 c_\psi^2-1)},   \quad \quad
&b_{Z_2}= \frac{c_\psi^4(2-7c_\psi^2+9 c_\psi^4-4 c_\psi^6)}{8s_\psi^2 (1-2 c_\psi^2)^2},\\
a_{Z_3}=  \frac{-2 c_\theta^4+5 c_\theta^6-4 c_\theta^8}{4(1-2 c_\theta^2)^2},\quad \quad
&b_{Z_3}=  \frac{\sqrt{2} s_\theta s_\psi c_\theta^6}{4 (2 c_\theta^2-1)(c_\theta^2-c_\psi^2) },\\
a_{Z_5}=  \frac{s_\theta}{2 \sqrt{2}(1-2 c_\theta^2)},\quad \quad
&b_{Z_5}= - \frac{s_\psi}{4(1-2 c_\psi^2)}.
\end{split}
\end{equation}

Because of the non-universality of the couplings of the neutral sector to the three generations of quarks we also need to present
the couplings of the $Z'$ to the top quark which will be relevant  for the processes we will deal with.
Due to the mixing of the top-quark with the new fermionic resonances that are coupled to the extra neutral gauge bosons (see eq. (\ref{eq:lagferm})),
after taking into account the mixing among the gauge and fermionic fields
the neutral current Lagrangian for the top (anti)quark is the following:

\begin{equation}
\mathcal{L}_{NC}^{top}=\frac{2}{3} e \bar \psi^t \gamma_{\mu} \psi^t A^{\mu}+\sum_{i=0}^5(g^L_{Z_i}(t) \bar\psi^t_L\gamma_{\mu}\psi^t_L+g^R_{Z_i}(t)\bar\psi^t_R\gamma^{\mu}\psi^t_R)Z_i^{\mu}
\label{eq:lag_nc_top}
\end{equation}

The expressions of the coefficients $g_{Z_i}^{L,R}(t)$ turn out to be quite complicated even at the leading order in $\xi$,
for this reason, we only show the terms originating from the elementary-composite mixing before EWSB ($\xi=0$):

\begin{equation}
\begin{split}
g_{Z_0}^L(t)=\frac{e}{s_\omega c_\omega}(\frac 1 2 -\frac 2 3 s^2_\omega),\quad 
&g_{Z_0}^R(t)=\frac{e}{s_\omega c_\omega}( -\frac 2 3 s^2_\omega),\\
g_{Z_1}^L(t)\sim 0,\quad \quad\quad\quad\quad&g_{Z_1}^R(t)\sim 0,\\
g_{Z_2}^L(t)=\frac{e}{6 c_\omega}\frac{s_\psi}{c_\psi}\frac{1}{(1+F_L)}(1-\frac{c_\psi^2}{s_\psi^2} F_L),\quad &g_{Z_2}^R(t)=\frac{2 e}{3 c_\omega}\frac{s_\psi}{c_\psi}\frac{1}{(1+F_R)}(1-\frac{c_\psi^2}{s_\psi^2} F_R),\\
g_{Z_3}^L(t)=-\frac{e}{2 s_\omega}\frac{s_\theta}{c_\theta}\frac{1}{(1+F_L)}(1-\frac{c_\theta^2}{s_\theta^2} F_L),\quad &g_{Z_3}^R(t)\sim 0,\\
g_{Z_4}^L(t)=g_{Z_4}^R(t)=0,\quad\quad&g_{Z_5}^L(t)\sim g_{Z_5}^R(t)\sim 0,
\end{split}
\end{equation}
with $F_{L,R}$  given in eq.~(\ref{FLR}) (for $\Delta_{b_L}=\Delta_{b_R}=0$).
Notice that, in the $\xi=0$ approximation, $\omega$ is equal to the Weinberg angle defined by: 
\begin{equation}
\label{weinberg}
s^2_W c^2_W=\frac{\sqrt{2}e^2}{8 M^2_Z G_F}.
\end{equation} 
In fact, the following relation holds in the 4DCHM:
\begin{equation}
s_\omega c_\omega \sim s_W c_W(1-g(\theta,\psi)\xi),\quad\quad g(\theta,\psi)=\frac 1 {32} (-6 s^2_\theta+4 s^4_\theta-3 s^2_\psi+2 s^4_\psi).
\end{equation}
Notice also that, before EWSB (that is for $\xi=0$), the $Z_0 t\bar t$ coupling is exactly the SM one (as it happens for leptons and for the first two-generation quarks)  and this is due to the unitarity of the rotation matrix in the fermionic sector.
For the exact values of the couplings $Z_iq\bar q$ with $i=0,2,3$ and $q=u,d$  and  of the ratio of these couplings with respect to
the SM ones we refer to \cite{Diboson}, while the exact values of the couplings of 
the top quark to $Z_0,~Z_2$ and $Z_3$ are listed in Tabs.~\ref{tab:zttcoup1}-\ref{tab:zttcoup2} of Appendix A.
{{ Finally, we would like to mention that the $Z_5$ state is actually not
accessible in the process considered here, so that we refrain here from presenting similar results for this case.}}

Despite the large number of parameters in the fermionic sector (both mixing and Yukawa ones), limited
to the analysis that we are going to perform in this paper, the characteristic of the latter can be easily summarised. As pointed out in
\cite{Barducci:2012kk}, it is sufficient to divide the composite fermion mass spectrum  in two different regimes, as follows.
\begin{itemize}
\item A regime where the mass of the lightest fermionic resonance is too heavy to allow for the decay of a $Z'$  in a pair of heavy fermions and,
consequently, the widths of the $Z'$  are small, typically well below $100$ GeV. This configuration of the 4DCHM {{is illustrated by the forthcoming
benchmarks (b), (d) and (f)  defined in Tabs. 20 and 21 of Ref. \cite{Barducci:2012kk}.}}
\item A regime where a certain number of masses of the new fermionic resonances are light enough to allow for the decay 
of a $Z'$ in a pair of heavy fermions and, consequently, the widths of the involved $Z'$  states are relatively large and can become
even comparable with the masses themselves. This configuration of the 4DCHM is illustrated by the forthcoming {\it colored} benchmarks 
 (green, 
magenta, 
yellow) as given in Tabs. 19 and 22 of Ref. \cite{Barducci:2012kk}.
\end{itemize}

In summary, the parameter  space of the 4DCHM is defined in terms of 13 independent variables, i.e.,
\begin{equation}
\label{eq:parpaper}
f,\; g_*,\; g_0,\; g_{0Y},\; m_*,\; \Delta_{t_L},\; \Delta_{t_R},\; Y_T,\; M_{Y_T},\; \Delta_{b_L},\; \Delta_{b_R},\; Y_B,\; M_{Y_B}.
\end{equation}
In order to constrain it, we have written a  Mathematica routine which considers $f$ and $g_*$ as free parameters and performs a scan over $m_*,\Delta_{t_L},\Delta_{t_R},Y_T,M_{Y_T}$,
$\Delta_{b_L},\Delta_{b_R},Y_B$, $M_{Y_B}$ that is able to find allowed points with respect to the physical constraints $e,M_Z,G_F,$ $m_t,m_b,v,m_H$, 
 the latter being consistent with the recent data coming from the ATLAS \cite{:2012gk} and CMS \cite{:2012gu} experiments: $124 ~{\rm GeV} \le m_H \le 126 ~{\rm GeV}$.
Further notice that we have compared the $W^-t\bar b$, $Zt\bar t$ and $Z b\bar b$ couplings as well to data. In particular
our program also checks that the left- and right-handed couplings of the $Z$ boson to the bottom (anti)quark are separately consistent with results 
of LEP and SLC \cite{Z-Pole}.

In scanning the 4DCHM parameter space, we have of course checked that the regions eventually investigated via our reference process
$pp(q\bar q,gg)\to t\bar t$ 
are compatible with  LHC direct searches for heavy gauge bosons, specifically with the data reported in
 \cite{Aad:2011fe,Chatrchyan:2012meb,Hayden:2012gc,Chatrchyan:2012it}. 
 
Extra gauge bosons give a positive contribution to the Peskin-Takeuchi $S$ parameter and the requirement of consistency 
with the EW Precision Test (EWPT) data generally gives a bound on the mass of these resonances around 
few TeV \cite{Marzocca:2012zn}. However, since we are dealing with a truncated theory describing only the lowest-lying resonances that may exist, 
we need to invoke an UV completion for the physics effects we are not including in our description. These effects could well compensate for $S$, 
albeit with some tuning. One example is given in \cite{Contino:2006qr} by considering the contribution of higher-order operators in the chiral expansion. 
Another scenario leading to a reduced $S$ parameter is illustrated in \cite{DeCurtis:2011yx}, by including non-minimal interactions in the 4DCHM. 
To be on the safe side, our analysis will consider gauge boson masses of the order of 2 TeV or larger, in order to avoid big contributions to the $S$ parameter.
Notice also that the fermionic sector of the 4DCHM
is quite irrelevant for the aforementioned EWPTs since the extra fermions are weakly coupled to the SM gauge bosons.
On the other hand,  these additional fermions, to which we collectively
refer as $t'$ and $b'$, can potentially be produced in hadron-hadron collisions.
The most stringent limits on their mass come presently from the LHC.
An analysis of the compatibility of the 4DCHM with these LHC direct measurements
has been performed. The pair production cross section $\sigma(pp(q\bar q,gg)\rightarrow t' \bar t'/b'\bar b')$ has been calculated according to the code described in \cite{Cacciari:2011hy}, which is essentially the one generally used
to emulate $t\bar t$ production. Herein, a rescaling of its cross section to take into account the non 100\% 
BRs of the $t'$ and $b'$ states into SM-like decay channels owing to the new ones specific to the 4DCHM has been taken into account.
%
%
{{Results for $t'$ and $b'$ essentially limit the heavy quark masses to values in excess of 500 GeV or so, }}
so that we have excluded such masses in the forthcoming parameter scans and in the definition of the benchmark points to be analysed.

\section{Calculation \label{sec:calculation}}
In this section, we present the details of the calculations performed, i.e., the tools used and  the kinematical variables that have been analysed.

\subsection{Tools}
\label{subsec:tools}

The numerical results obtained in the previous section for the 4DCHM spectrum generation and tests against experimental data were based on two
codes, one exploiting Mathematica and the other using the LanHEP/CalcHEP environment \cite{Semenov:1996es}, cross-checked against each other 
where overlapping\footnote{{{These modules have been described in detail in Ref.~\cite{Barducci:2012kk} so we do not dwell on them here.
Further, altough limited to the LanHEP/CalcHEP part, they are available on the High Energy Physics Data-Base~\cite{Brooijmans:2012yi}: see
https://hepmdb.soton.ac.uk/.}}}.

The code exploited for our study of the asymmetries is based on helicity amplitudes, defined through the HELAS
subroutines~\cite{HELAS}, and built up by means of MadGraph~\cite{MadGraph}. Initial state quarks have been taken
as massless whereas for the final state top (anti)quarks we have taken $m_t$ as obtained following the description in the previous section.
The Parton Distribution Functions (PDFs) exploited
were CTEQ6L1~\cite{cteq}, with factorisation/renormalisation
scale set to $Q=\mu\sim M_{Z_{2,3}}$. VEGAS~\cite{VEGAS} was used for the multi-dimensional numerical integrations.

\subsection{Asymmetries}
\label{subsec:asymmetries}

The charge/spin variables that we are going to study have been described in Refs.~\cite{Basso:2012sz} 
(see also \cite{Basso:2012ad})
and we summarise here their salient features. The dependence on the chiral couplings of the asymmetries can be expressed analytically, using helicity formulae from Ref.~\cite{Arai:2008qa} (also derived independently here with the guidance of~\cite{Hagiwara:1985yu}), for a neutral gauge boson exchanged in the $s$-channel.

\subsubsection{Charge asymmetry}\label{subsubsec:charge}
Charge (or spatial) asymmetry is a measure of the symmetry of a process under charge conjugation. Due to the Charge/Parity ({CP}) invariance of the neutral current interactions, this translates into an angular asymmetry at the matrix element level. It can only be generated from the $q\bar{q}$ initial state due to the symmetry of the gluon-gluon system and it can occur via both subtle Next-to-Leading Order (NLO)
QCD effects, as described in detail in~\cite{QCDasymmetry}, as well as more standard EW ones.

The symmetric $pp$ initial state at the LHC necessitates a more suitable definition of such an observable compared to, e.g., the well-known top quark forward-backward asymmetry employed at the Tevatron. 
Several possibilities exist, though, as investigated in~\cite{Basso:2012sz,Basso:2012ad}, the spatial asymmetry that delivers the higher sensitivity is the rapidity dependent forward-backward asymmetry, $A_{RFB}$. It uses the rapidity difference of the final state fermion pair, $\Delta y=|y_{t}|-|y_{\bar{t}}|$, and enhances the $q\bar{q}$ initial state parton luminosity via a cut on the rapidity of the fermion--anti-fermion system, $y_{t\bar{t}}$,
\begin{equation}\label{eqn:asy_AFBSTAR}
    A_{RFB}=\frac{N(\Delta y > 0)-N(\Delta y < 0)}{N(\Delta y > 0)+N(\Delta y < 0)}\Bigg |_{|y_{t\bar{t}}|>y_{t\bar t}^{\rm cut}}
\end{equation}
with, hereafter,  $y^{\rm cut}_{t\bar{t}}=0.5$.

Another charge asymmetry which will turn out to be of relevance (especially to resolve adjacent $Z'$ peaks, which are a characteristic of the 4DCHM) is 
\begin{equation}\label{eqn:asy_AFB}
    A^{\ast}_{FB}=\frac{N(\cos\theta^{\ast} > 0)-N(\cos\theta^{\ast} < 0)}{N(\cos\theta^{\ast} > 0)+N(\cos\theta^{\ast} < 0)},
\end{equation}
where $\cos\theta^{\ast}$ is defined  with $z$-axis in the direction of  $y_{t\bar{t}}$, so that
$\theta^*$ is the polar angle in the $t\bar t$ rest frame, i.e., the Centre-of-Mass (CM) system at parton level, 
to which the entire event can generally be boosted to, no matter the actual final state produced by the $t\bar t$ pair after it decays~\cite{ttreco}. 

These observables can only be generated by a $Z'$ boson if its vector and axial couplings to both the initial ($i$) and final ($t$) state fermions are non-vanishing. 

\subsubsection{Spin asymmetries}\label{subsubsec:spin}
Spin asymmetries focus on the helicity structure of the final state fermions and, when such properties are measurable, display interesting dependences on the chiral structure of the $Z^\prime$ boson couplings. The helicity of a final state can only be experimentally determined for a decaying final state, where the asymmetries are extracted as coefficients in the angular distribution of its decay products. This is well described for the case of top quarks in Ref.~\cite{tt-pol}. As such, our parton level implementation does not represent the full reconstruction and extraction chain of such observables but highlights their potential strength while estimating the reconstruction efficiencies from recent experimental publications. We elaborate on this point in Sec.~\ref{subsec:reco}.
We define two such asymmetries. 

The first observable we consider is the polarisation $A_L$, or single spin asymmetry, defined as follows:
\begin{equation}\label{eqn:asy_AL}
 A_{L}=\frac{N(-,-) + N(-,+) - N(+,+) - N(+,-)}{N_{\rm Total}},
\end{equation}
where  $N$ denotes the number of observed events and its first(second) argument corresponds to the helicity of the final state particle(antiparticle)
whereas ${N_{\rm Total}}$ is the total number of events. It singles out one final state particle, comparing the number of its positive and negative helicities, while summing over the helicities of the other antiparticle (or vice versa).
This observable is proportional to the product of the vector and axial couplings of the final state only and is 
therefore additionally sensitive to their relative sign, a unique feature among asymmetries and cross section observables. 
In other words, it is a measure of the relative `handedness' of the $Z^\prime$ couplings to the final state.

In the case of appreciably massive final states, like the top quark, the spin correlation $A_{LL}$, or double spin asymmetry, is accessible. 
This observable relies on the helicity flipping of either of the final state particles, whose amplitude is proportional to $m_{t}/\sqrt{\hat{s}}$, 
where $\sqrt{\hat{s}}$ is the partonic CM energy, and gives the proportion of like-sign final states against the opposite-sign ones,
\begin{equation}\label{eqn:asy_ALL}
      A_{LL}=\frac{N(+,+) + N(-,-) - N(+,-) - N(-,+)}{\rm N_{Total}}.
\end{equation} 
This observable depends only on the square of the couplings in a similar way to the total cross section. 
In the massless limit $A_{LL}$ becomes maximal, making it a relevant quantity to measure only in the $t\bar{t}$ final state. 
\section{Results \label{sec:results}}
\noindent
{{We split this section in two parts. Firstly, we perform a scan of the 4DCHM parameter space by searching for 
regions  where at least one $Z'\to t\bar t$ signal can be established, assuming $\sqrt s =7,8$ and 14 TeV
for the LHC energy.}}
Secondly, we will study the aforementioned benchmarks defined over such a parameter space,  although limited to the case of maximum energy and
luminosity of the CERN machine.

 \subsection{Reconstruction}\label{subsec:reco}
 While the $t\bar{t}$ channel offers a wide choice of observables that 
 are sensitive to new physics, one of the primary complications of such 
 analyses is the difficulty in reconstructing the 6-body final state 
 that results from the pair production of tops. Ideally, one would 
 perform a full chain of event generation, showering and hadronisation, 
 culminating in a detector simulation to get an accurate representation 
 of the reconstruction process for observables of interest. The 
 associated efficiencies will depend on the information required for the 
 observable and the particular decay channel of the $t\bar{t}$ system. Since our analysis is limited to be at parton level, without subsequent 
 decay of the tops, it is necessary for us to employ reasonable 
 estimates of reconstruction efficiencies such that our qualitative 
 predictions correspond better to the reality of a detector environment. 
 We estimate this quantity in a conservative manner by gauging the 
 selection efficiencies from the requirements of each observable in 
 each decay channel. We take the sum of these and define a net 
 efficiency to reconstruct a $t\bar{t}$ event coming from a high mass 
 ($\sim 2$ TeV) $Z^{\prime}$, weighting by the branching fractions.

 The common experimental requirement between the two asymmetry 
 observables of interest and also the invariant mass distribution is a 
 full reconstruction of the $t\bar{t}$ system. The only extra   
 information needed comes in the form of the angular distributions of 
 the decay products of one or two the tops when extracting the top spin 
 observables. The other important point concerning the analysis of new 
 physics at several TeV is the likely boosted nature of the final states, which will have an impact on 
 the reconstruction process. The collimation of decay products means 
 that many traditionally reliable measurements such as $b$-tagging, 
 invariant mass reconstruction and isolation become hampered and must be 
 adjusted. A variety of pruning and jet substructure methods are applied 
 at the LHC~\cite{boosted} and quote efficiencies of about 30-40\% to 
 tag a hadronic top and a number of analyses have used such methods in 
 recent resonance searches~\cite{boosted-resonance}, showing that 
 including the boosted methods increases sensitivity to higher $Z^{\prime}$ 
 masses. In the ATLAS analyses, for the hadronic case, a weighted 
 efficiency of around 5\% is quoted while in the semi leptonic case, 
 the 
 net reconstruction efficiency plateaus at around 15\% which, when 
 scaled by the 46\% branching fraction, gives around 6\%. As yet, we 
 are not aware of any asymmetry measurements nor analyses in the 
 dilepton channel using these techniques. We therefore choose a total 
 10\% efficiency as a conservative estimate to reconstruct high mass 
 $t\bar{t}$ events, not counting any extra contribution that might come 
 from dilepton analyses.

 The charge asymmetry measurement can be made in any of the three 
 $t\bar{t}$ decay channels and a reconstruction of the top four 
 momenta, after potential top-tagging using boosted methods, is 
 sufficient to 
 obtain the quantity and nothing extra is needed beyond sufficient 
 statistics to represent it as a function of $M_{t\bar{t}}$. We 
 therefore use the same reconstruction efficiency estimate for this 
 observable as used in the resonance searches. The top polarisation 
 asymmetry and spin correlation are more complicated due to the need for reconstructing the 
 angular distributions of decay products. What is clear is that the 
 boosted systems will inhibit the measurement of such quantities as the 
 collimation of the decay products approaches the angular resolution of 
 the calorimeters. At this stage, a lack of experimental analyses makes 
 it difficult to estimate how well such quantities can be measured at 
 high $p_{T}$ although a number of papers discuss the problem and pose 
 potential solutions moving away from the requirement of fully 
 reconstructing the decay products~\cite{tt-pol,boosted-pol}. For this study, 
 we further halve estimate to 5\%, bearing in mind that this is somewhat crude since we 
 cannot assess the process of extracting the coefficient from the decay product 
 kinematics. We feel an estimate of this order is reasonable in that most 
 measurment techniques proposed involve using the semi-leptonic or di-leptonic 
 channels which should not suffer as much dilution to the 100\% spin analysing 
 power of the daughter lepton. However, the question of associated systematic 
 uncertainties is not addressed in this paper.
 It is therefore a given that the results shown here 
 remain of an illustrative nature, showing that this model has the 
 potential to be investigated in the ttbar channel. 
 
\subsection{Parameter scan\label{subsec:scan}}

In order to completely define the parameter space of the 4DCHM where at least one $Z'\rightarrow t\bar t$ signal can be established we have performed a scan over the fermionic parameters of the model\footnote{We remind the reader that 
these are the mass $m_*$, the mixings
$\Delta_{{t,b}_{L,R}}$ and the Yukawas $Y_{T,B},\;M_{Y_{T,B}}$.}
for various choices of the model scale $f$ and gauge coupling constant $g_*$ that, as stressed in Sect.~\ref{sec:model}, completely determine the neutral (and charged) gauge boson mass spectrum. 
The parameters of the model have been varied in a range from $0.5$ TeV to $5$ TeV except, in the spirit of partial compositeness, the mixing parameters of the bottom sector that have been varied
between $0.05$ TeV to $0.5$ TeV (the Yukawa parameters $M_{Y_{T,B}}$ have been varied in a negative range with the same absolute values of the others).

{{Scans have been performed with the use of both MadGraph and CalcHEP for 7, 8 and 14 TeV in presence of the following selection cuts:
\begin{equation}
\frac{M_{Z_2}+M_{Z_3}}{2}-3 \frac{\Gamma_{Z_2}+\Gamma_{Z_3}}{2}<M_{t\bar t}< \frac{M_{Z_2}+M_{Z_3}}{2}+3 \frac{\Gamma_{Z_2}+\Gamma_{Z_3}}{2}
\label{eq:scan-cut1}
\end{equation}
for the lightest $Z'$ bosons
where $M_{t \bar t}=\sqrt{(p_t+p_{\bar t})^2}$ is the invariant mass of the $t\bar t$ final 
state\footnote{Due to the large widths of these $Z'$ in certain region of the parameters space lower/upper bounds on the selection cut
have been imposed to be the maximum/minimum between the ones of eq. (\ref{eq:scan-cut1})
 and $2m_{t}/\sqrt s$.}.}}

The signal $S$ has been defined as the difference between the total cross section $T$ in the 
full 4DCHM and the SM background $B$,
so that interference effects in the $q\bar q$ channel between the $Z'$s and $\gamma,Z$ are taken in account in
the former, and the dimensional significance $\sigma$ has been defined as $S/\sqrt{B}$ (with unit $\sqrt{\rm fb}$). However, the
actual statistical significance is obtained from this through multiplying it by $\sqrt{\epsilon {\cal L}}$, where $\epsilon=10\%$ is
the estimated selection efficiency
for the $t\bar t$ final state, as discussed in the previous section
and 
${\cal L}=5, 20$ and 300 fb$^{-1}$ (for $\sqrt s=7,8$ and 14 TeV, respectively) for
the integrated collider luminosity. 

\begin{figure}[!t]
\centering
\includegraphics[angle=0,width=0.48\textwidth]{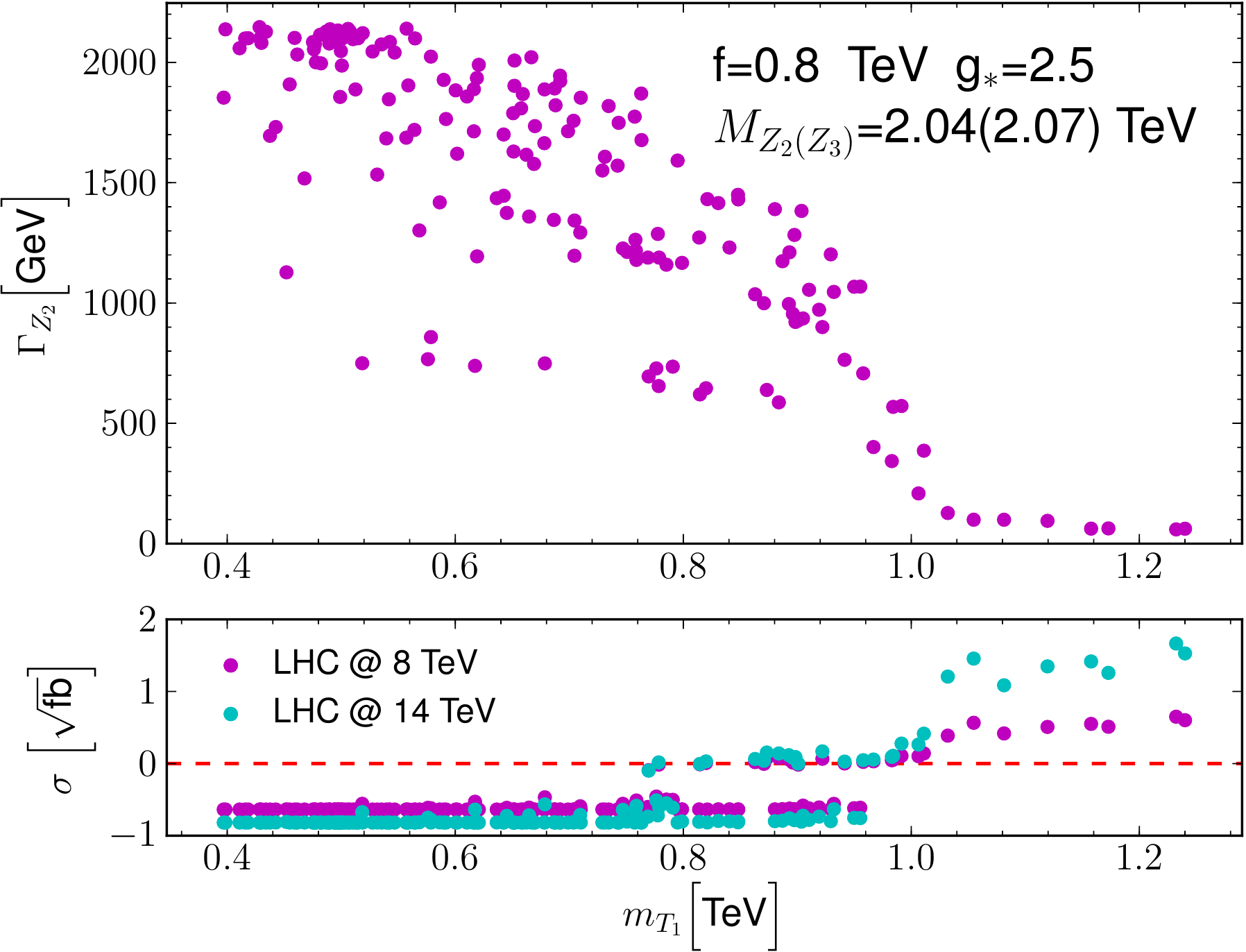}
\includegraphics[angle=0,width=0.48\textwidth]{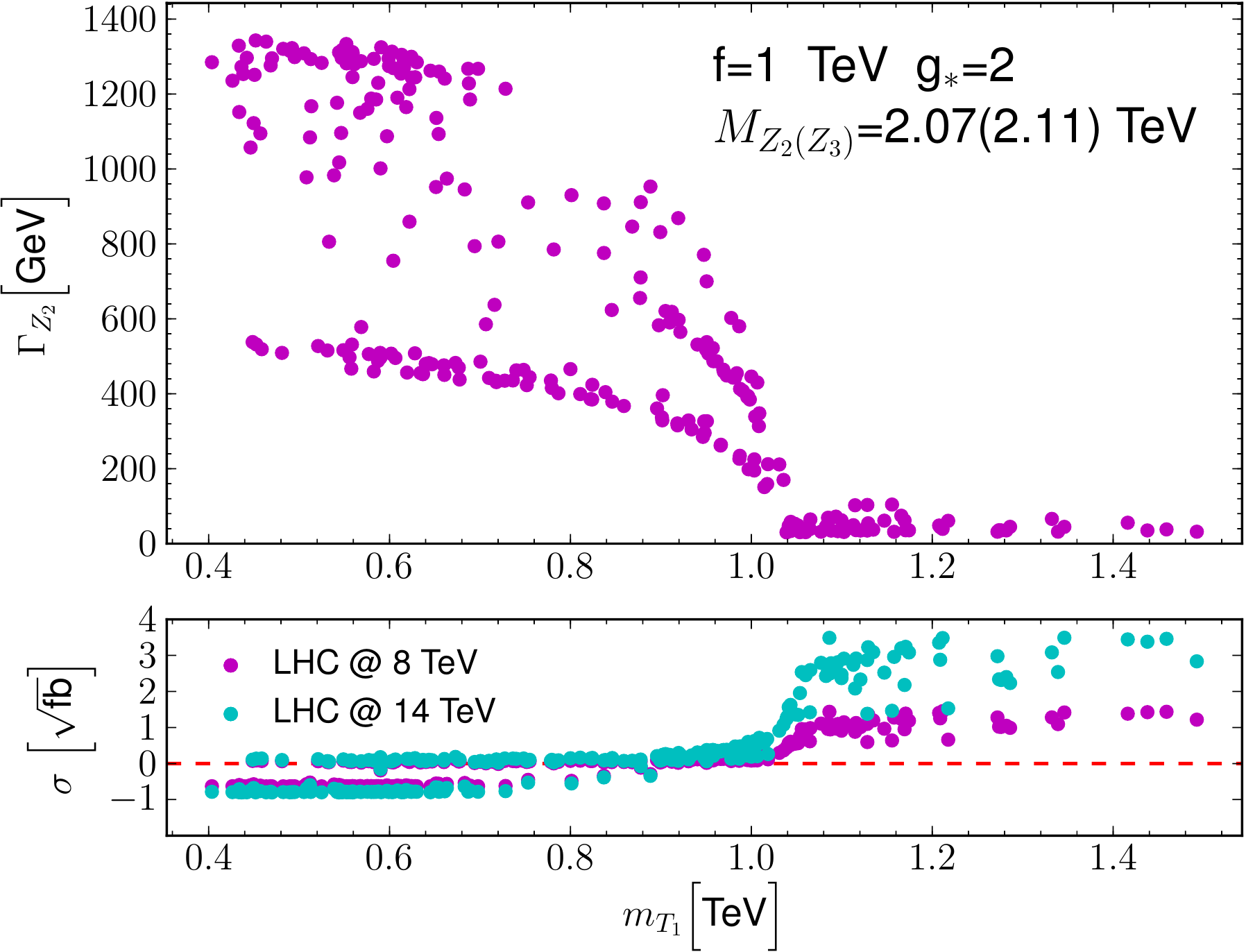}\\
\includegraphics[angle=0,width=0.48\textwidth]{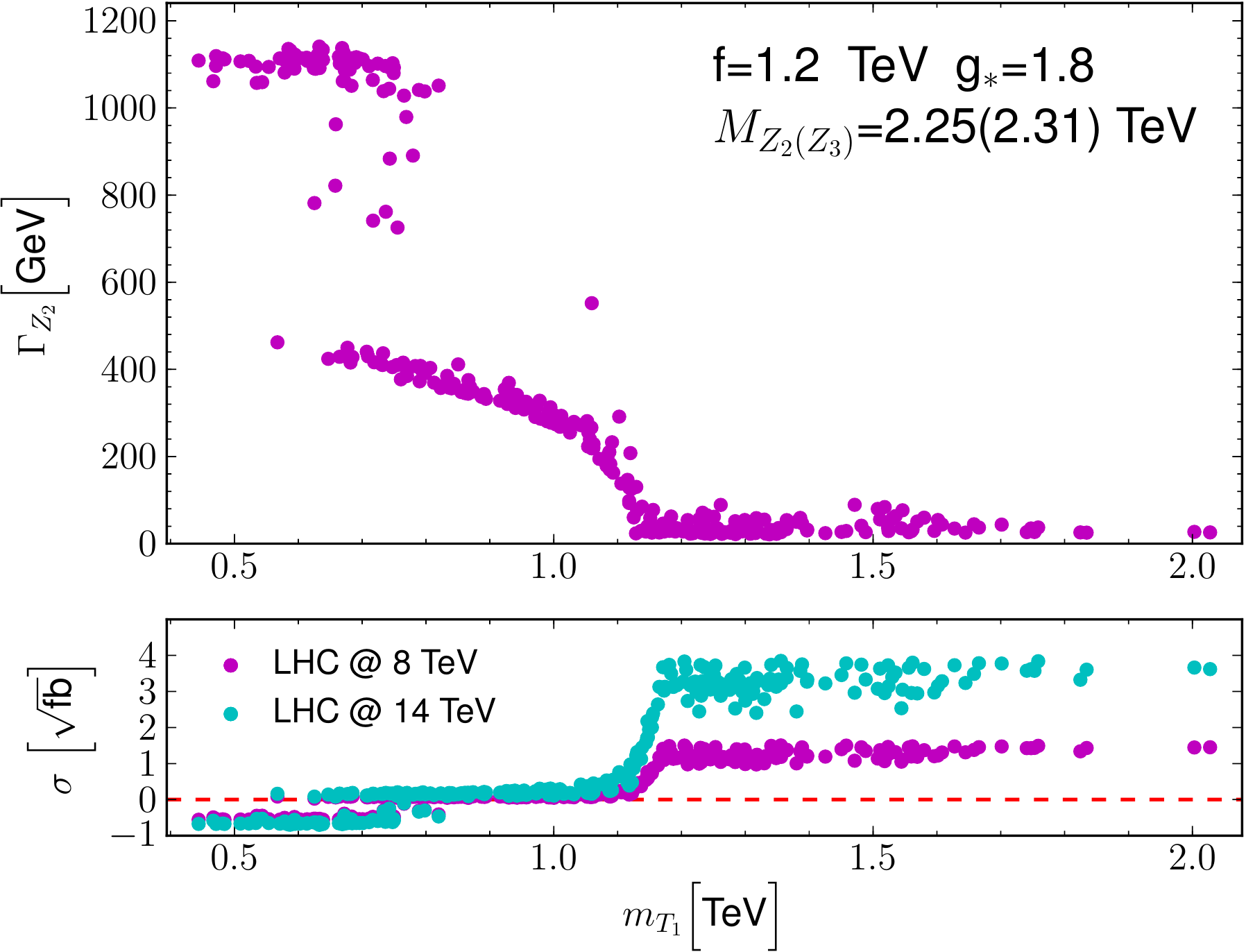}
 \begin{minipage}[b]{0.48\textwidth}
       \hspace{0.05\linewidth}
       \begin{minipage}[b]{0.9\linewidth}      
 \caption{\label{fig:low} \emph{(color online)} Scatter plot in the plane $m_{T_1}/\Gamma_{Z_2}$ for the choice of 
$f=0.8$ TeV and $g_*=2.5$ (top),
$f=1$ TeV and $g_*=2$ (middle)  plus
$f=1.2$ TeV and $g_*=1.8$ (bottom).
We show in the lower frames the relative dimensional significance for $\sqrt s=8$ (purple/dark) and 14 (cyan/light) TeV.}
      \end{minipage}
      \vspace{1cm}
  \end{minipage}
\end{figure}

In Fig. \ref{fig:low} we present the results of the scans for three choices of the model scale $f$ and the coupling
constant $g_*$ in terms of scatter  in the $m_{T_1}$/$\Gamma_{Z_2}$ plane
(results for $\Gamma_{Z_3}$ are very similar), with the corresponding dimensional significance $\sigma$, for the 8 and 14 TeV stages. Notice that in some cases the 
latter is negative, owing to the
fact that, for very large widths, the selection cuts in eqs. (\ref{eq:scan-cut1}) sample large
interference effects which are not positive definite. Results of the scan for the 7 TeV stage of the LHC are not presented since the resulting 
dimensional significances are rather similar to the ones for the 8 TeV stage yet the statistical significances are smaller by a factor of 2 or so.

\begin{figure}[h]
	\centering
	\includegraphics[width=0.45\linewidth]{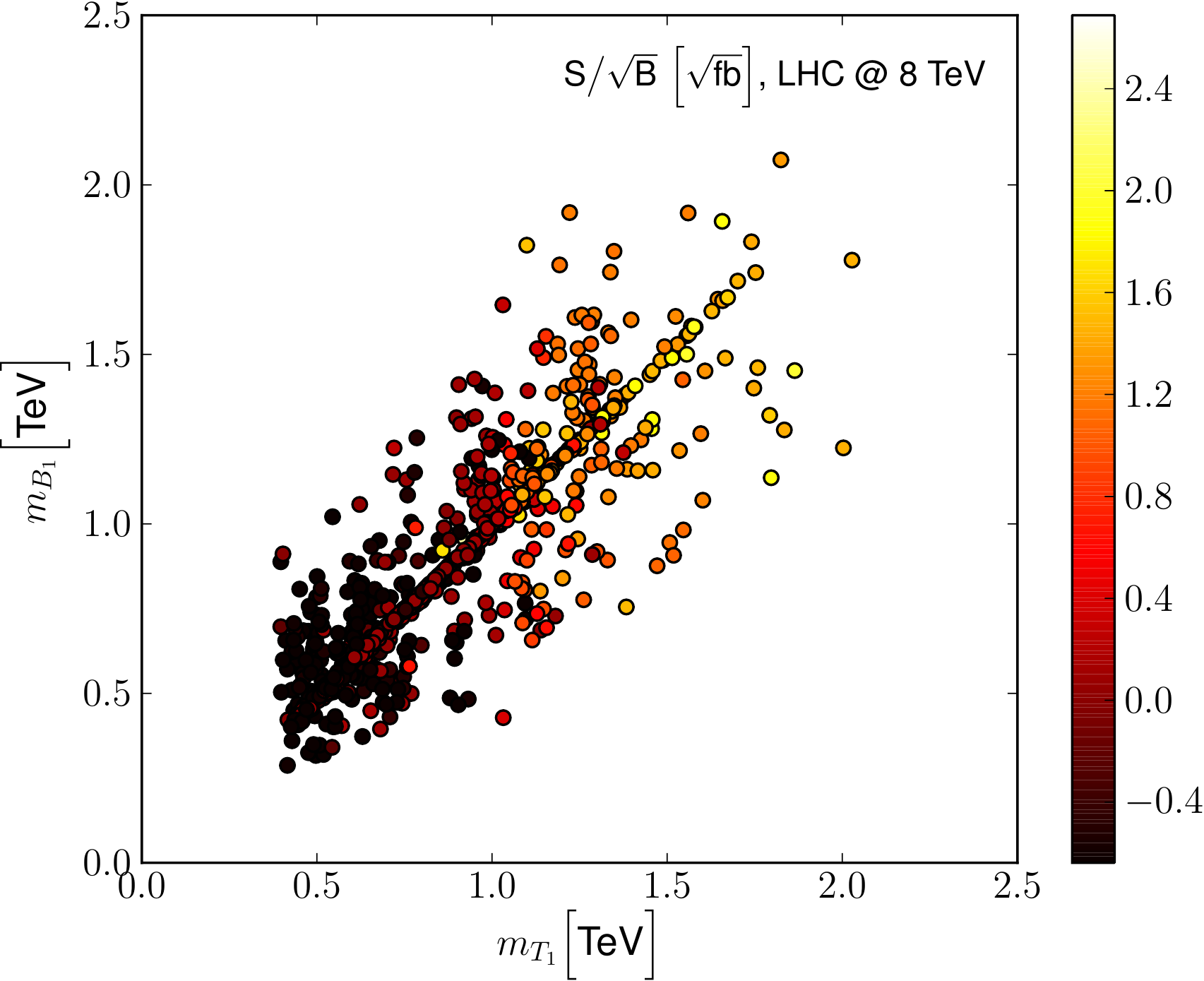}
	\includegraphics[width=0.45\linewidth]{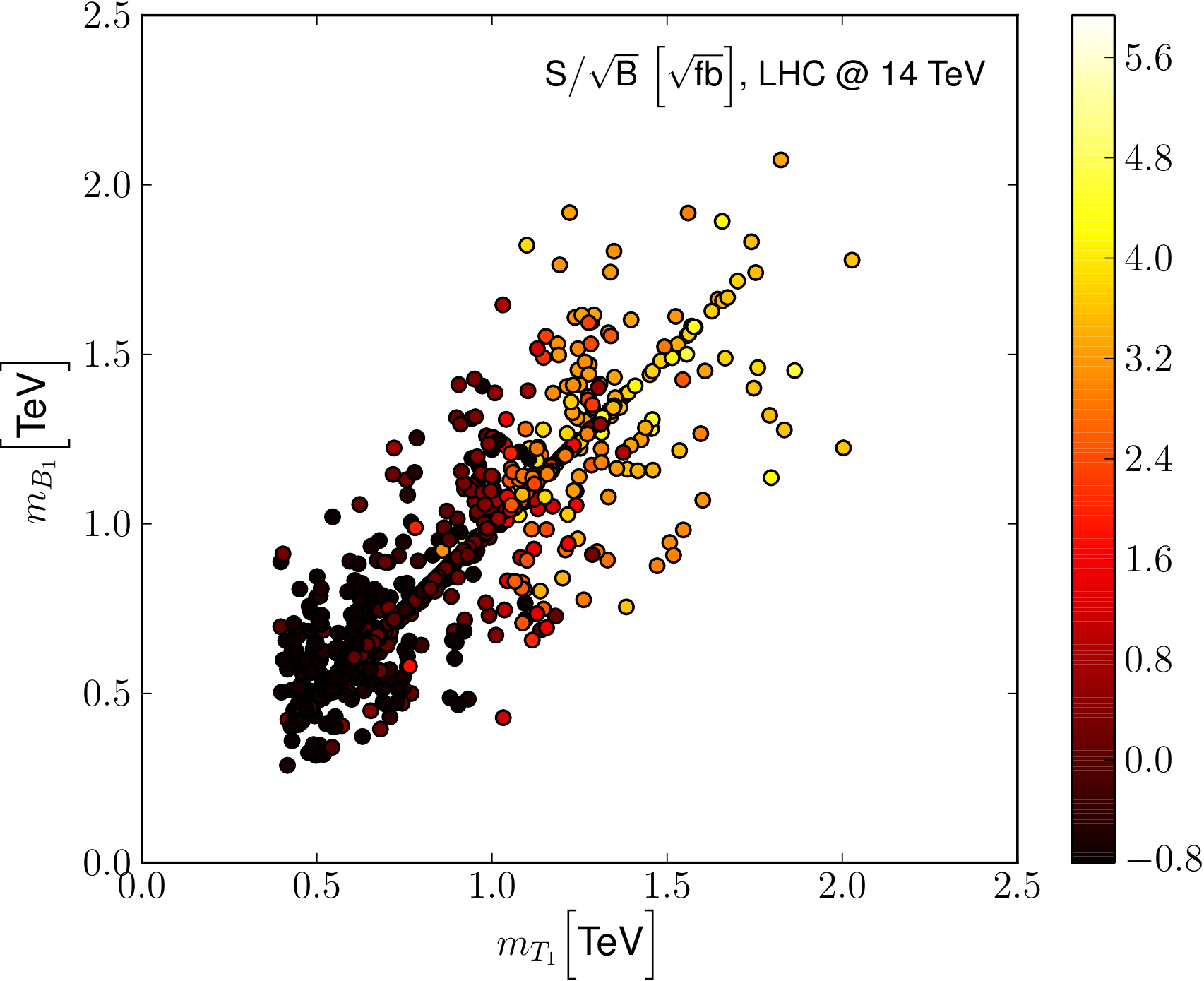}
	\caption{\emph{(color online)} Scatter plot in the plane $m_{T_1}/m_{B_1}$ with the dimensional significance of the collection of all scanned points shown by the color bar
	 for the low mass cut singling out $Z_2$ and $Z_3$ for all benchmarks collected together for the LHC at 8 TeV (left) and 14 TeV (right).}
	\label{fig:all_signif}
\end{figure}

One can see the clear relationship between the mass scale of the heavy third generation partners and the visibility of the resonances 
in that, once their $Z^\prime$ decay channel becomes kinematically accessible, the widths grow substantially and prevent any significant deviations from the SM background. Some branch structure is also apparent in this region, corresponding roughly to when the decay channels into either one or two `generations' of heavy quarks are open. Since $m_{T_1}$ and $m_{B_1}$ are strongly correlated (see Fig.~\ref{fig:all_signif}), it is quite rare for only one out of the pair of lightest top and bottom partners to contribute to the decay width.
In such cases, one is essentially probing the interference effects 
which generically yield negative significances, which we leave this way 
to remind ourselves of this fact.
The off peak effects of such widths of order the gauge boson masses 
themselves have consequences down to very low invariant masses, 
perhaps even near the $t\bar{t}$ threshold. These may not only already 
be constrainable with current LHC data but would certainly require 
analyses with background estimates beyond leading order to have a more 
precise prediction of the overall shape and normalisation of the 
invariant mass spectrum. 
Without this, it is difficult to make meaningful statements about such 
deficits in the production cross section over a large $M_{t\bar{t}}$ 
range and we do not associate any physical meaning to such negative significances.
It is evident that our intended resonant analyses become difficult 
beyond the limit in which the $Z'$s are 
narrow and cannot decay into the heavy fermions. This is further 
emphasised by Fig.~\ref{fig:all_signif}, collecting all scanned points, 
where the correlation of the dimensional significance
with $m_{B_1}$ is also shown, other than with $m_{T_1}$ (as previously 
established). Herein, the reliance of such a significance of the 
$Z^\prime$ signal on a narrow 
resonance hypothesis is evident. 

\subsection{Benchmark studies\label{subsec:benchmarks}}

In this subsection, we will concentrate in some detail on the scope of the $t\bar t$ final state in profiling
$Z'$ bosons of the 4DCHM for the case of the LHC at 14 TeV in energy and 300 fb$^{-1}$ in luminosity, as
the results in the previous subsection clearly highlighted only some limited scope in this respect 
at lower values of $\sqrt s$ and ${\cal L}$. Moreover, the parameter scan has prompted us to focus primarily on 
cases where the resonances remain narrow, although we do show the effects of allowing them to become very wide later. In order to get a feel for the strength of the observables studied, we also define an illustrative measure of `theoretical significance' of an asymmetry 
prediction for the signal $A_{S}$ as the number of standard deviations 
it lies away from the background prediction, $A_{B}$,
\begin{equation}
	s=\frac{|A_{S}-A_{B}|}{\sqrt{\delta A^{2}_{S}+\delta
     A^{2}_{B}}}.\label{eqn:signif}
\end{equation}
which, of course, remains within the scope of our parton-level analysis. As such, they should not be interpreted as true LHC significances but be indicative of the strength of a particular observable.

We will concentrate in particular on the invariant mass distribution of the top-antitop pair, $M_{t\bar t}$, which we will
sample in terms of the cross section as well as of the asymmetries introduced in Subsect.~\ref{subsec:asymmetries}.
In fact, we can anticipate here that the spectra obtained in the case of $A_{RFB}$ are in shape essentially
identical to those displayed by $A_{FB}^*$. Furthermore, on the one hand, the actual value of the asymmetry is
larger in the former case than in the latter, whereas, on the other hand, the total number of
events at disposal to construct it is larger in the latter case than in the former. Overall though, the significance is
always larger for $A_{FB}^*$ than for $A_{RFB}$, so that, hereafter, we will only show  $A_{FB}^*$.

{{Figs.~\ref{fig:b}, \ref{fig:c} and \ref{fig:f} show the differential values of the cross section ($\sigma$) and three asymmetries
($A_L$, $A_{LL}$ and $A_{FB}^*$) as a function of $M_{t\bar t}$, with each figure
referring to the following three benchmarks of \cite{Barducci:2012kk}, respectively:
  (b) $f=0.8$ TeV and $g^*=2.5$;
  (c) $f=1$ TeV and $g^*=2$;
  (f) $f=1.2$ TeV and $g^*=1.8$.
Recall that the mass scale of the two lightest (and nearly degenerate in mass) gauge boson resonances, 
$Z_2$ and $Z_3$, is given by
$
M_{\rm lightest}=f g_*
$ and notice that the heaviest one, $Z_5$, has a mass between 600 GeV and 1 TeV above such a value, depending on the
benchmark. Furthermore, the mass difference between $M_{Z_2}$ and $M_{Z_3}$ is at most 60 GeV or so. }}

These points in parameter space correspond, as intimated in Sect.~\ref{sec:model}, to the case of
small $Z'$ widths, i.e., 
the threshold for the gauge boson decays in pairs of heavy fermions has not been reached, so that
the resonances are rather narrow. Therefore, one may hope to resolve the individual $Z_2$, $Z_3$ and $Z_5$ peaks
in the cross section already. Unfortunately, this is not the case. For a start, once should note that the invariant mass resolution of $t\bar t$ pairs is realistically of order 100 GeV
or so (somewhat better for semileptonic decay channels and somewhat worse for fully hadronic/leptonic ones) so that
it is not generally possible to separate the $Z_2$ and $Z_3$ peaks (they do however cluster together in what looks like
a single wider resonance). Then, it should be appreciated that the (isolated) $Z_5$ peaks never emerges over the background. These two points are made explicit for all benchmarks by the two top frames in Figs.~\ref{fig:b}, \ref{fig:c}  and  \ref{fig:f}.
Herein, the left frames show the differential cross sections binned over (artificially) narrow $M_{t\bar t}$ bins,
of 5 GeV, whereas the right ones use a much larger (and more realistic) 100 GeV resolution.  
Despite this, in most cases, a significance $S/\sqrt B$ larger than 5 can be achieved after an event sample of
${\cal L}=300$ fb$^{-1}$ has been collected for signal ($S$) and background ($B$), i.e., for benchmarks (c) and (f).
For benchmark (b), instead, the significance is only above 3.

Under these circumstances, wherein a detection cannot either be established with enough significance or otherwise it cannot
resolve nearby resonances, the ability to exploit the three asymmetries introduced previously
is crucial. In fact, all of these complement the scope of the cross section, as in all cases they offer a similar level of significance for the signal, 
so the contribution of the former and the latter can be combined to increase significance (where needed), albeit for the case of $Z_2$ and $Z_3$ only, 
not $Z_5$. Furthermore, amongst the asymmetries, $A_{L}$ is unique in offering the chance to separate (in presence of resolution and
efficiency estimates) $Z_2$ and $Z_3$, as the two objects contribute to the asymmetry in opposite directions, unlike the case of $A_{LL}$ and $A_{FB}^\ast$, which establish an excess in the same direction, so that the result is here indistinguishable 
from the case of a lone wider resonance. Such dynamics is exemplified in last three rows of  in  Figs.~\ref{fig:b}, \ref{fig:c} and \ref{fig:f}. 
As discussed in Subsect.~\ref{subsec:asymmetries}, and referring to~\cite{Basso:2012sz}, this distinguishability is owed to the sensitivity 
of $A_{L}$ and $A^{\ast}_{FB}$ to the relative `handedness' of the $Z^\prime$ couplings. For the latter observable, however, 
this sensitivity extends to both the initial and final state which does not give it the same distinguishing power as $A_L$. 
All this is particularly relevant if one notices that it appears a generic feature of this model from the tables in appendix~\ref{sec:appendix} that the $Z_2$ and $Z_3$ have 
predominantly right- and left-handed top couplings, respectively.

As illustrated in Ref.~\cite{Barducci:2012kk}, if one allows for the heavy fermion masses to be lighter than half the 
mass of the $Z'$ states, their widths grow substantially. The aforementioned {\it colored} benchmarks are representative of this
phenomenological situation. They are modifications of the $f=1.2$ TeV and $g_*=1.8$ point.  The corresponding
cross section and asymmetry distributions are found in Figs.~\ref{fig:green}, \ref{fig:magenta} and \ref{fig:yellow}. With a growing width,
the ability to resolve the presence of the $Z_2$ and $Z_3$ resonances degrades substantially and with it also the 
discriminative power of $A_L$  between the two nearby peaks. This is not surprising, as in this case the effects induced
by the two gauge bosons, $Z_2$ and $Z_3$, which are opposite in sign, start overlapping in invariant mass hence
cancelling each other. In contrast, for the cross section and $A_{LL}$ as well as $A_{FB}^*$ this is not the case, so
that these observables are more robust in comparison. Altogether though, $Z_2$ and $Z_3$ signals should 
remain accessible so long that their widths are less than ${\cal O}(10\%)$ of their masses, see the first two {\it colored} benchmarks
(green and magenta). For the other one (yellow), the case in which the masses and widths
are comparable, which is also when the $M_{t\bar t}$ resolution is actually less than $\Gamma_{Z'}$, 
any discovery power unfortunately vanishes.

\section{Conclusions \label{sec:summary}}

In this paper, we have emphasised that the $t\bar t$ final state can be an efficient LHC probe of the neutral gauge sector 
of the 4DCHM that represents a complete framework for the physics of a composite Higgs boson as a 
pseudo-Nambu-Goldstone boson and
which is based on the mechanism of partial compositeness. The latter implies that, alongside the SM gauge bosons, only the third generation quarks (unlike the first two and all leptons)
of the SM are mixed with their composite counterparts, so that the $pp(q\bar q)\to Z'\to t\bar t$ 
process emerges as an obvious discovery channel. 

In fact, we have been able to show that such a mode can enable the detection of two of the three accessible (i.e.,
sufficiently coupled to the initial quarks) $Z'$ bosons of the 4DCHM already by data taken at 7 and 8 TeV,
albeit in limited regions of parameter space, i.e., those with the smallest possible $Z'$ masses, yet compatible with
all current experimental data. Once the CERN machine will reach the 14 TeV stage, detection will be guaranteed essentially up to the kinematical limit of the machine itself, so long that the $Z'$ boson of the 4DCHM are sufficiently narrow, i.e., with
widths being at most 10\% of the masses.

Other than discovering such possible new states, the LHC (at maximal energy and luminosity) could afford one,
under the same width conditions, with the
possibility of profiling the $Z'$s, i.e., of measuring their quantum numbers, thanks to the fact that one can use
$t\bar t$ samples  to define charge and spin asymmetries, which are particularly sensitive to the chiral couplings
of the new gauge bosons. Furthermore, these observables, unlike the cross section, once mapped in
invariant mass, also enable one to separate the two resonances, $Z_2$ and $Z_3$,
that the 4DCHM predicts to be
very close in mass, in fact closer than the standard 
mass resolution afforded by top-antitop pairs. In fact, one could exploit the cancellation effect observed in $A_{L}$ but not in $A_{FB}^{\ast}$ to deduce the presence of nearly degenerate resonances without relying on the appearance of their distributions in $M_{t\bar{t}}$ by correlating the two observed asymmetries and comparing to predictions from single resonances as shown in~\cite{AADD}.
  
We have reached these conclusions after including  both EW and QCD backgrounds (the latter dominated by
$pp(gg)\to t\bar t$), including interference effects (where applicable, i.e.,
in the $pp(q\bar q)\to t\bar t$ subprocess), through a parton level simulation, in presence 
of realistic detector resolution and statistical error estimates. In this connection,
before closing, we should acknowledge that in drawing our conclusions we have
neglected systematic uncertainties \cite{syst}. However, we do not expect these to undermine our results. The reason is twofold. 
On the one hand, the already wide mass resolutions exploited here are expected to
milden their actual phenomenological impact and furthermore one could equally exploit transverse
momentum spectra, which are less sensitive to such effects in comparison to the invariant mass. 
On the other hand, by the time the LHC will reach the 14 TeV energy stage
and accrue 300 fb$^{-1}$ of luminosity, where our most interesting results are
applicable, the systematics will be much better understood than at present. 
Under any circumstances though, the inclusion of systematics would require detailed detector simulations
which were clearly beyond the scope of this paper\footnote{A validation of our results using a simulation
of $t\bar t$ decays, followed by parton shower and hadronisation, also in presence of detector effects, is
currently in progress.}. 

\section*{Acknowledgements}
We are grateful to A. Belyaev and G.M. Pruna for discussions.
The work of DB, KM and SM is partially supported through the NExT Institute.

\appendix
\section{4DCHM Benchmark points}\label{sec:appendix}
In this Appendix we list the exact numerical values of the masses, widths (limited to the new resonances) and couplings of the $Z, Z_2$ and $Z_3$ gauge bosons to the light quarks ($u,d,c,s$) and also to the top quark in terms of left- and right-handed coefficients defined in eq.~(\ref{LNC}) and eq.~(\ref{eq:lag_nc_top}) for the benchmark points of \cite{Barducci:2012kk} adopted in this work. We neglect here the case of the $Z_5$ state, as we have shown this to
be inaccessible in the process studied in this paper. Such values are reported in Tabs.~\ref{tab:mass1}-\ref{tab:zttcoup2}.
\begin{table}[htb]
\captionsetup[subfloat]{labelformat=empty,position=top}{(b)}
\centering
\begin{tabular}{||c||c||c||}
\hline
\hline 
 		&	$M_{Z_i}$(GeV)	&	$\Gamma_{Z_i}$(GeV)	\\
\hline
\hline
$Z_2$		&			2048 &	61			\\
$Z_3$		&			2068 &	98			\\
\hline
\hline
\end{tabular}\\
\captionsetup[subfloat]{labelformat=empty,position=bottom}{(c)}
\centering
\begin{tabular}{||c||c||c||}
\hline
\hline
 		&	$M_{Z_i}$(GeV)	&	$\Gamma_{Z_i}$(GeV)	\\
\hline
\hline
$Z_2$		&			2066 &	39			\\
$Z_3$		&			2111 &	52			\\
\hline
\hline
\end{tabular}
\captionsetup[subfloat]{labelformat=empty,position=top}{(f)}
\centering
\begin{tabular}{||c||c||c||}
\hline
\hline
 		&	$M_{Z_i}$(GeV)	&	$\Gamma_{Z_i}$(GeV)	\\
\hline
\hline
$Z_2$		&			2249 &	32			\\
$Z_3$		&			2312 &	55			\\
\hline
\hline
\end{tabular}\\
\caption{Table of the masses and widths of the neutral gauge resonances limited to $Z_2$ and $Z_3$ for the benchmarks of \cite{Barducci:2012kk} with
$f=0.8$ TeV,  $g_*=2.5$ (b), $f=1$ TeV, $g_*=2$ (c)  and $f=1.2$ TeV, $g_*=1.8$ (f).}
\label{tab:mass1}
\end{table}
\begin{table}[htb]
\captionsetup[subfloat]{labelformat=empty,position=top}{(green)}
\centering
\begin{tabular}{||c||c||c||}
\hline
\hline
 		&	$M_{Z_i}$(GeV)	&	$\Gamma_{Z_i}$(GeV)	\\
\hline
\hline
$Z_2$		&			2249 &	48			\\
$Z_3$		&			2312 &	86			\\
\hline
\hline
\end{tabular}\\
\captionsetup[subfloat]{labelformat=empty,position=bottom}{(magenta)}
\centering
\begin{tabular}{||c||c||c||}
\hline
\hline
 		&	$M_{Z_i}$(GeV)	&	$\Gamma_{Z_i}$(GeV)	\\
\hline
\hline
$Z_2$		&			2249 &	75			\\
$Z_3$		&			2312 &	104			\\
\hline
\hline
\end{tabular}
\captionsetup[subfloat]{labelformat=empty,position=top}{(yellow)}
\centering
\begin{tabular}{||c||c||c||}
\hline
\hline
 		&	$M_{Z_i}$(GeV)	&	$\Gamma_{Z_i}$(GeV)	\\
\hline
\hline
$Z_2$		&			2249 &	1099			\\
$Z_3$		&			2312 &	822			\\
\hline
\hline
\end{tabular}\\
\caption{Table of the masses and widths of the neutral gauge resonances limited to $Z_2$ and $Z_3$ for the $colored$ benchmark points of \cite{Barducci:2012kk} in green, magenta and yellow.}
\label{tab:mass2}
\end{table}
\begin{table}[htb]
\captionsetup[subfloat]{labelformat=empty,position=top}{(b)}
\centering
\begin{tabular}{||c||c||c||c||c||}
\hline
\hline
 		&	$g^L_{Z_i}(u,c)$	&	$g^R_{Z_i}(u,c)$ & $g^L_{Z_i}(d,s)$	&	$g^R_{Z_i}(d,s)$ 	\\
\hline
\hline
$Z$		&	0.256		      &	$-0.115$         &	$-0.313$		& 	0.057				\\
$Z_2$		&	0.0075		&	 0.048         &	0.017			&	$-0.024$			\\
$Z_3$		&	$-0.086$		&	$-0.004$	   &	0.084			&	0.002			\\
\hline
\hline
\end{tabular}\\
\captionsetup[subfloat]{labelformat=empty,position=bottom}{(c)}
\centering
\begin{tabular}{||c||c||c||c||c||}
\hline
\hline
 		&	$g^L_{Z_i}(u,c)$	&	$g^R_{Z_i}(u,c)$ & $g^L_{Z_i}(d,s)$	&	$g^R_{Z_i}(d,s)$ 	\\
\hline
\hline
$Z$		&	0.256			&	$-0.115$         &	$-0.313$		&	0.057			\\
$Z_2$		&	0.012	    		&	0.061	         &	0.019			&	$-0.031$			\\
$Z_3$		&	-$0.110$		&	$-0.003$	   &	0.109			&	0.002			\\
\hline
\hline
\end{tabular}\\
\captionsetup[subfloat]{labelformat=empty,position=top}{(f)}
\centering
\begin{tabular}{||c||c||c||c||c||}
\hline
\hline
 		&	$g^L_{Z_i}(u,c)$	&	$g^R_{Z_i}(u,c)$ & $g^L_{Z_i}(d,s)$	&	$g^R_{Z_i}(d,s)$ 	\\
\hline
\hline
$Z$		&	0.256		      &	$-0.115$         &	$-0.313$		&   0.057				\\
$Z_2$		&	0.015		      &	0.069	         &	0.020			&   $-0.034$			\\
$Z_3$		&	$-0.125$		&	$-0.002$	   &	0.123			&	0.001			\\
\hline
\hline
\end{tabular}\\
\caption{Table of the couplings of the up and down quark to the neutral sector limited to $Z,Z_2$ and $Z_3$ for the benchmarks of \cite{Barducci:2012kk} with
$f=0.8$ TeV,  $g_*=2.5$ (b), $f=1$ TeV, $g_*=2$ (c)  and $f=1.2$ TeV, $g_*=1.8$ (f).}
\label{tab:light1}
\end{table}
\begin{table}[htb]
\captionsetup[subfloat]{labelformat=empty,position=top}{($colored$)}
\centering
\begin{tabular}{||c||c||c||c||c||}
\hline
\hline
 		&	$g^L_{Z_i}(u,c)$	&	$g^R_{Z_i}(u,c)$ & $g^L_{Z_i}(d,s)$	&	$g^R_{Z_i}(d,s)$ 	\\
\hline
\hline
$Z$		&	0.256		      &	$-0.115$     	   &	$-0.313$		&   0.057					\\
$Z_2$		&	0.015			&	 0.069     	   &	0.020			&  $-0.034$				\\
$Z_3$		&	$-0.125$		&	$-0.002$   	   &	0.123			&   0.001				\\
\hline
\hline
\end{tabular}\\
\caption{Table of the couplings of the up and down quark to the neutral sector limited to $Z,Z_2$ and $Z_3$ for the $colored$ benchmark points of \cite{Barducci:2012kk} in green, magenta and yellow. Since these couplings depend only on the two parameters $f$ and $g_*$ they are equal for all the three  $colored$ benchmark considered.}
\label{tab:light2}
\end{table}
\begin{table}[htb]
\captionsetup[subfloat]{labelformat=empty,position=top}{(b)}
\centering
\begin{tabular}{||c||c||c||}
\hline
\hline
 		&	$g^L_{Z_i}(t)$	&	$g^R_{Z_i}(t)$	\\
\hline
\hline
$Z$		&			0.248 &	$-0.123$			\\
$Z_2$		&			$-0.108$ &	$-0.603$			\\
$Z_3$		&			0.481 &	0.009			\\
\hline
\hline
\end{tabular}\\
\captionsetup[subfloat]{labelformat=empty,position=bottom}{(c)}
\centering
\begin{tabular}{||c||c||c||}
\hline
\hline
 		&	$g^L_{Z_i}(t)$	&	$g^R_{Z_i}(t)$	\\
\hline
\hline
$Z$		&	0.251   &	$-0.120$	\\
$Z_2$		&	$-0.091$  &	$-0.571$	\\
$Z_3$		&	0.377   &	0.006		\\
\hline
\hline
\end{tabular}
\captionsetup[subfloat]{labelformat=empty,position=top}{(f)}
\centering
\begin{tabular}{||c||c||c||}
\hline
\hline
 		&	$g^L_{Z_i}(t)$	&	$g^R_{Z_i}(t)$	\\
\hline
\hline
$Z$		&			0.252	& $-0.118$			\\
$Z_2$		&			$-0.106$ &	$-0.486$			\\
$Z_3$		&			0.427	 & 0.006			\\
\hline
\hline
\end{tabular}
\caption{Table of the couplings of the top quark to the neutral sector limited to $Z,Z_2$ and $Z_3$ for the benchmarks of \cite{Barducci:2012kk} with
$f=0.8$ TeV,  $g_*=2.5$ (b), $f=1$ TeV, $g_*=2$ (c)  and $f=1.2$ TeV, $g_*=1.8$ (f).}
\label{tab:zttcoup1}
\end{table}
\begin{table}[htb]
\captionsetup[subfloat]{labelformat=empty,position=top}{(green)}
\centering
\begin{tabular}{||c||c||c||}
\hline
\hline
 		&	$g^L_{Z_i}(t)$	&	$g^R_{Z_i}(t)$	\\
\hline
\hline
$Z$		&			0.251	& $-0.117$			\\
$Z_2$		&			$-0.143$ &	$-0.617$			\\
$Z_3$		&			0.591	& 0.010			\\
\hline
\hline
\end{tabular}\\
\captionsetup[subfloat]{labelformat=empty,position=top}{(magenta)}
\centering
\begin{tabular}{||c||c||c||}
\hline
\hline
 		&	$g^L_{Z_i}(t)$	&	$g^R_{Z_i}(t)$	\\
\hline
\hline
$Z$		&			0.251	& $-0.117$			\\
$Z_2$		&			$-0.162$ &	$-0.694$			\\
$Z_3$		&			0.666	& 0.0118			\\
\hline
\hline
\end{tabular}
\captionsetup[subfloat]{labelformat=empty,position=top}{(yellow)}
\centering
\begin{tabular}{||c||c||c||}
\hline
\hline
 		&	$g^L_{Z_i}(t)$	&	$g^R_{Z_i}(t)$	\\
\hline
\hline
$Z$		&			0.248 &	$-0.120$			\\
$Z_2$		&			$-0.190$ &	$-0.790$			\\
$Z_3$		&			0.795 &	0.027			\\
\hline
\hline
\end{tabular}
\caption{Table of the couplings of the top quark to the neutral sector limited to $Z,Z_2$ and $Z_3$ for the $colored$ benchmark points of \cite{Barducci:2012kk} in green, magenta and yellow.}
\label{tab:zttcoup2}
\end{table}

\clearpage\thispagestyle{empty}


\clearpage\thispagestyle{empty}

\begin{figure}[h!]
\centering
\includegraphics[angle=0,width=0.425\linewidth ]{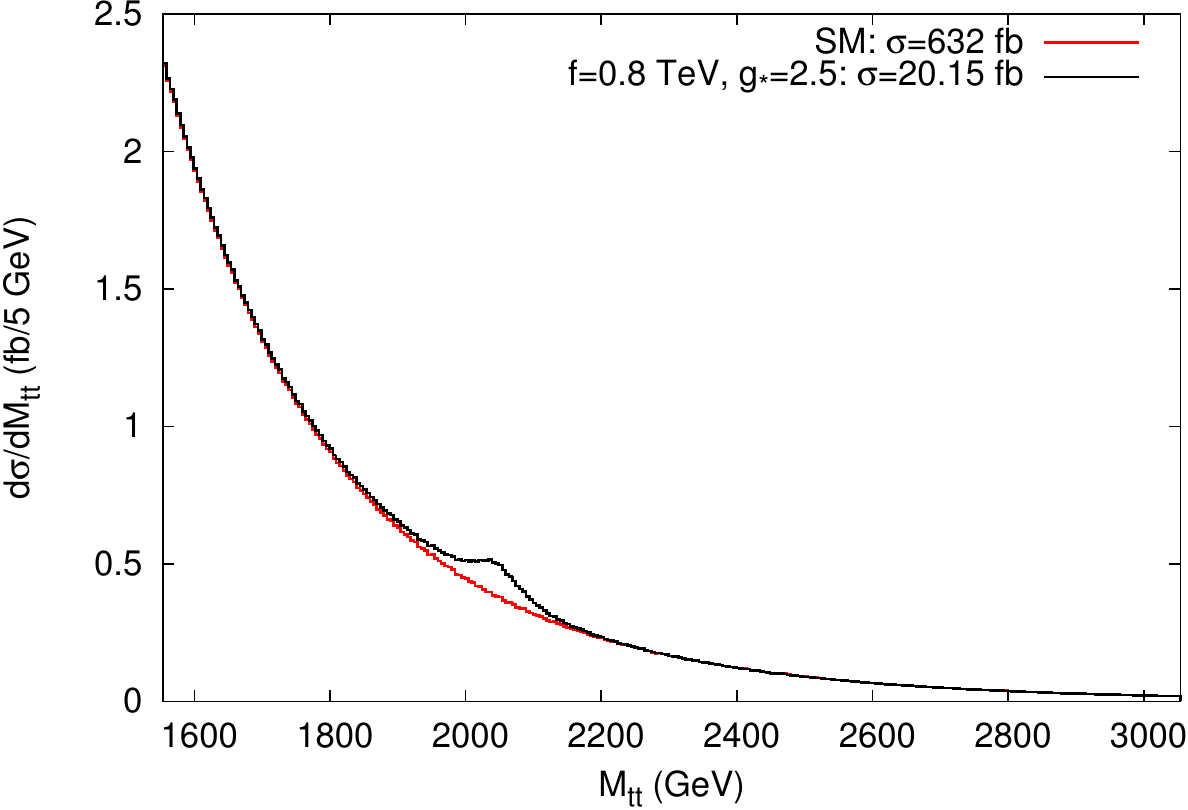}
\includegraphics[angle=0,width=0.45\linewidth ]{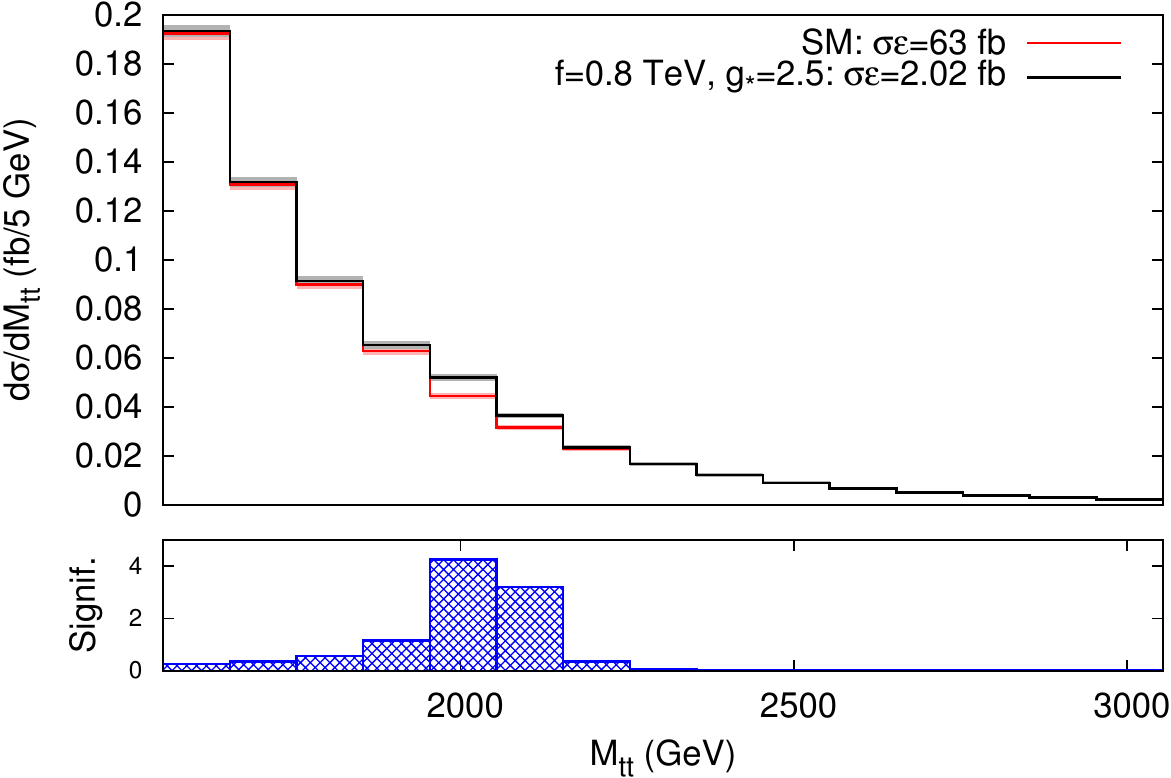}\\
\includegraphics[angle=0,width=0.425\linewidth ]{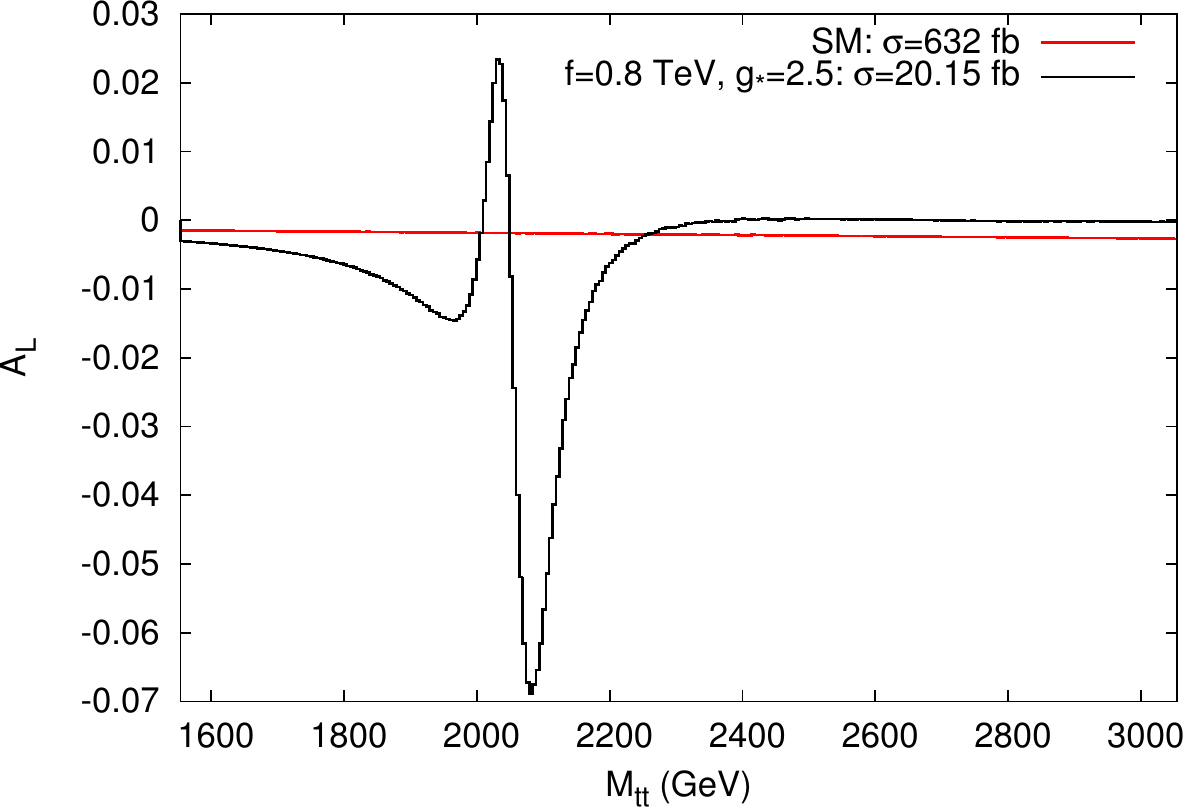}
\includegraphics[angle=0,width=0.45\linewidth ]{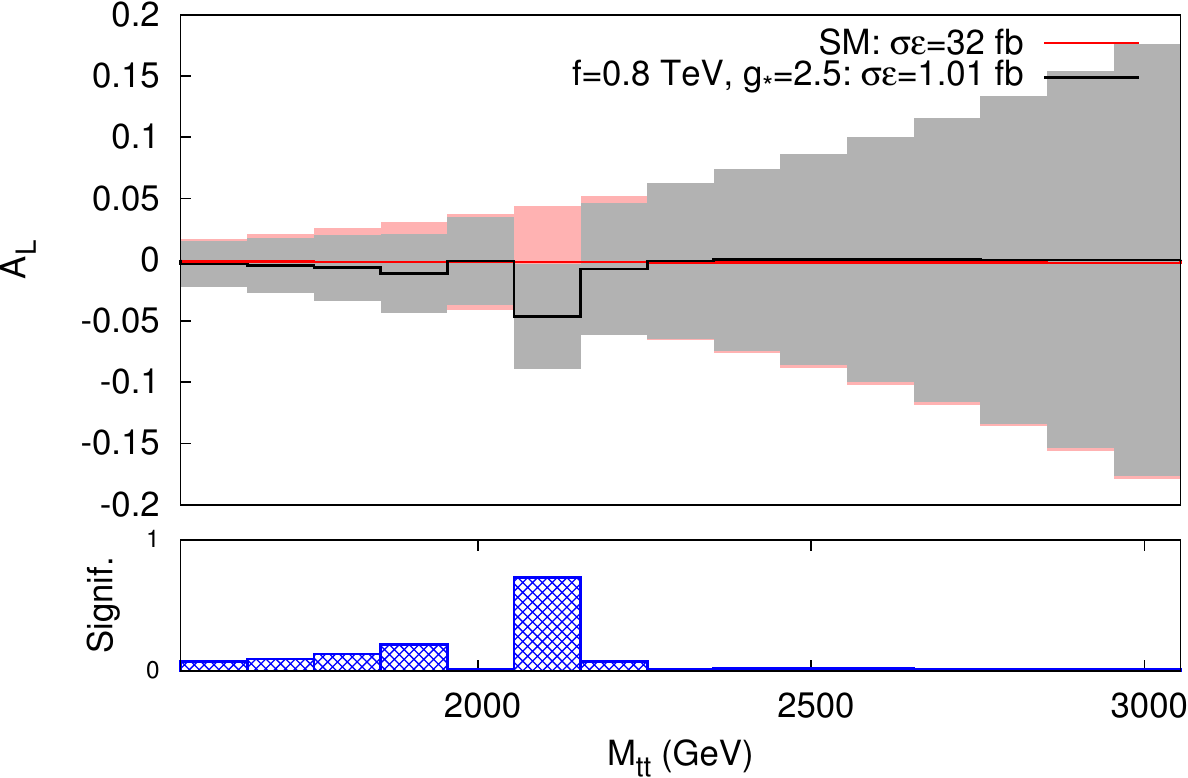}\\
\includegraphics[angle=0,width=0.425\linewidth ]{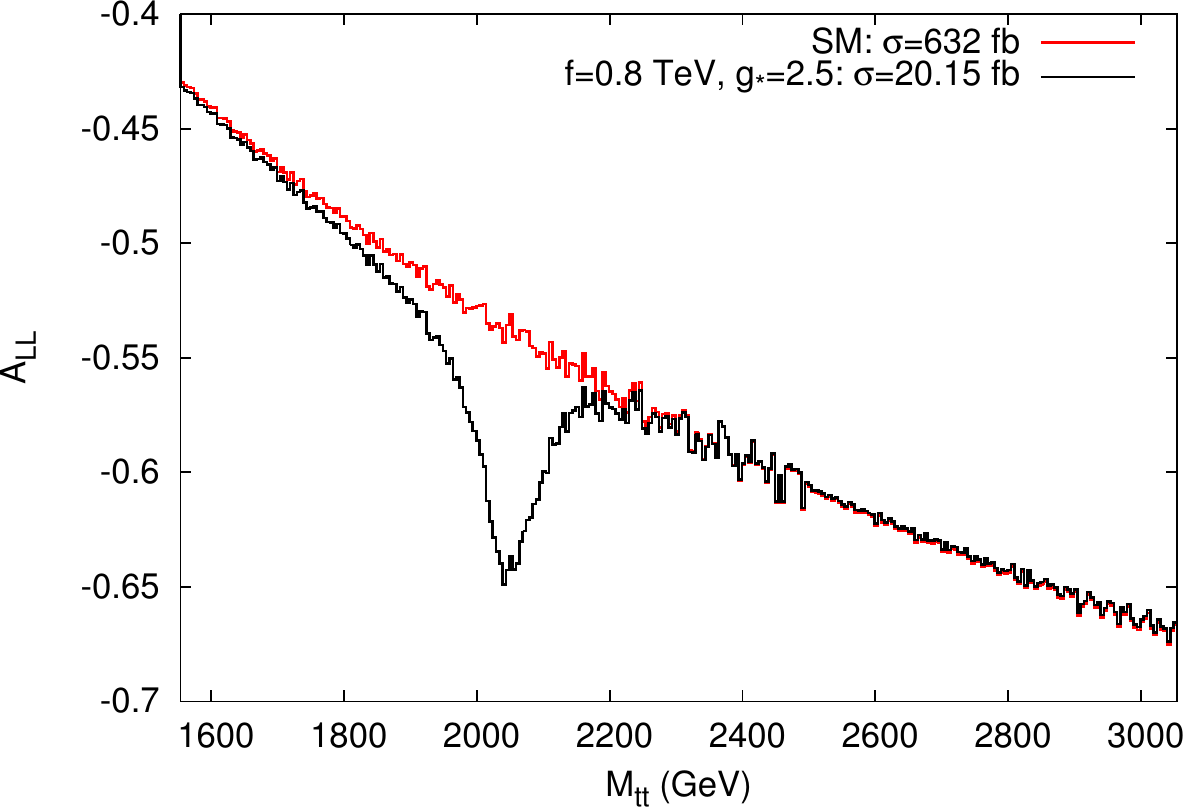}
\includegraphics[angle=0,width=0.45\linewidth ]{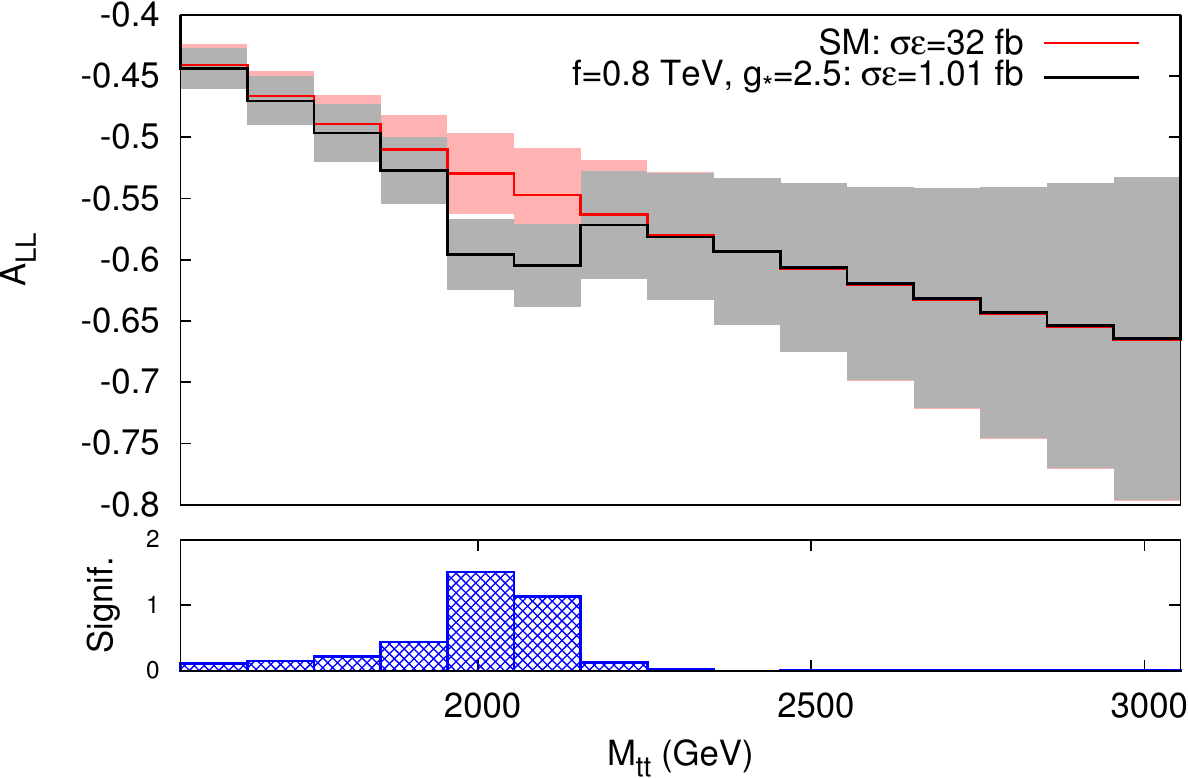}\\
\includegraphics[angle=0,width=0.425\linewidth ]{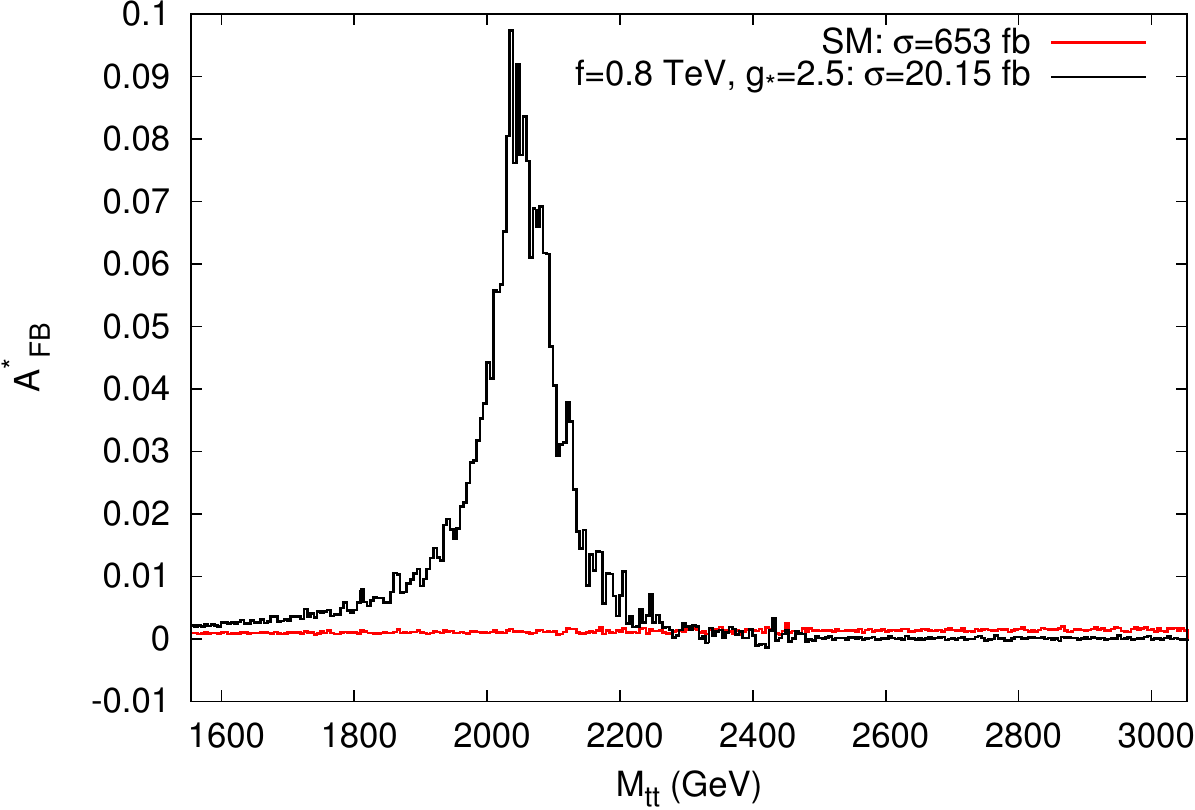}
\includegraphics[angle=0,width=0.45\linewidth ]{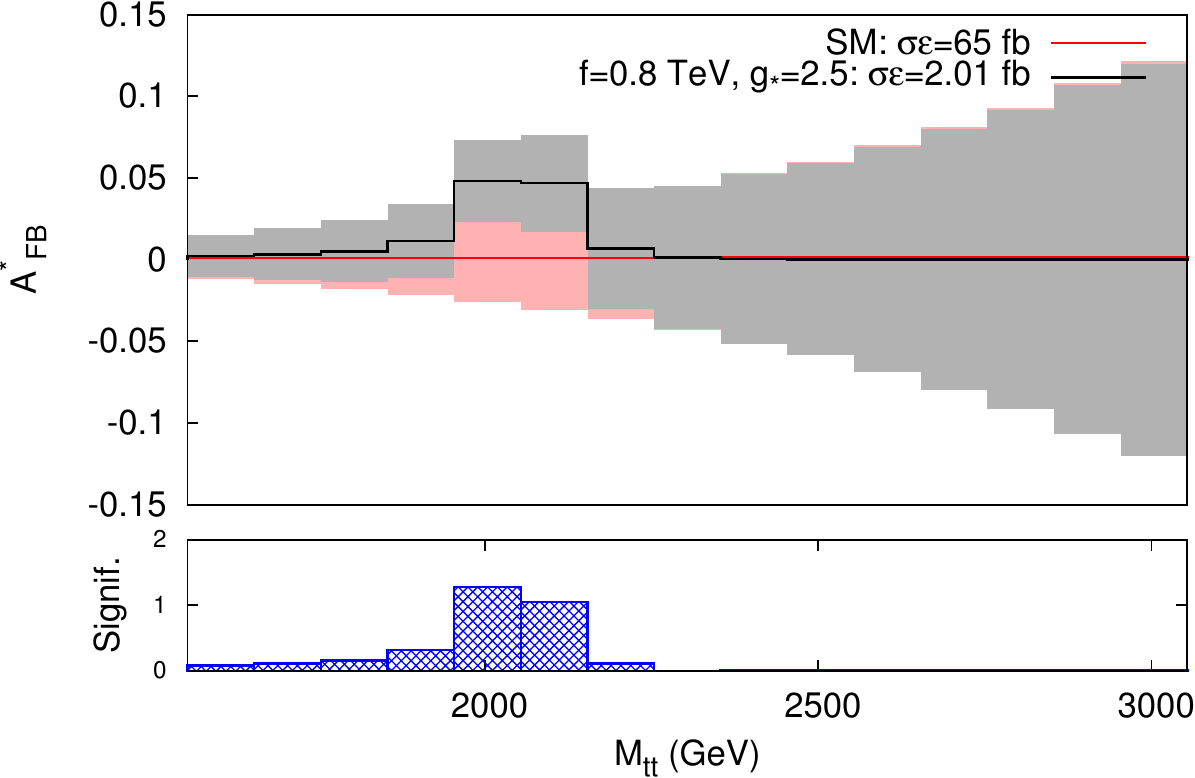}
\caption{\emph{(color online)} Cross section and asymmetries as a function of the $t\bar t$ invariant mass   for the $f=$0.8 TeV, $g_*=$2.5 benchmark at the 14 TeV LHC with 300 fb$^{-1}$.
The left column shows the fully differential observable. 
Right plots (upper frames) include estimates of statistical uncertainty assuming a realistic 
100 GeV mass resolution and also display (lower frames) the theoretical significance assuming a 10\% reconstruction efficiency. Grey(Pink) shading represents the (statistical) error on the 4DCHM(SM) rates, in black(red) solid lines. Masses and widths of the gauge bosons are $M[\Gamma]_{Z_2,Z_3}=2048[61]~{\rm GeV},2068[98]~{\rm GeV}$.}
\label{fig:b}
\end{figure}

\clearpage\thispagestyle{empty}


\clearpage\thispagestyle{empty}

\begin{figure}[h!]
\centering
\includegraphics[angle=0,width=0.425\linewidth ]{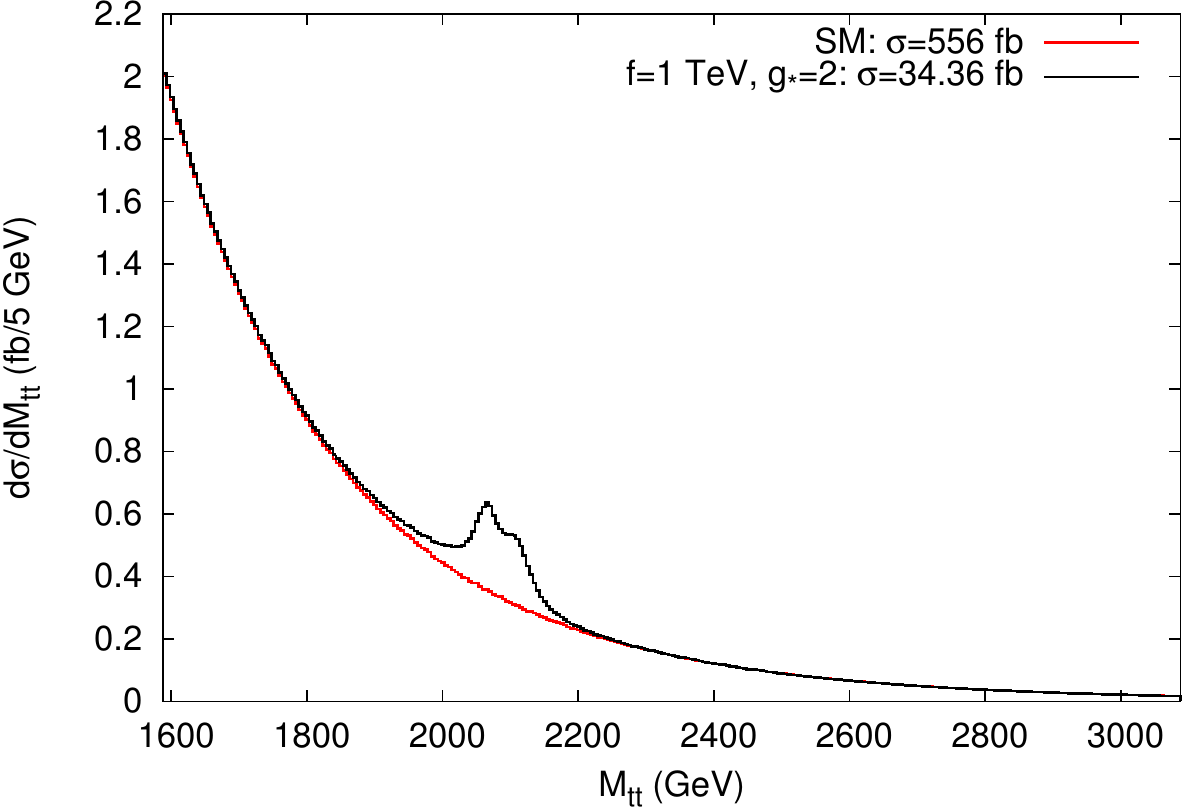}
\includegraphics[angle=0,width=0.45\linewidth ]{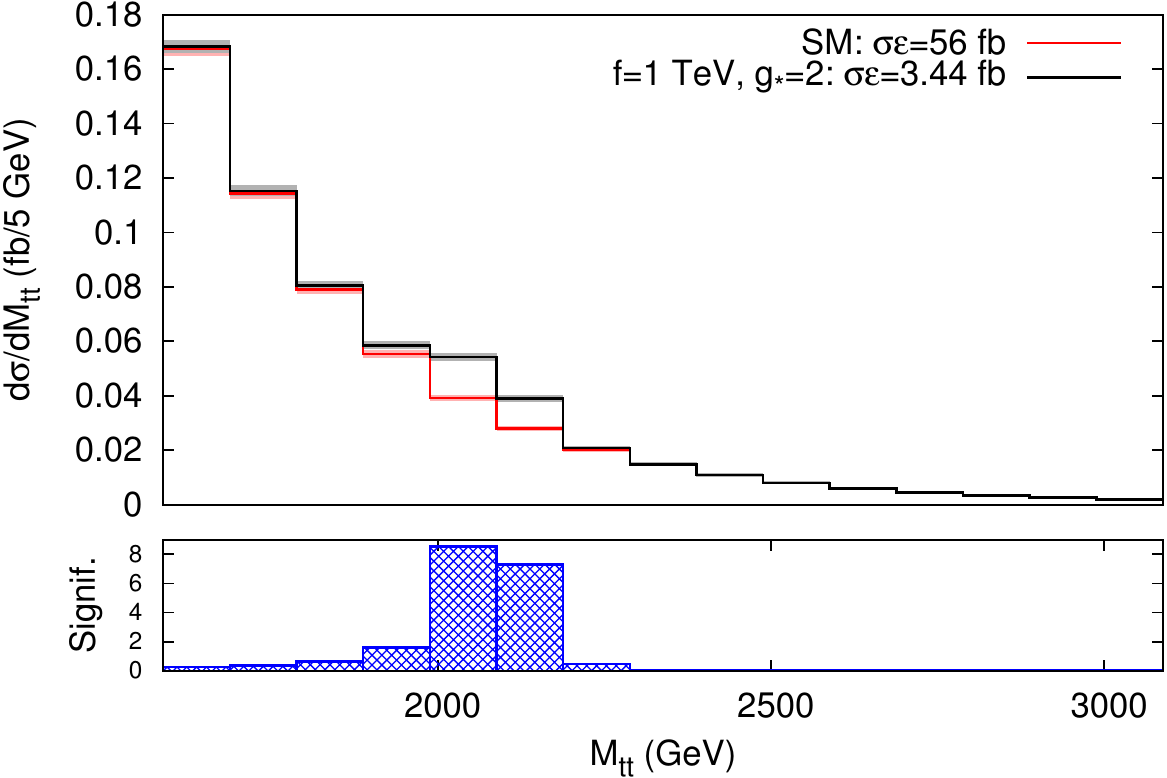}\\
\includegraphics[angle=0,width=0.425\linewidth ]{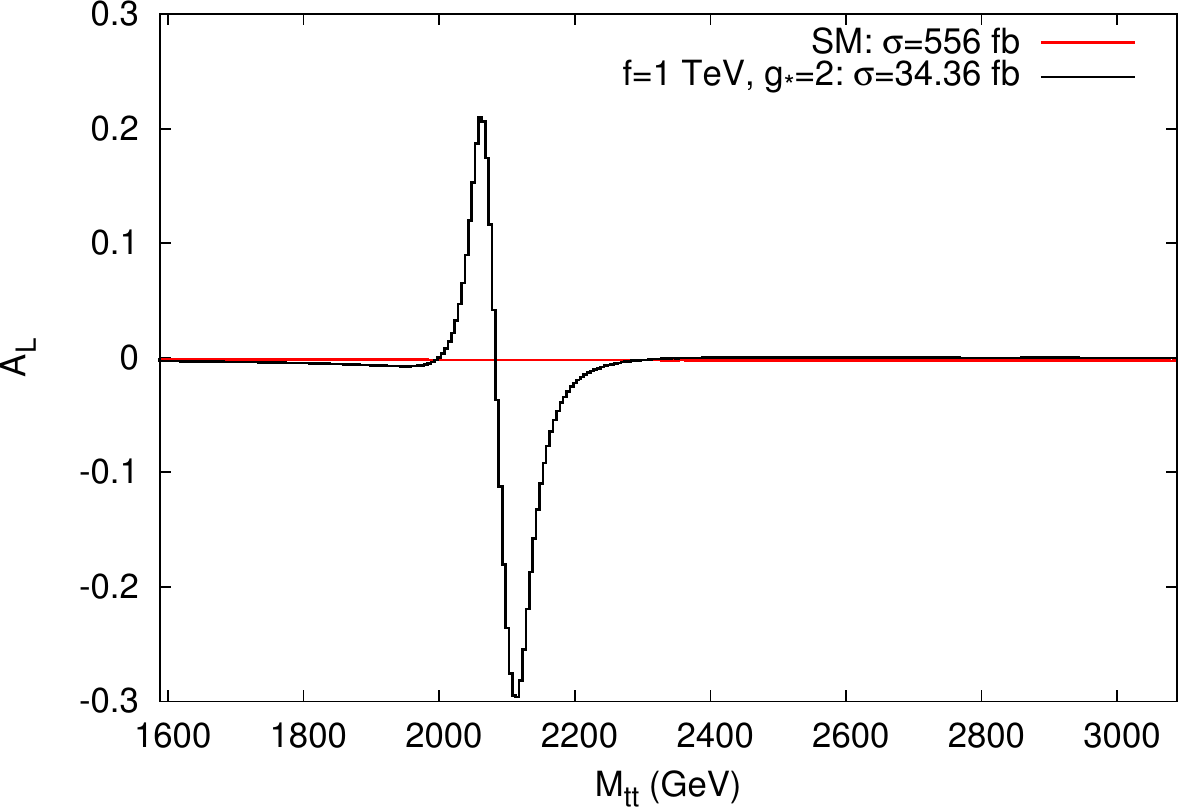}
\includegraphics[angle=0,width=0.45\linewidth ]{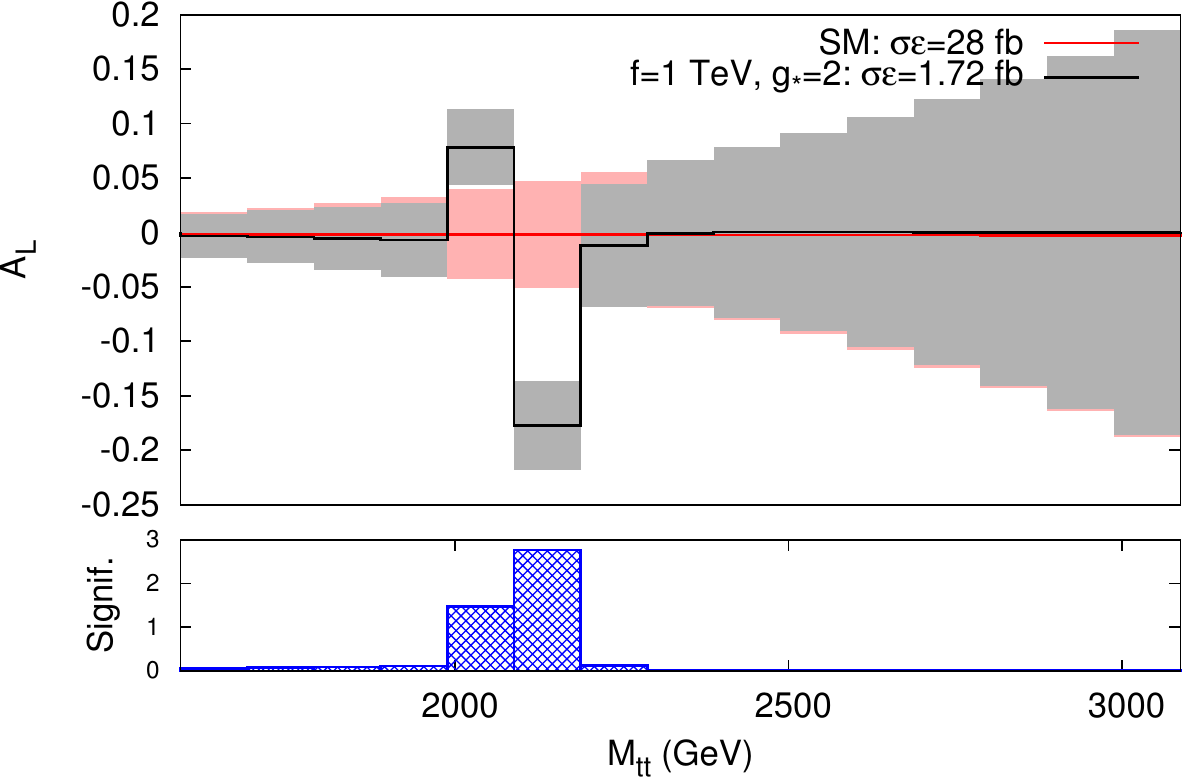}\\
\includegraphics[angle=0,width=0.425\linewidth ]{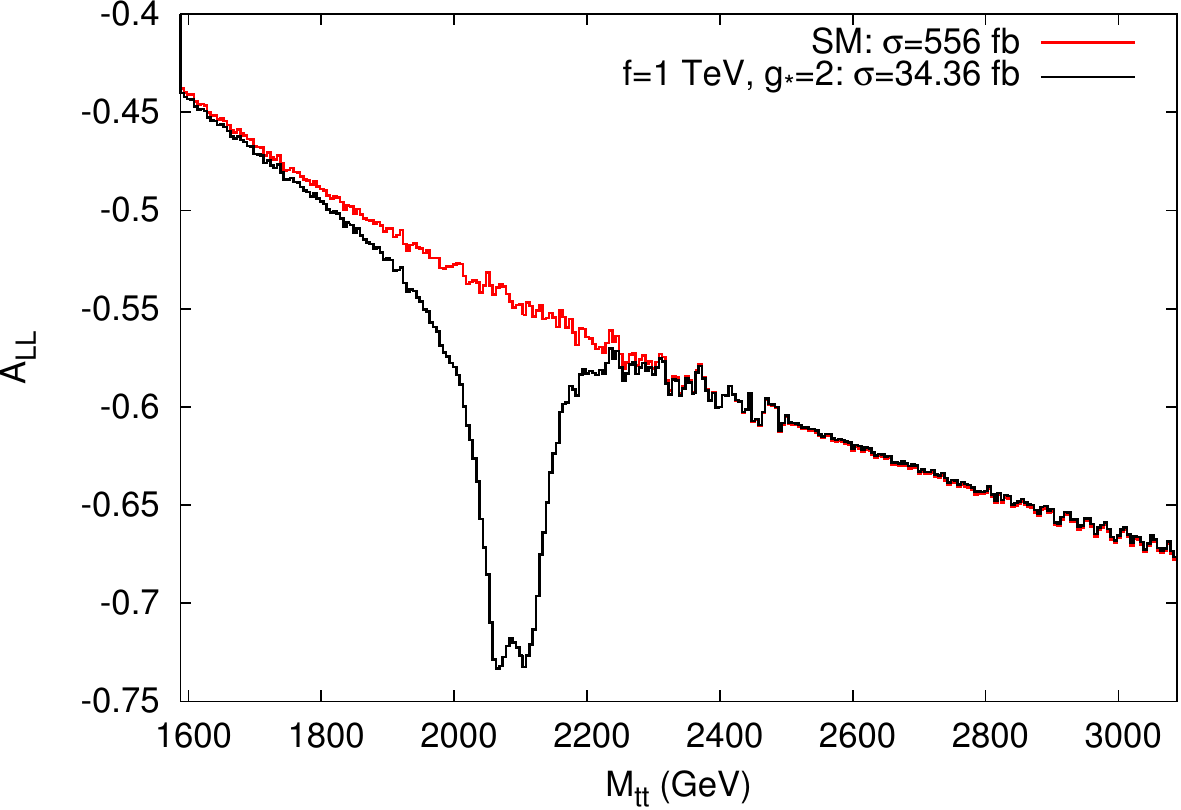}
\includegraphics[angle=0,width=0.45\linewidth ]{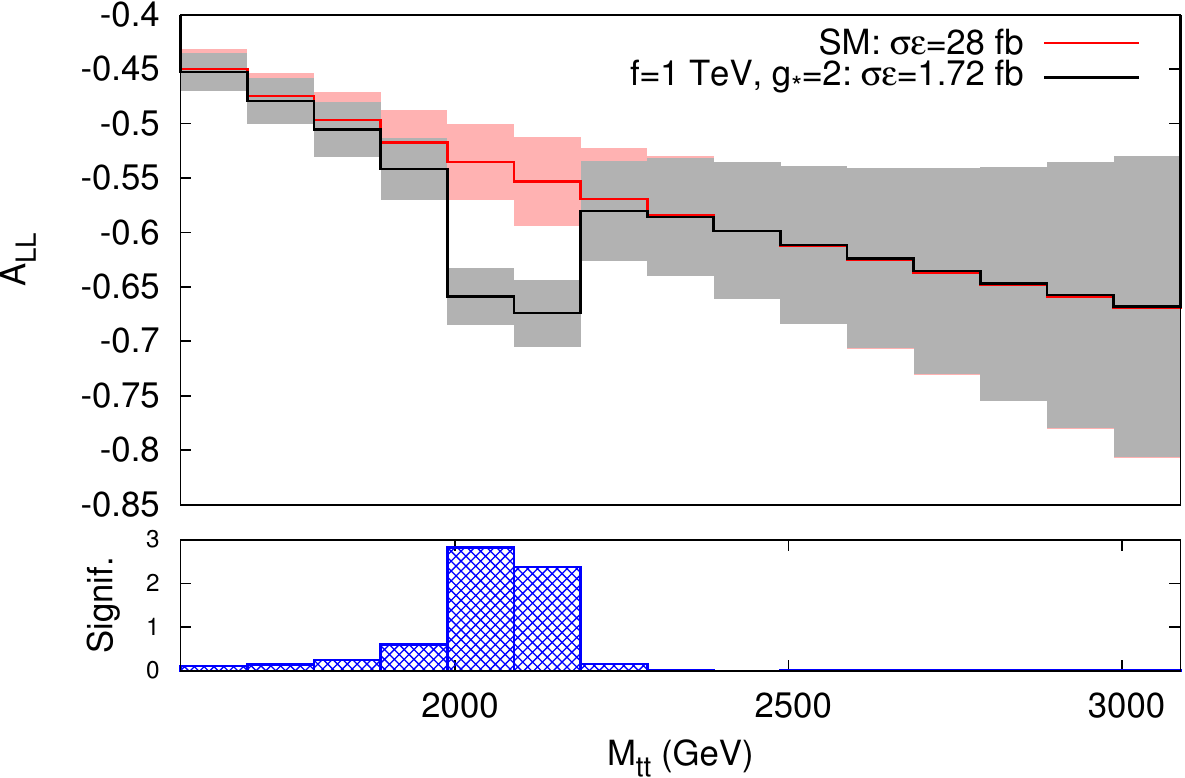}\\
\includegraphics[angle=0,width=0.425\linewidth ]{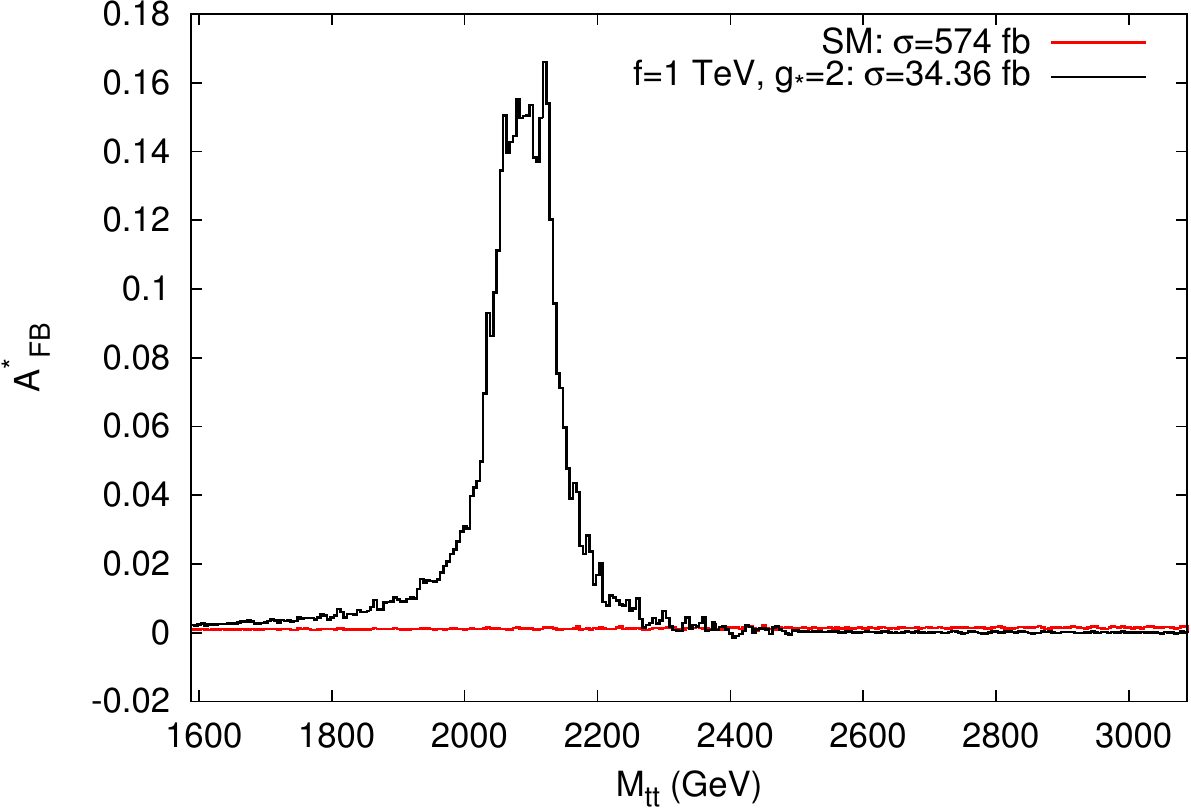}
\includegraphics[angle=0,width=0.45\linewidth ]{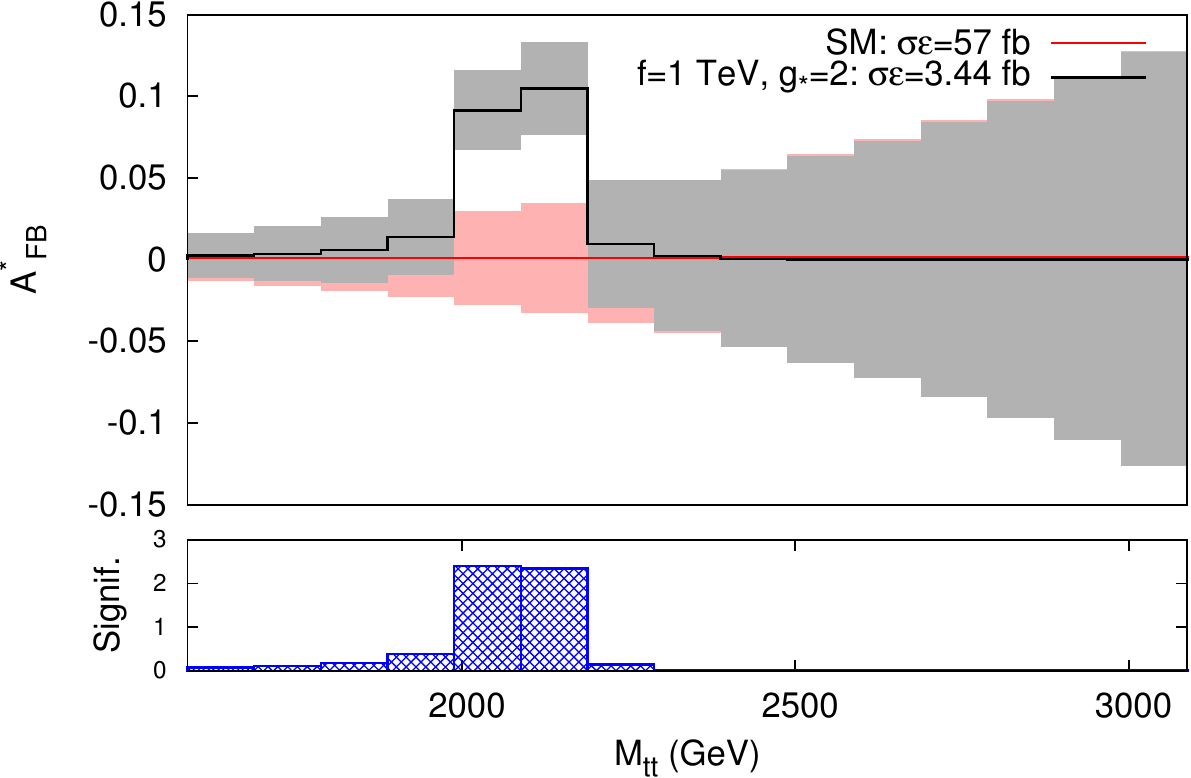}
\caption{\emph{(color online)} Cross section and asymmetries as a function of the $t\bar t$ invariant mass   for the $f=$ 1 TeV, $g_*=$2 benchmark at the 14 TeV LHC with 300 fb$^{-1}$.
The left column shows the fully differential observable. 
Right plots (upper frames) include estimates of statistical uncertainty assuming a realistic 
100 GeV mass resolution and also display (lower frames) the theoretical significance assuming a 10\% reconstruction efficiency. Grey(Pink) shading represents the (statistical) error on the 4DCHM(SM) rates, in black(red) solid lines. Masses and widths of the gauge bosons are $M[\Gamma]_{Z_2,Z_3}=2066[39]~{\rm GeV},2111[52]~{\rm GeV}$.}
\label{fig:c}
\end{figure}

\clearpage\thispagestyle{empty}


\clearpage\thispagestyle{empty}

\begin{figure}[h!]
\centering
\includegraphics[angle=0,width=0.425\linewidth ]{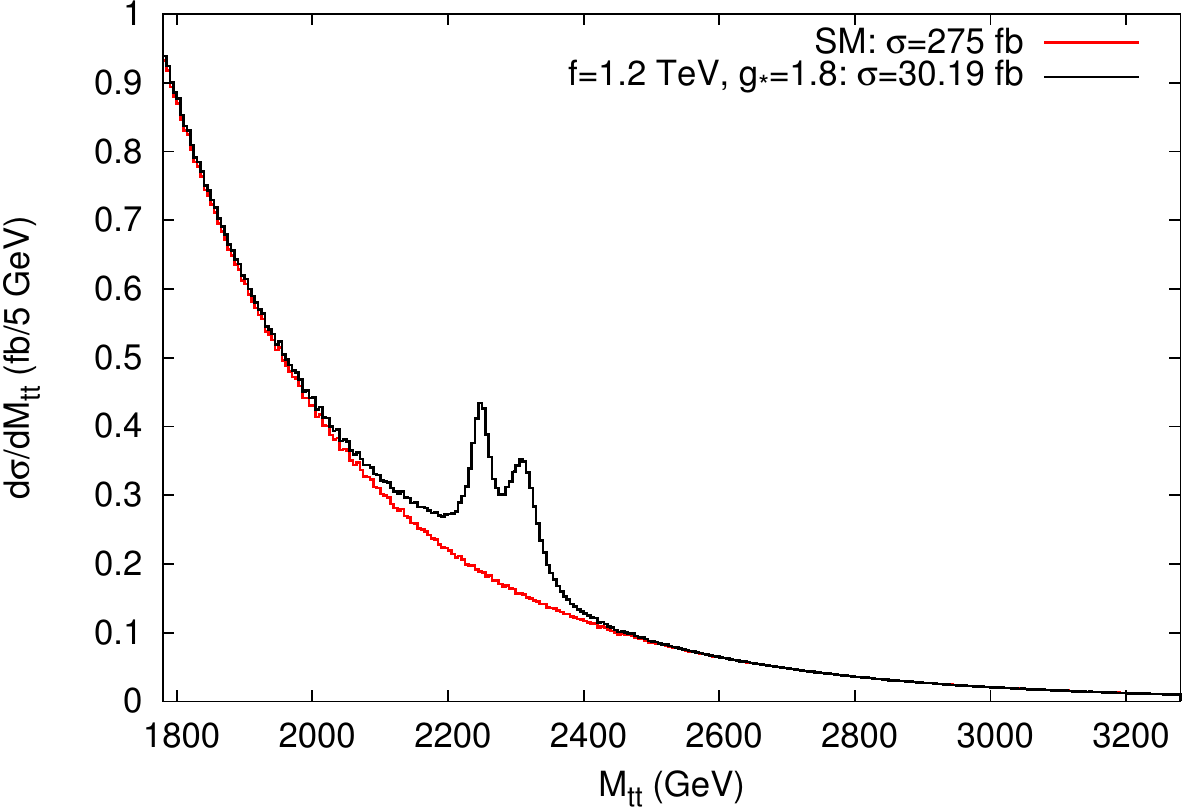}
\includegraphics[angle=0,width=0.45\linewidth ]{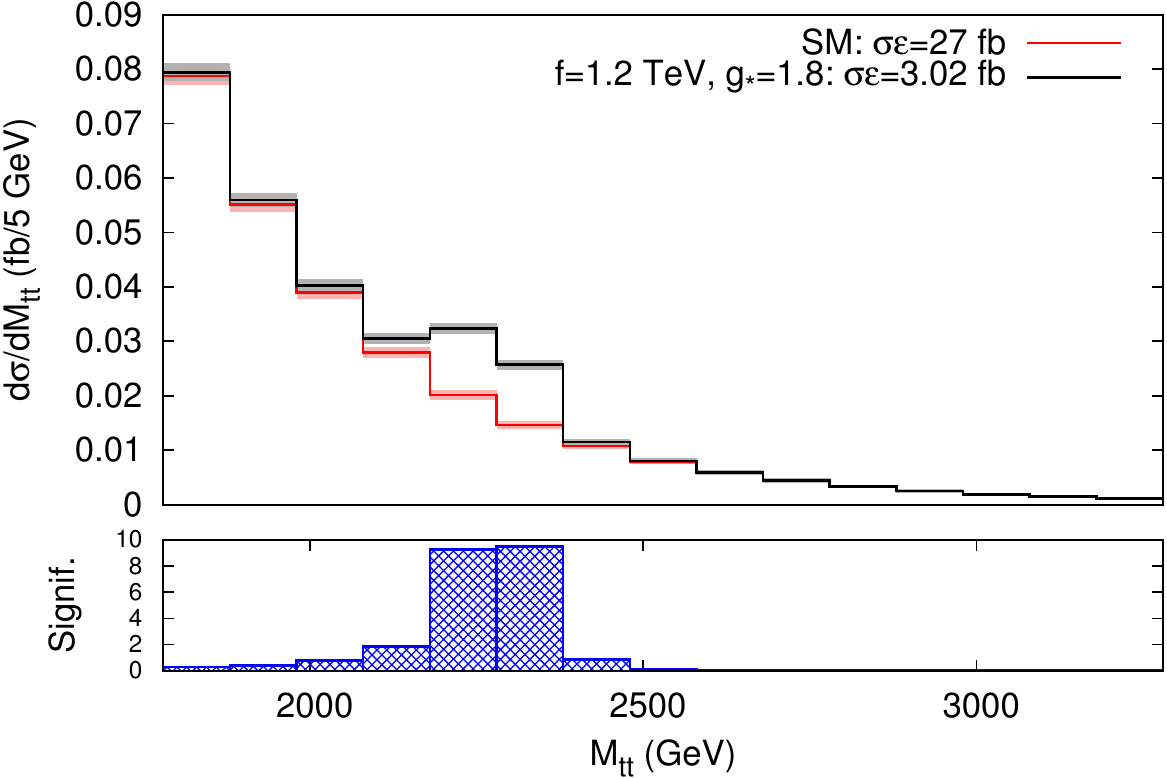}\\
\includegraphics[angle=0,width=0.425\linewidth ]{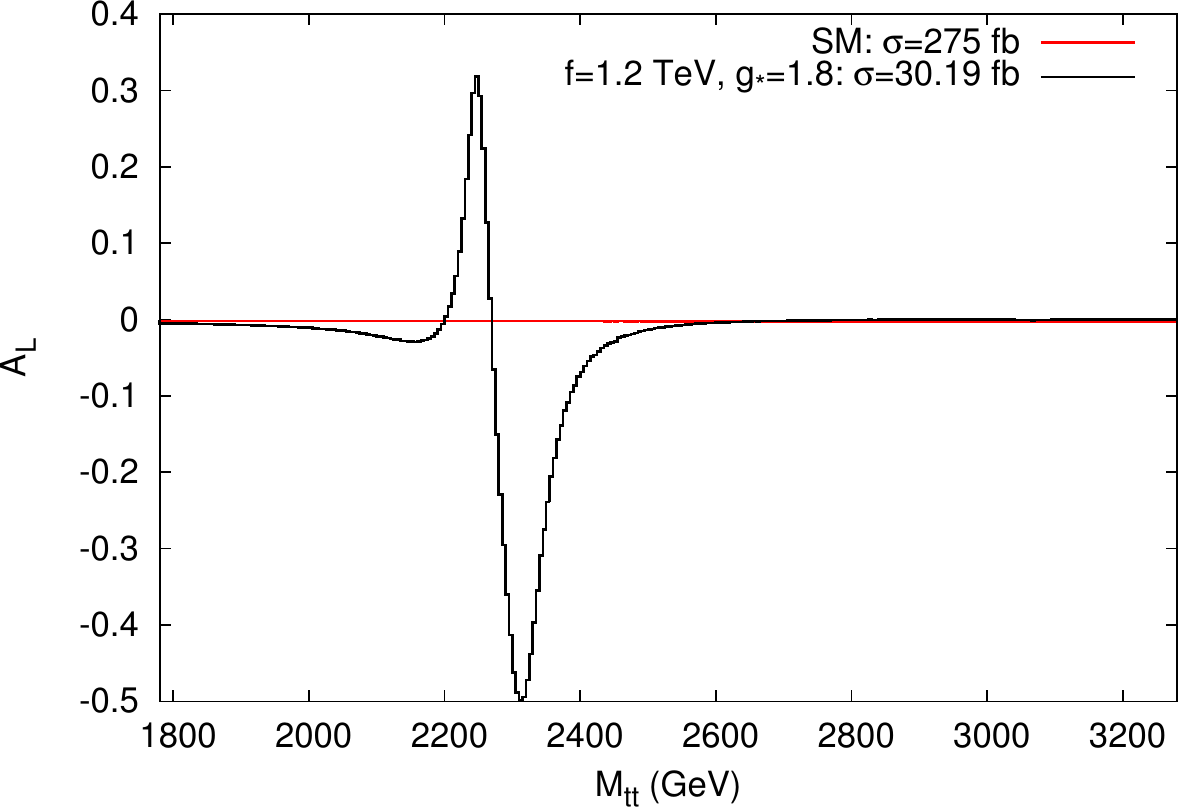}
\includegraphics[angle=0,width=0.45\linewidth ]{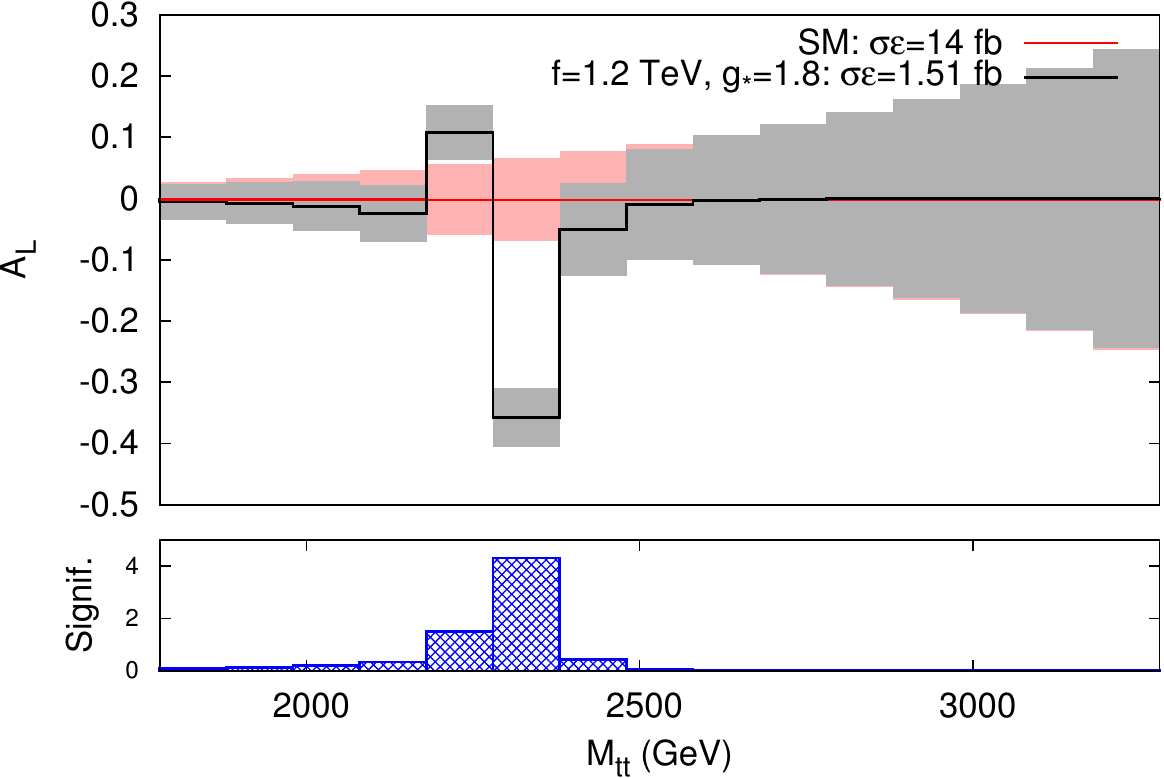}\\
\includegraphics[angle=0,width=0.425\linewidth ]{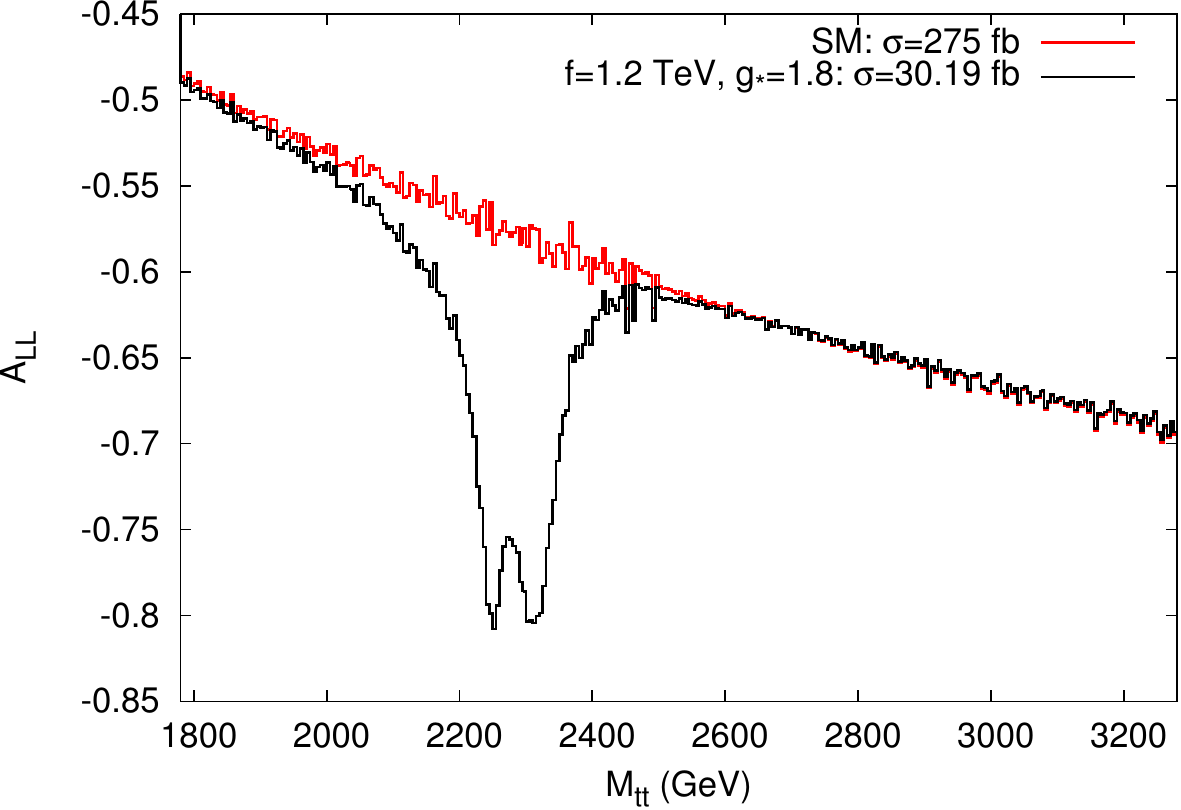}
\includegraphics[angle=0,width=0.45\linewidth ]{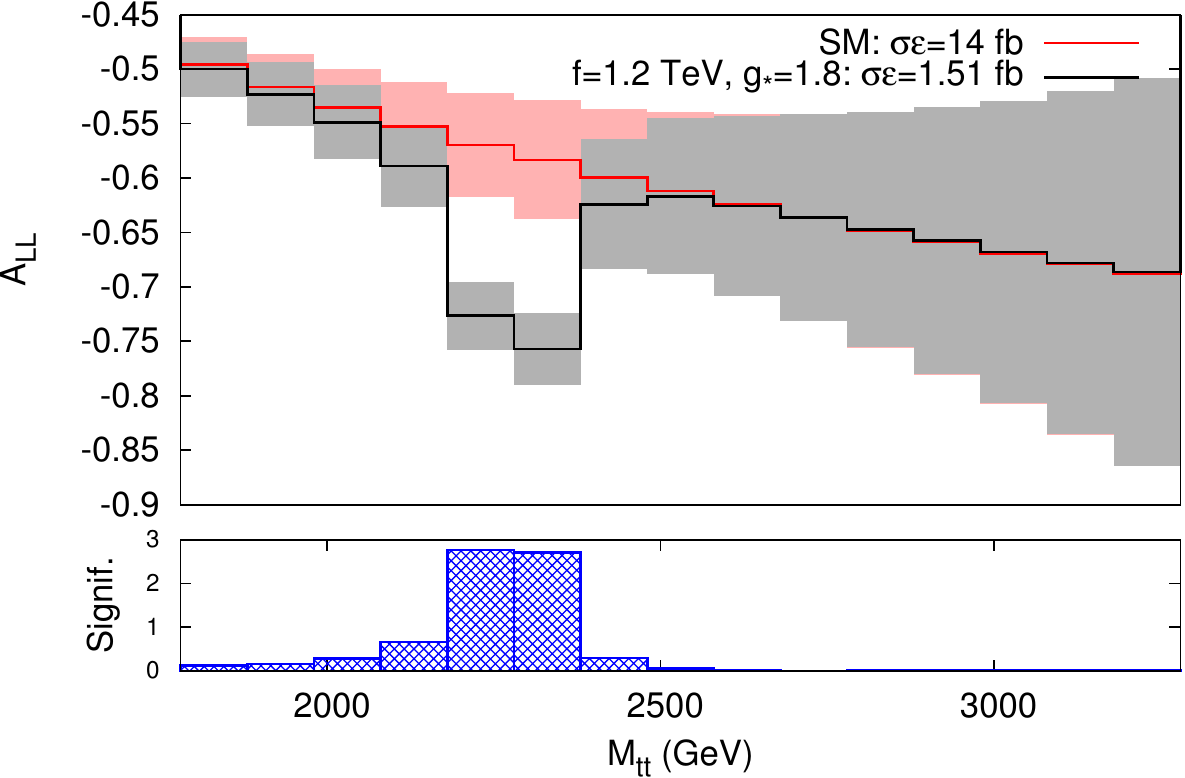}\\
\includegraphics[angle=0,width=0.425\linewidth ]{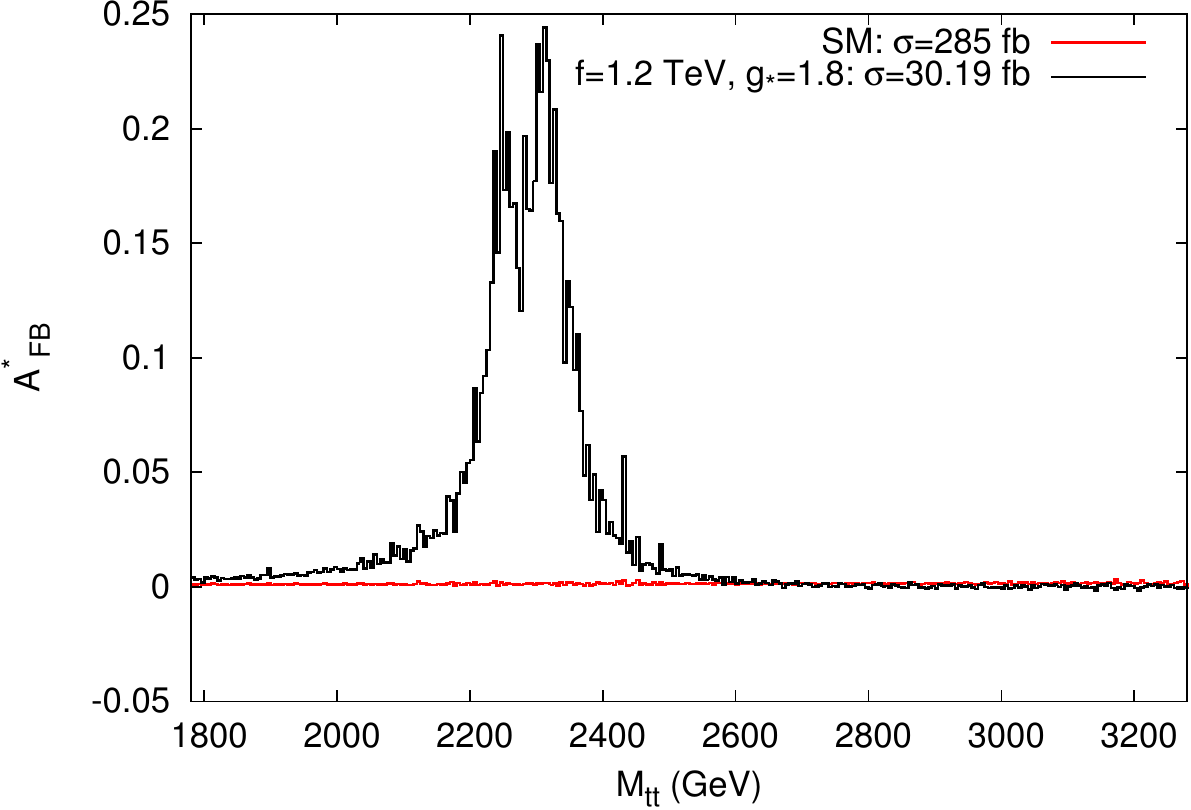}
\includegraphics[angle=0,width=0.45\linewidth ]{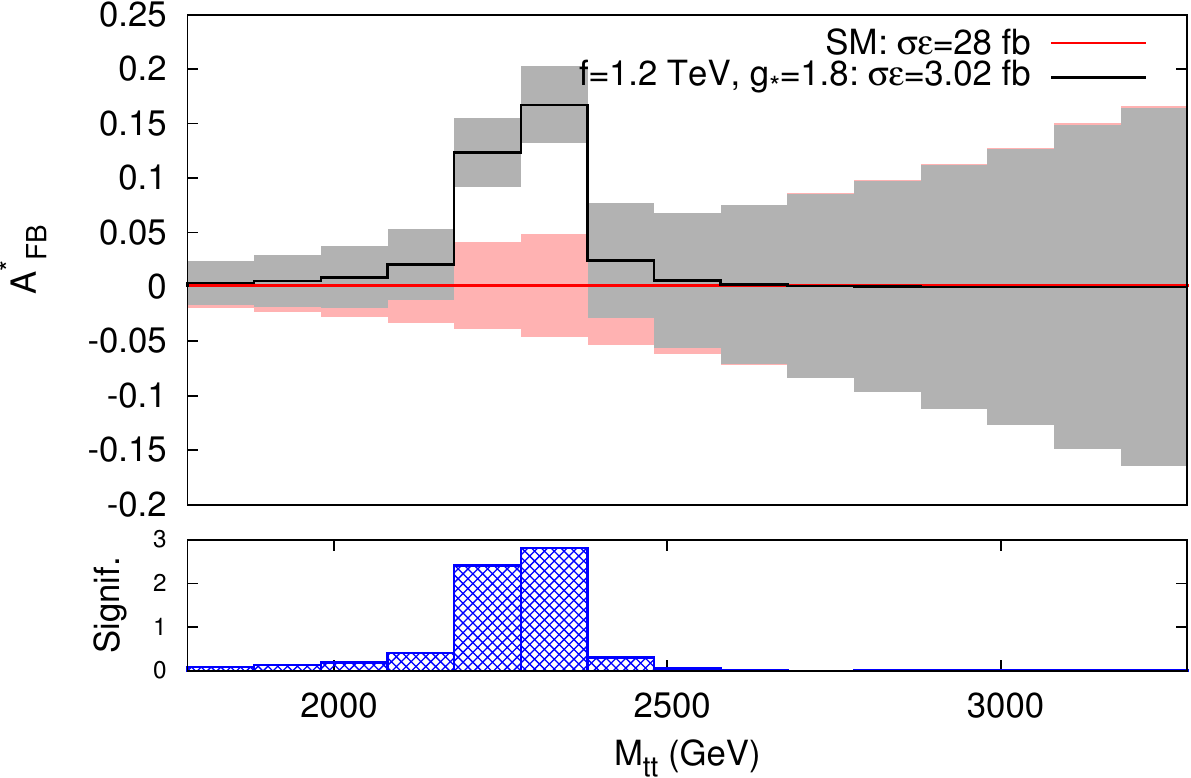}
\caption{\emph{(color online)} Cross section and asymmetries as a function of the $t\bar t$ invariant mass   for the $f=$1.2 TeV, $g_*=$1.8 benchmark at the 14 TeV LHC with 300 fb$^{-1}$.
The left column shows the fully differential observable. 
Right plots (upper frames) include estimates of statistical uncertainty assuming a realistic 
100 GeV mass resolution and also display (lower frames) the theoretical significance assuming a 10\% reconstruction efficiency. Grey(Pink) shading represents the (statistical) error on the 4DCHM(SM) rates, in black(red) solid lines. Masses and widths of the gauge bosons are $M[\Gamma]_{Z_2,Z_3}=2249[32]~{\rm GeV},2312[55]~{\rm GeV}$.}
\label{fig:f}
\end{figure}

\clearpage\thispagestyle{empty}

\begin{figure}[h!]
\centering
\includegraphics[angle=0,width=0.425\linewidth ]{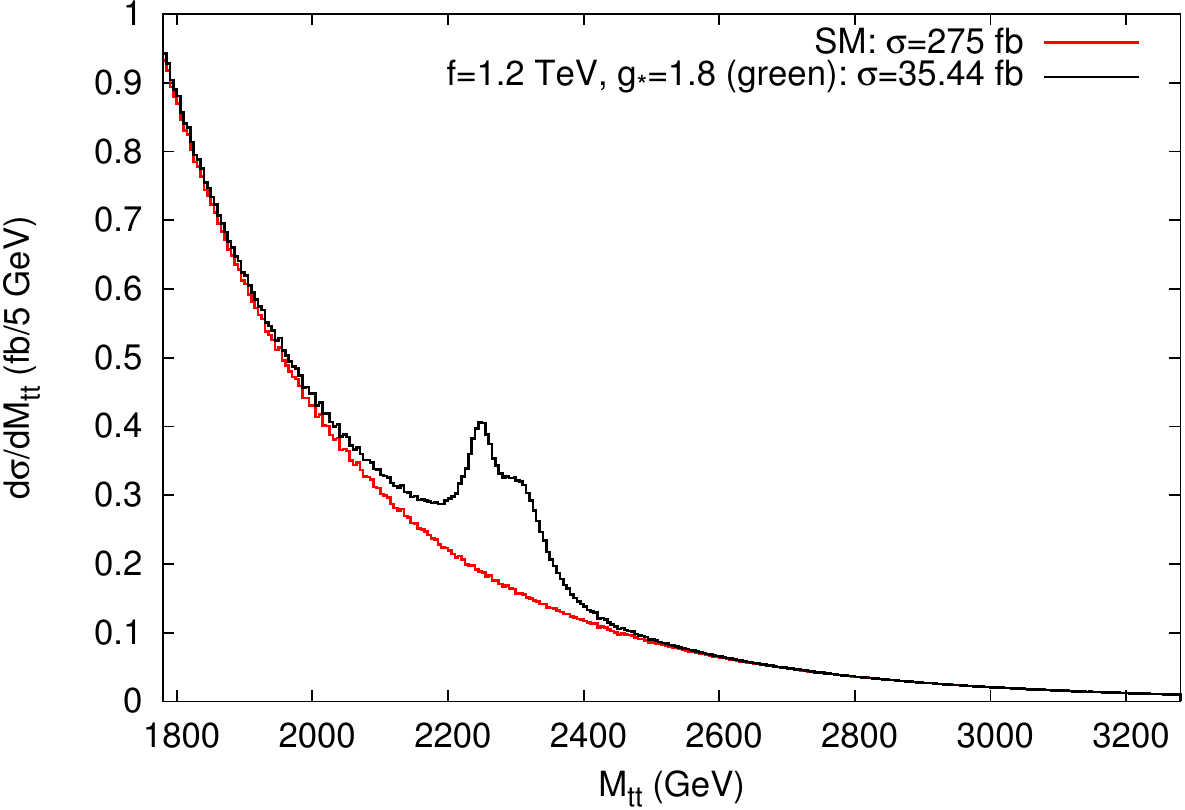}
\includegraphics[angle=0,width=0.45\linewidth ]{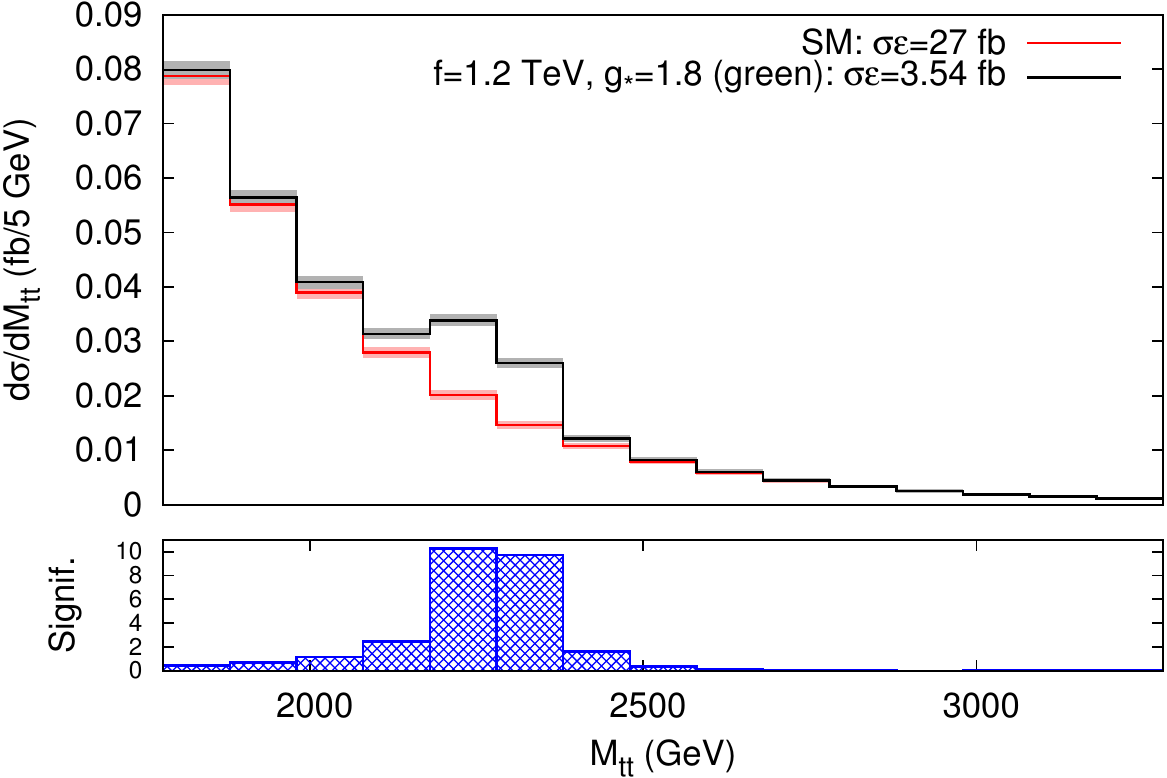}\\
\includegraphics[angle=0,width=0.425\linewidth ]{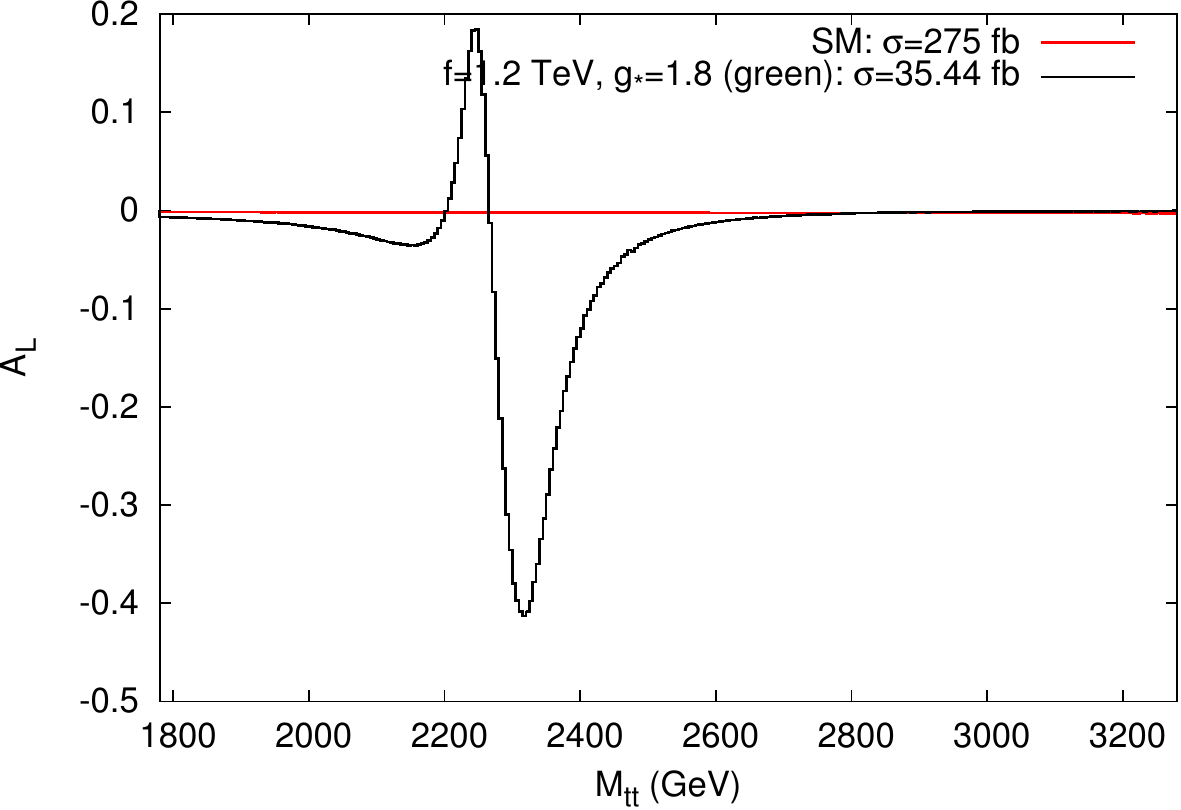}
\includegraphics[angle=0,width=0.45\linewidth ]{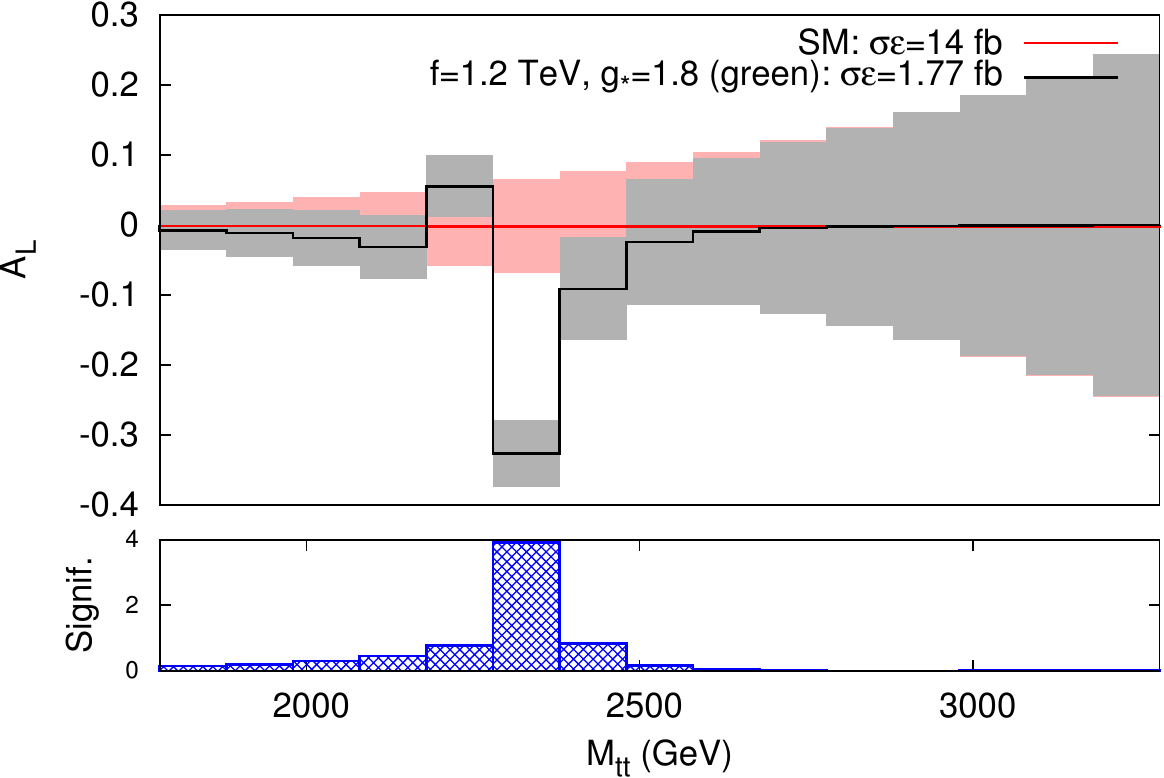}\\
\includegraphics[angle=0,width=0.425\linewidth ]{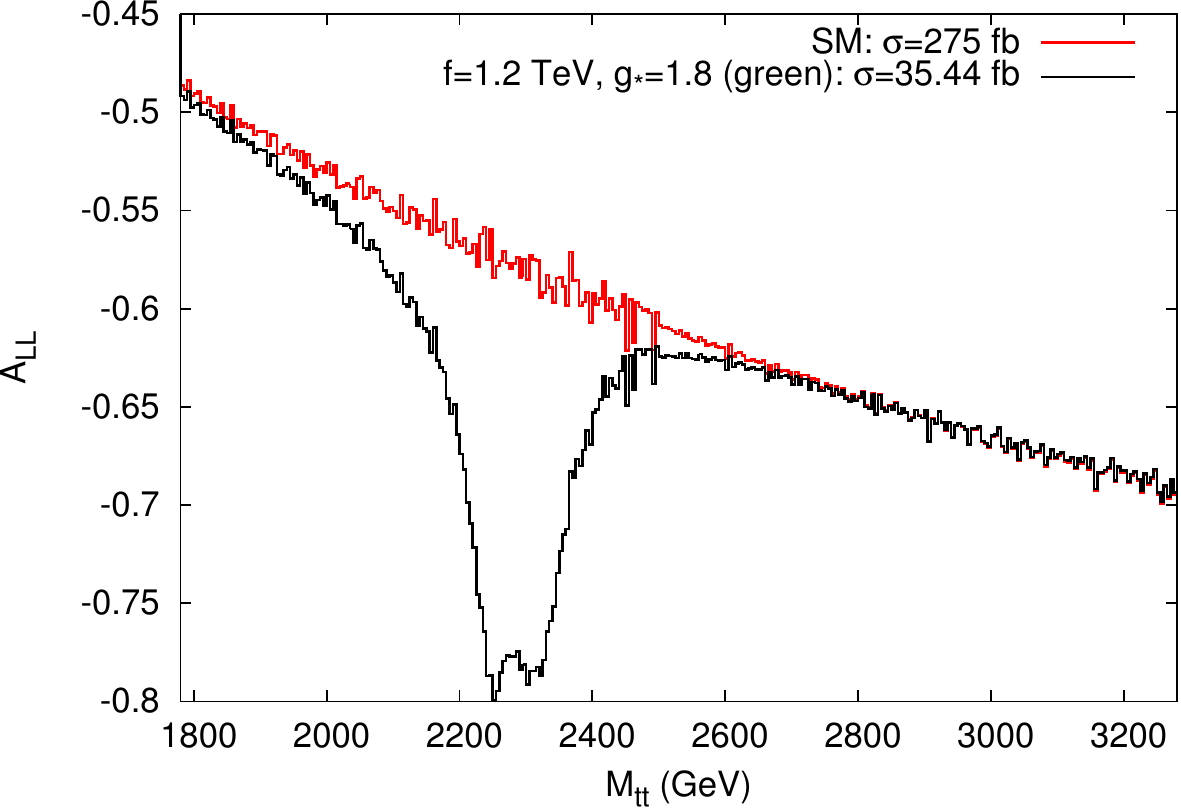}
\includegraphics[angle=0,width=0.45\linewidth ]{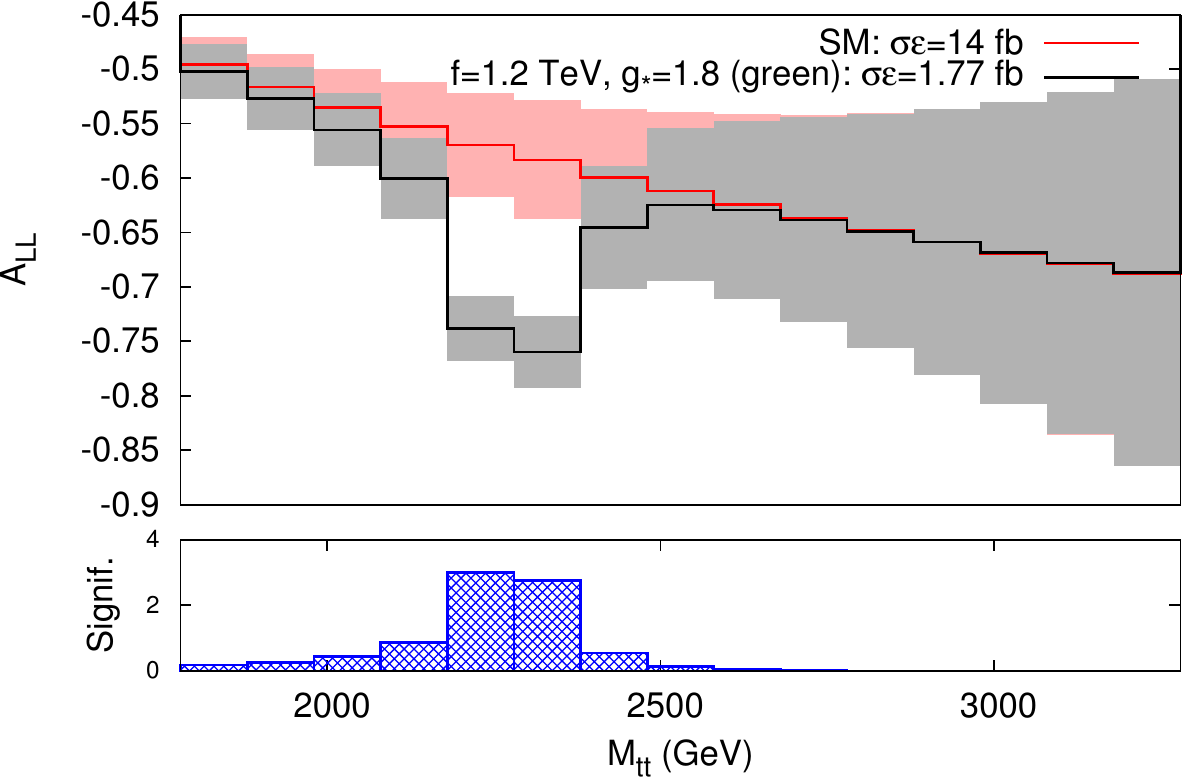}\\
\includegraphics[angle=0,width=0.425\linewidth ]{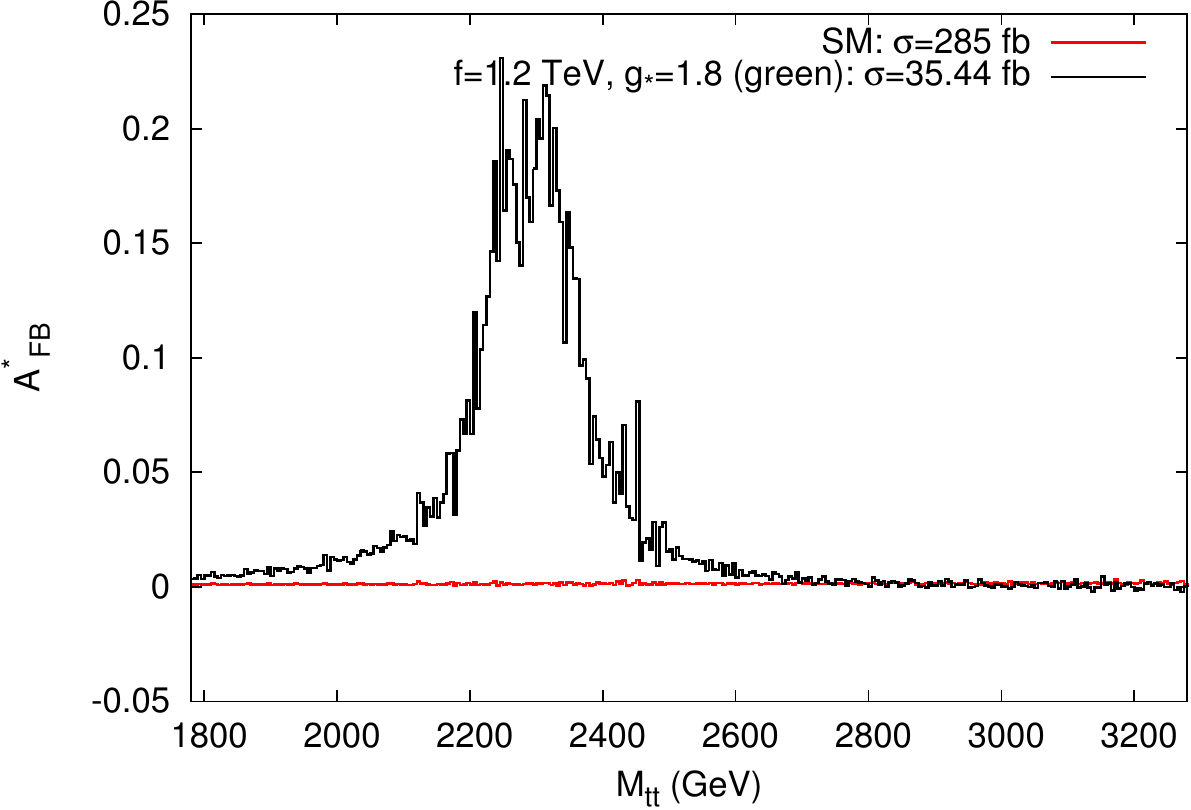}
\includegraphics[angle=0,width=0.45\linewidth ]{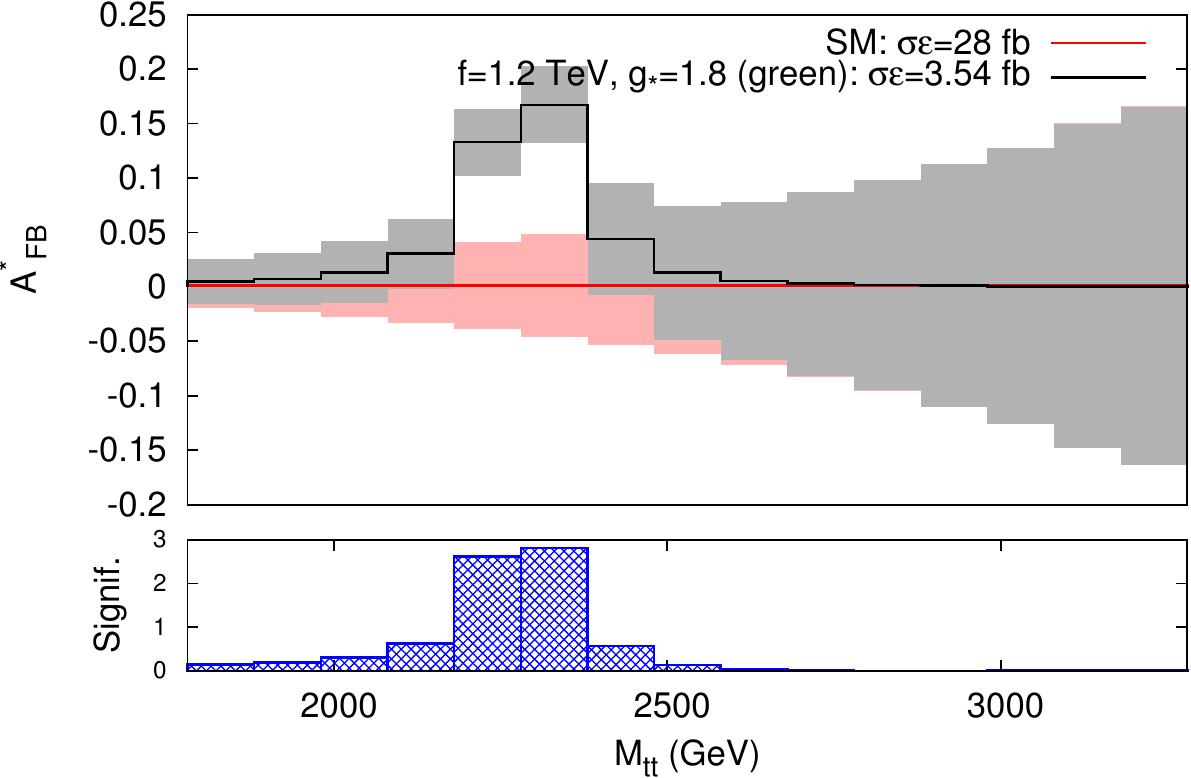}
\caption{\emph{(color online)} Cross section and asymmetries as a function of the $t\bar t$ invariant mass   for the $f=$1.2 TeV, $g_*=$1.8 (green) benchmark at the 14 TeV LHC with 300 fb$^{-1}$.
The left column shows the fully differential observable. 
Right plots (upper frames) include estimates of statistical uncertainty assuming a realistic 
100 GeV mass resolution and also display (lower frames) the theoretical significance assuming a 10\% reconstruction efficiency. Grey(Pink) shading represents the (statistical) error on the 4DCHM(SM) rates, in black(red) solid lines. Masses and widths of the gauge bosons are $M[\Gamma]_{Z_2,Z_3}=2249[48]~{\rm GeV},2312[86]~{\rm GeV}$.}
\label{fig:green}
\end{figure}

%

\clearpage\thispagestyle{empty}

\begin{figure}[h!]
\centering
\includegraphics[angle=0,width=0.425\linewidth ]{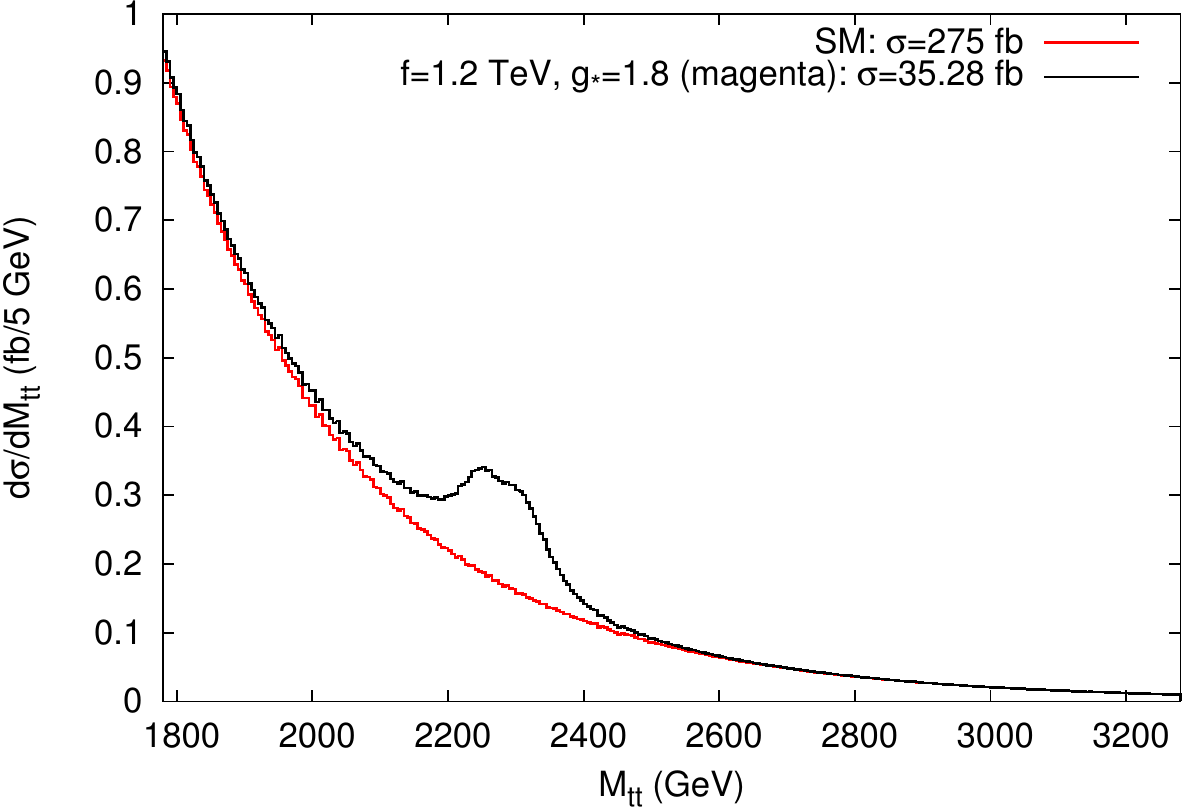}
\includegraphics[angle=0,width=0.45\linewidth ]{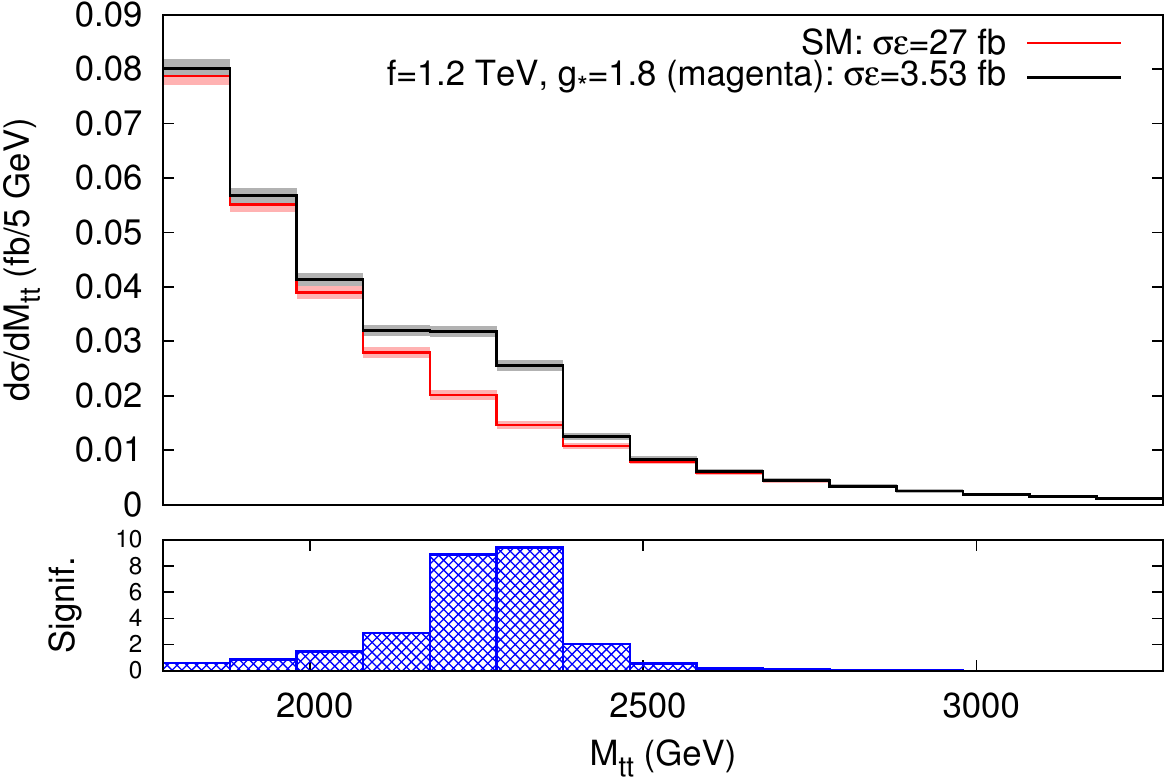}\\
\includegraphics[angle=0,width=0.425\linewidth ]{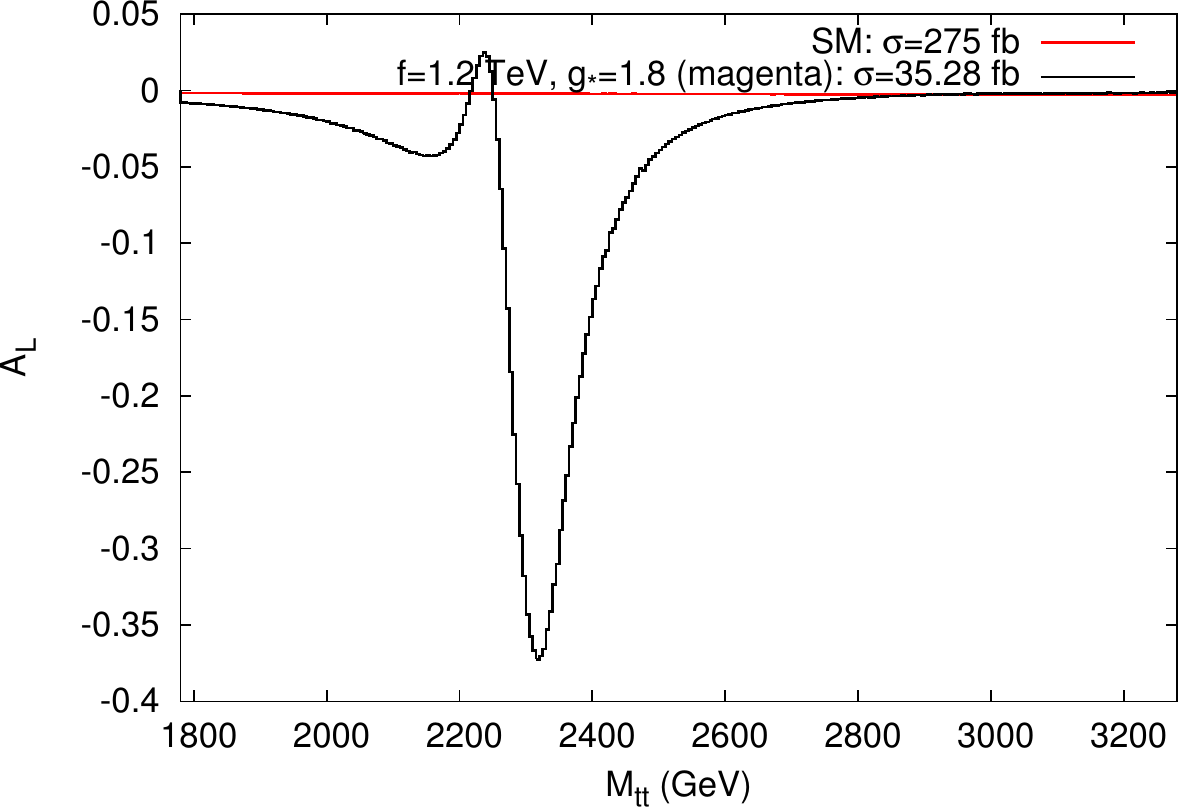}
\includegraphics[angle=0,width=0.45\linewidth ]{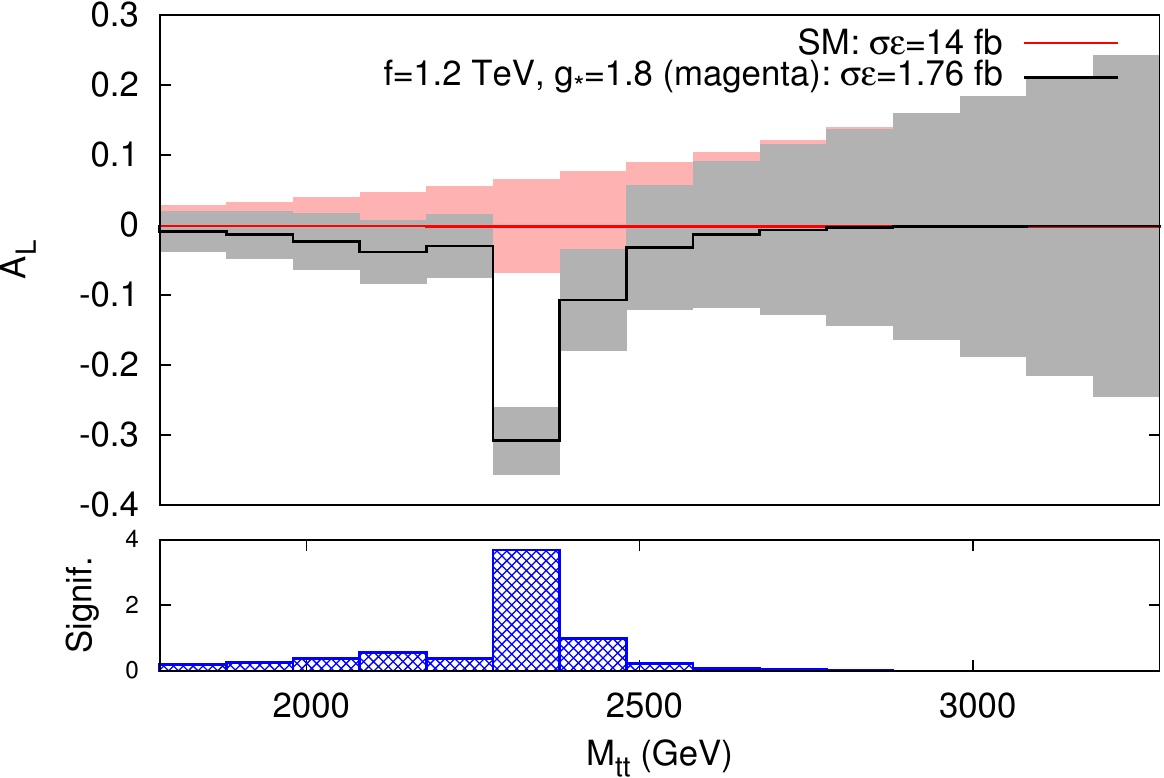}\\
\includegraphics[angle=0,width=0.425\linewidth ]{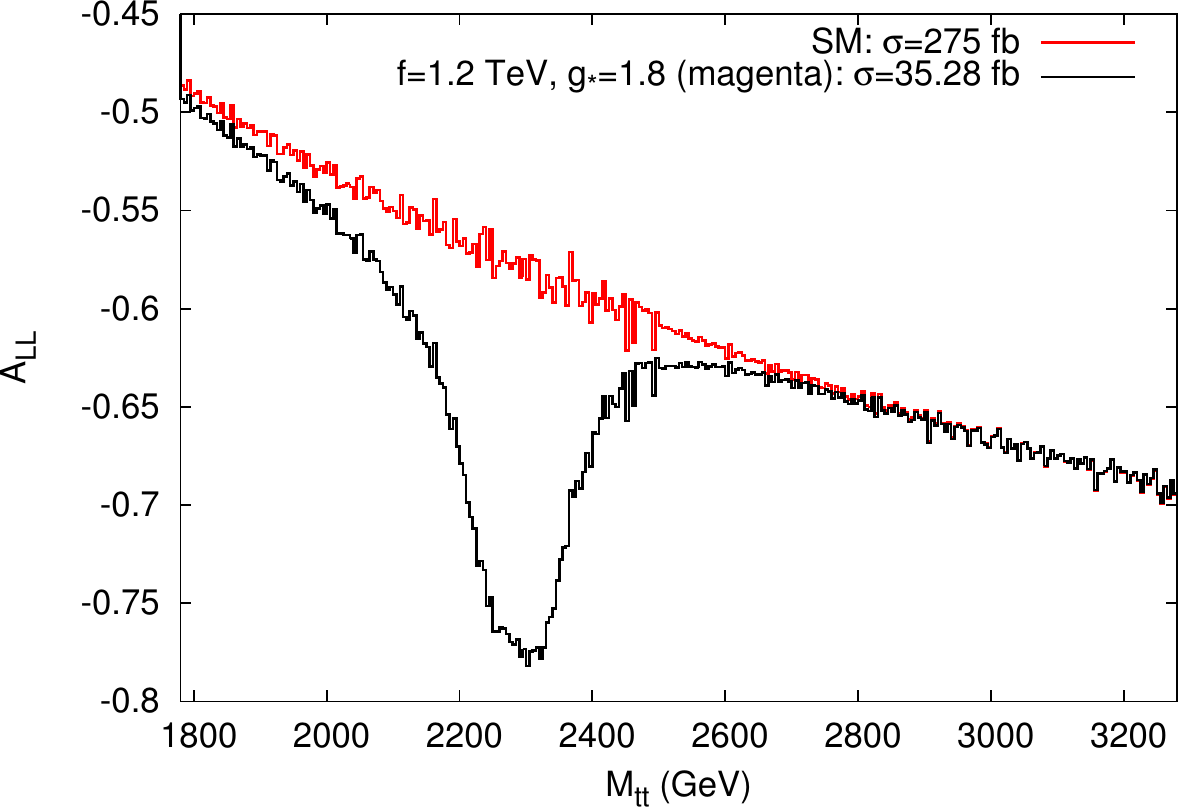}
\includegraphics[angle=0,width=0.45\linewidth ]{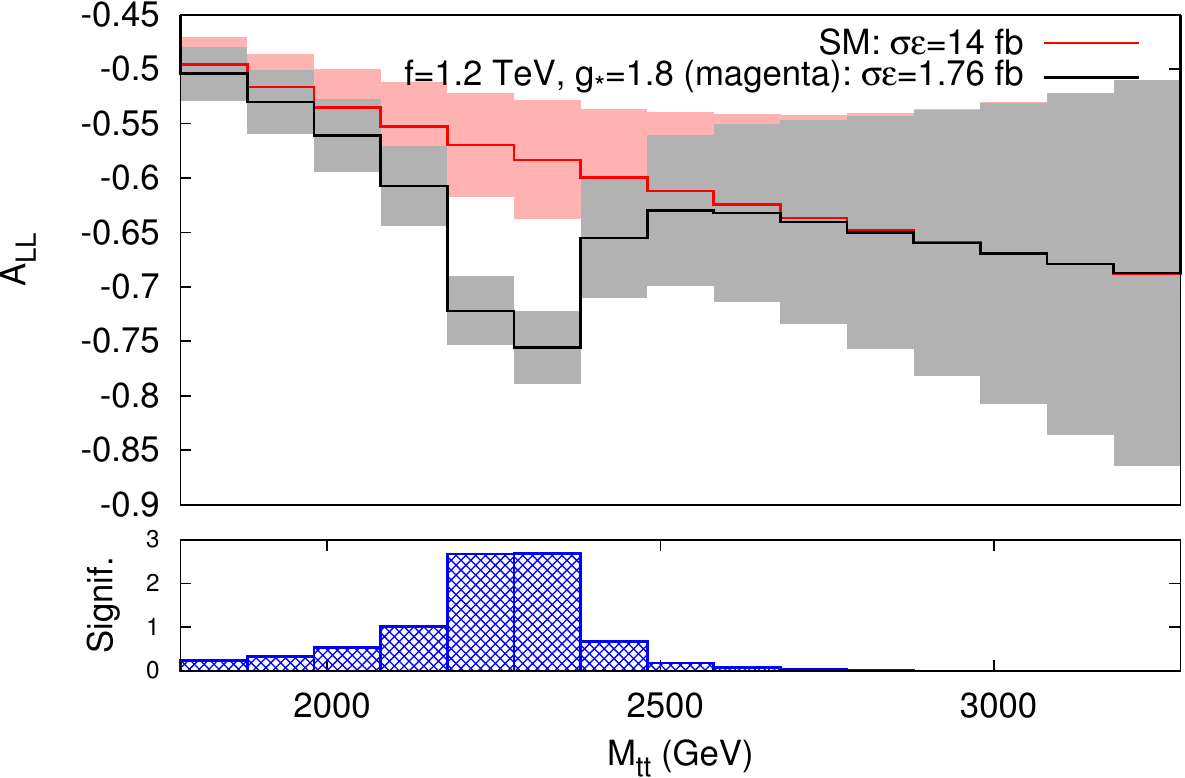}\\
\includegraphics[angle=0,width=0.425\linewidth ]{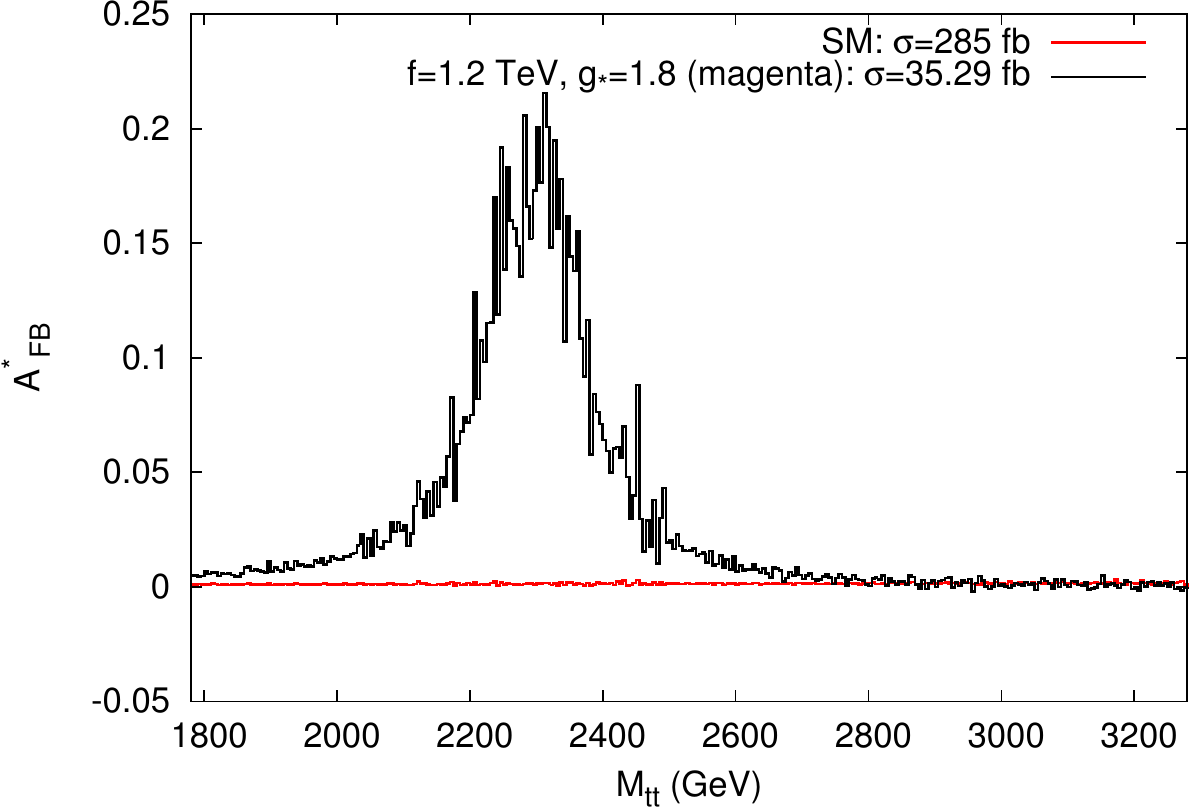}
\includegraphics[angle=0,width=0.45\linewidth ]{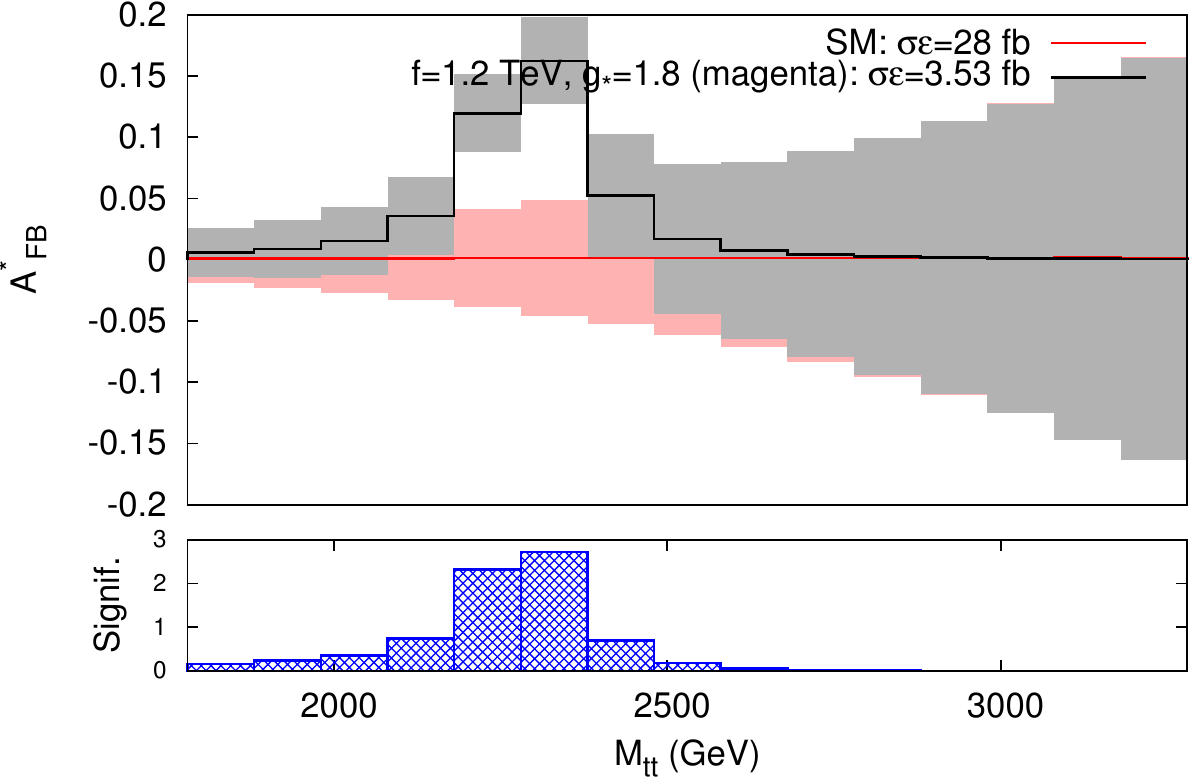}
\caption{\emph{(color online)} Cross section and asymmetries as a function of the $t\bar t$ invariant mass   for the $f=$1.2 TeV, $g_*=$1.8 (magenta) benchmark at the 14 TeV LHC with 300 fb$^{-1}$.
The left column shows the fully differential observable. 
Right plots (upper frames) include estimates of statistical uncertainty assuming a realistic 
100 GeV mass resolution and also display (lower frames) the theoretical significance assuming a 10\% reconstruction efficiency. Grey(Pink) shading represents the (statistical) error on the 4DCHM(SM) rates, in black(red) solid lines. Masses and widths of the gauge bosons are $M[\Gamma]_{Z_2,Z_3}=2249[75]~{\rm GeV},2312[104]~{\rm GeV}$.}
\label{fig:magenta}
\end{figure}

%

\clearpage\thispagestyle{empty}

\begin{figure}[h!]
\centering
\includegraphics[angle=0,width=0.425\linewidth ]{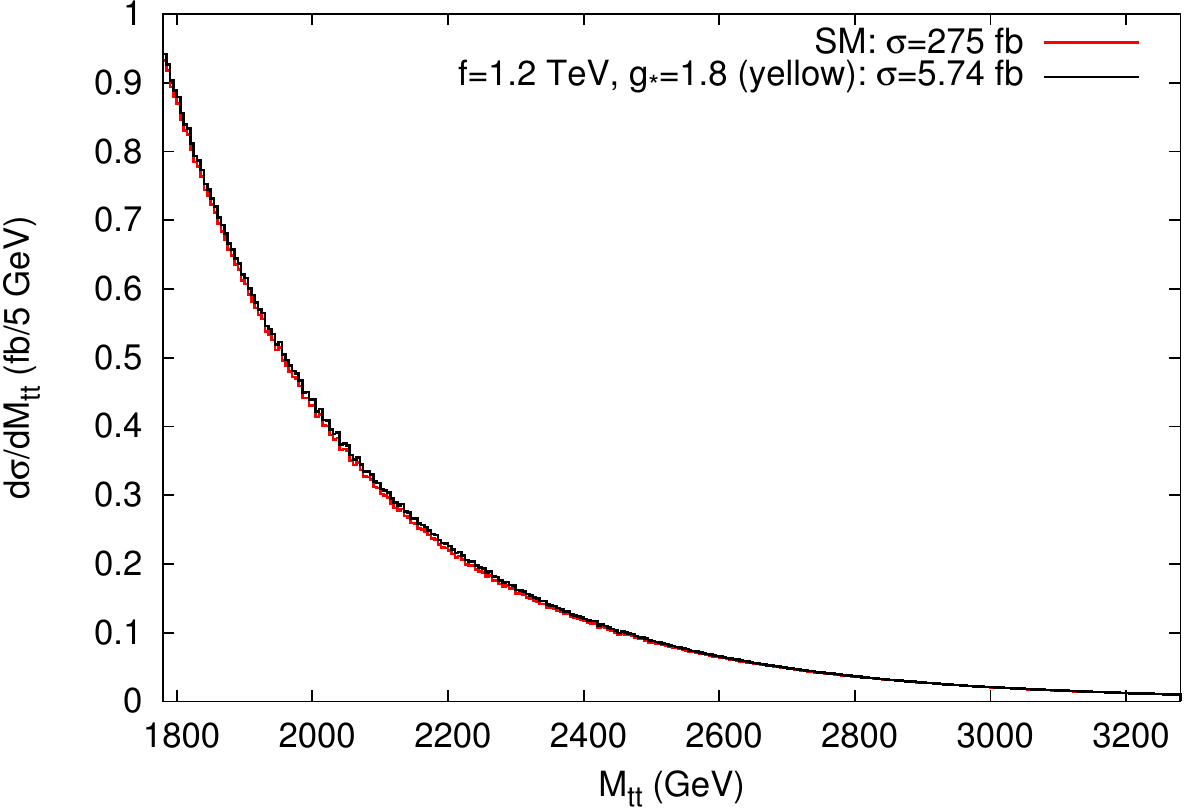}
\includegraphics[angle=0,width=0.45\linewidth ]{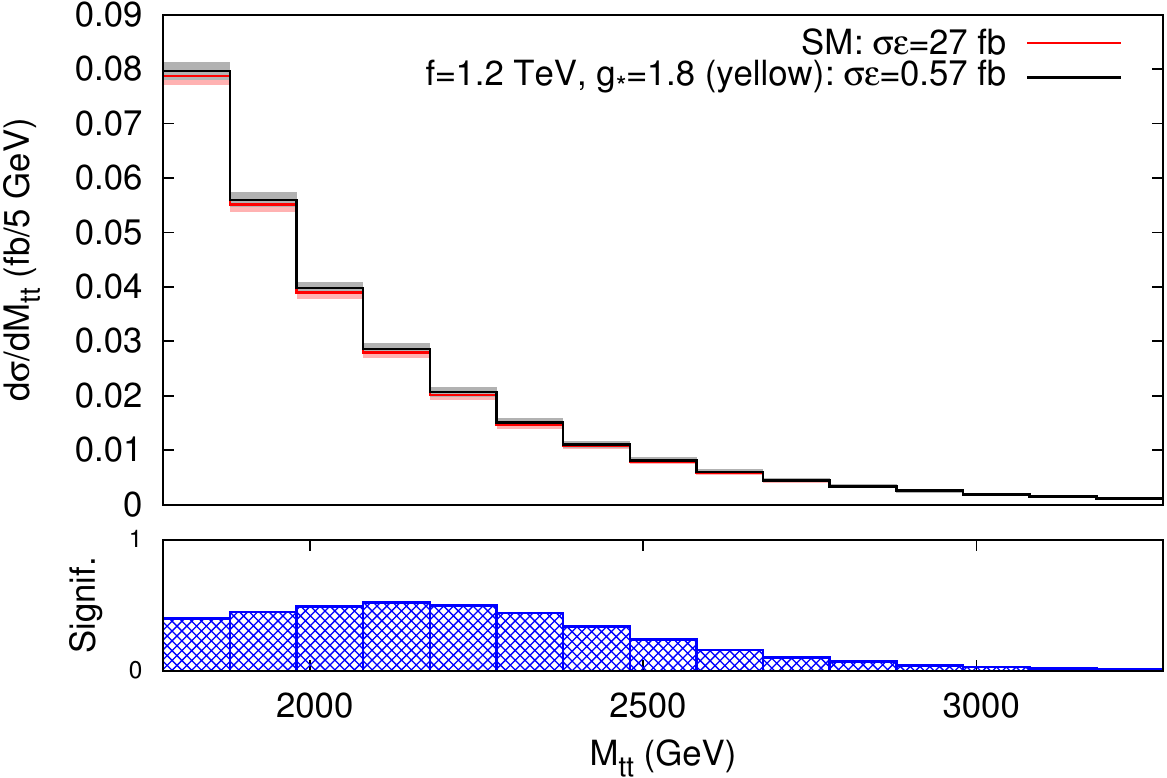}\\
\includegraphics[angle=0,width=0.425\linewidth ]{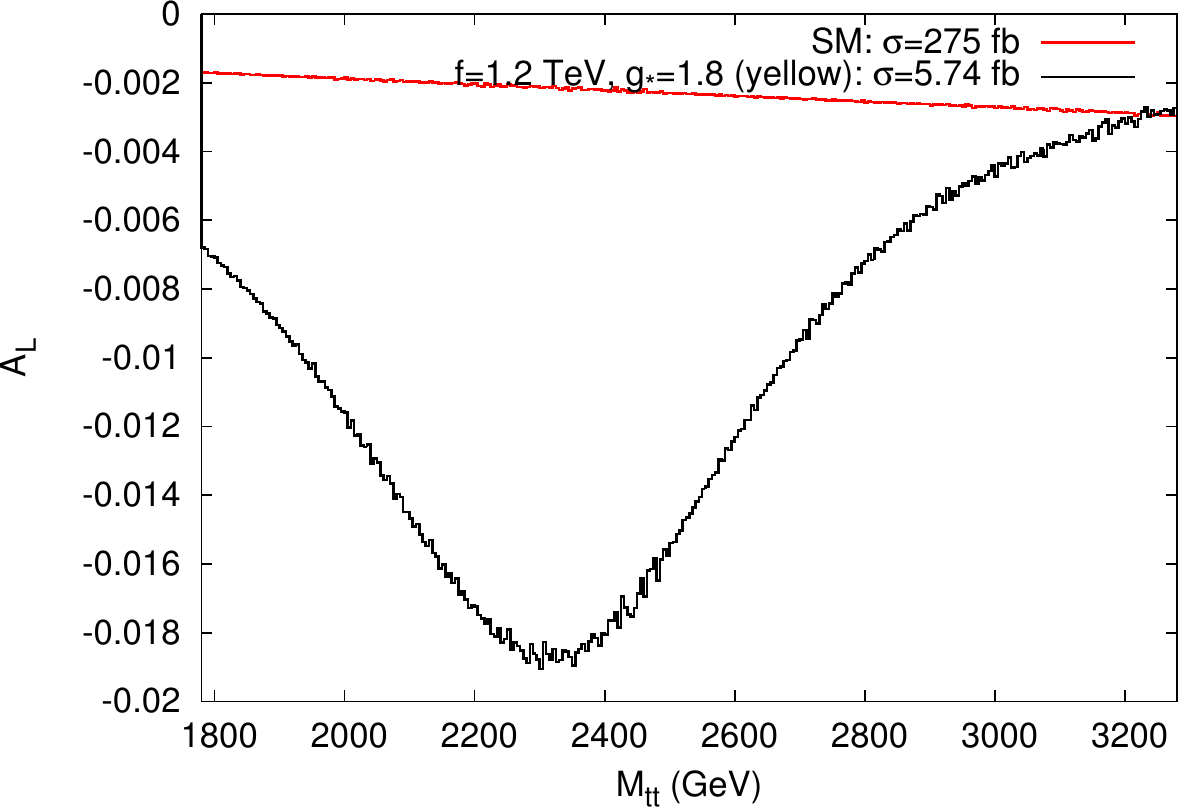}
\includegraphics[angle=0,width=0.45\linewidth ]{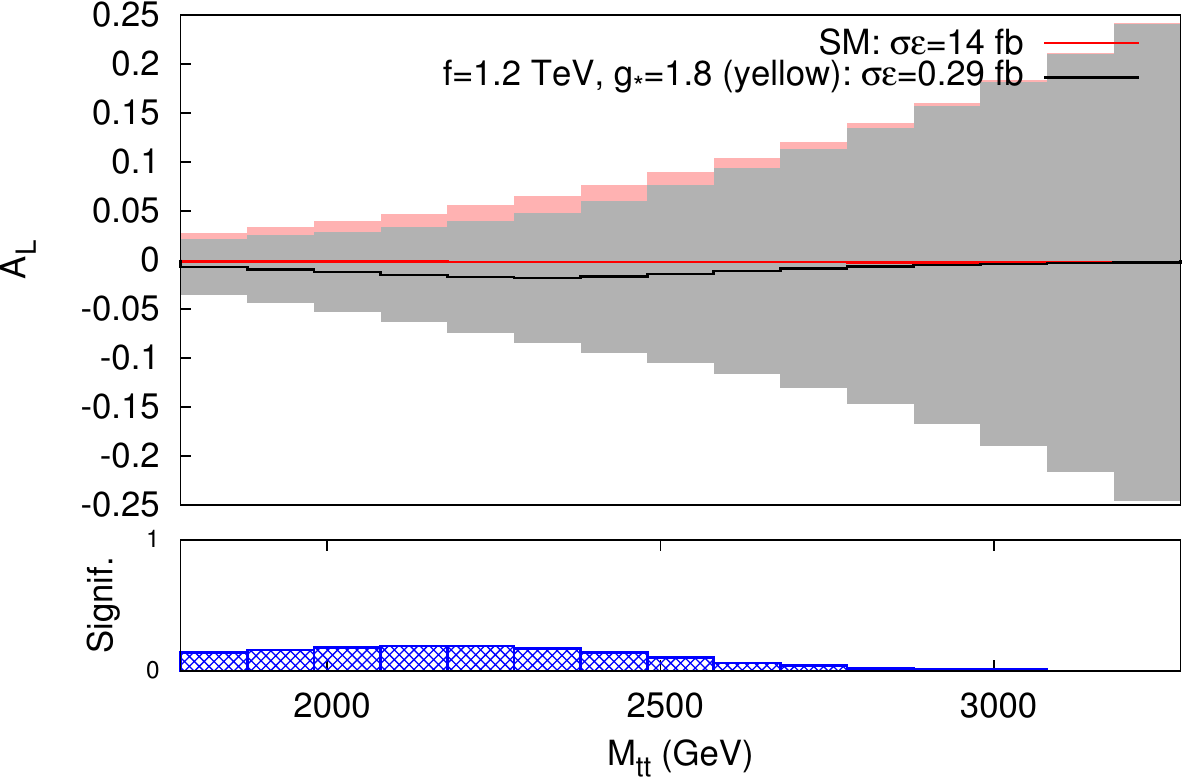}\\
\includegraphics[angle=0,width=0.425\linewidth ]{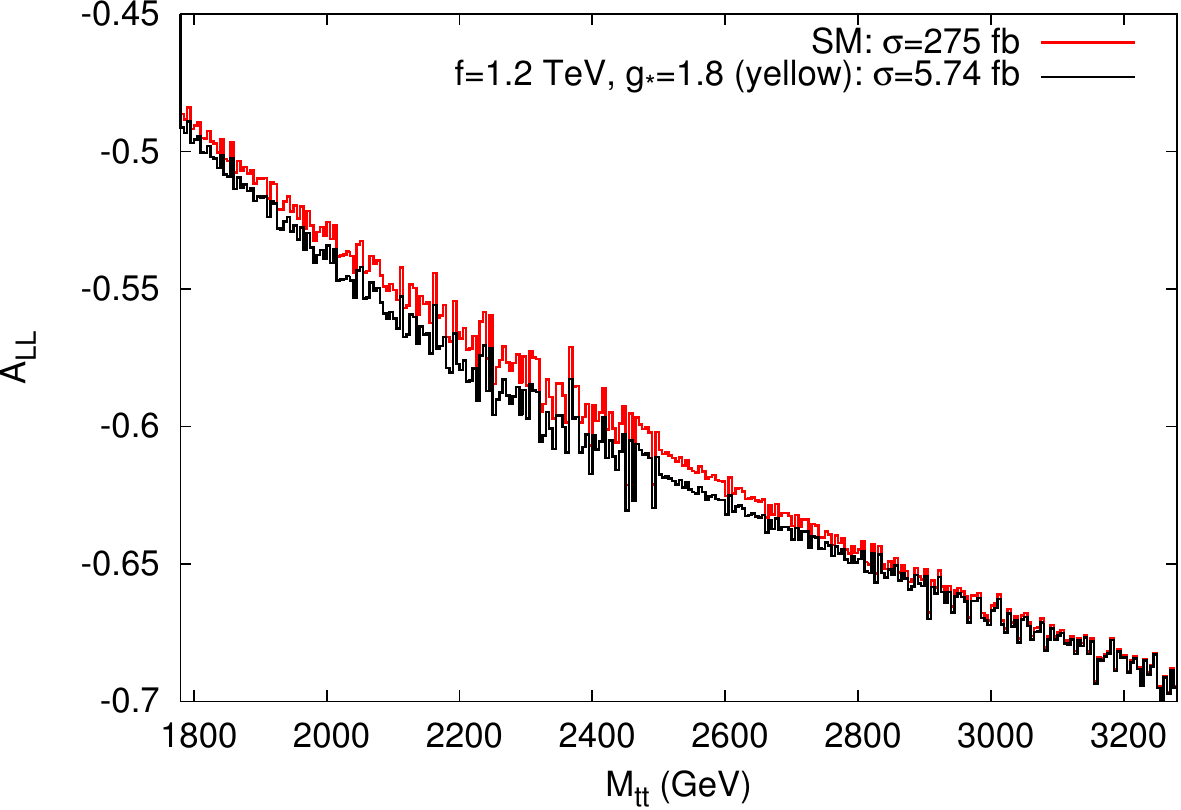}
\includegraphics[angle=0,width=0.45\linewidth ]{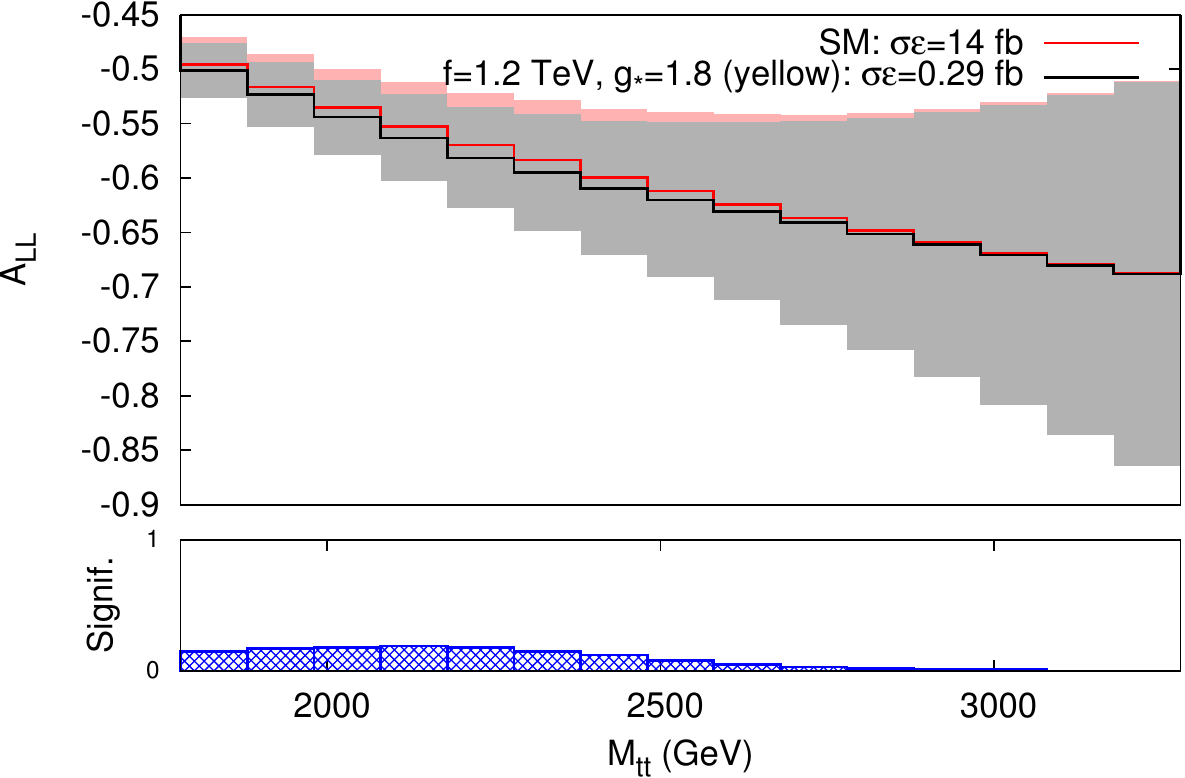}\\
\includegraphics[angle=0,width=0.425\linewidth ]{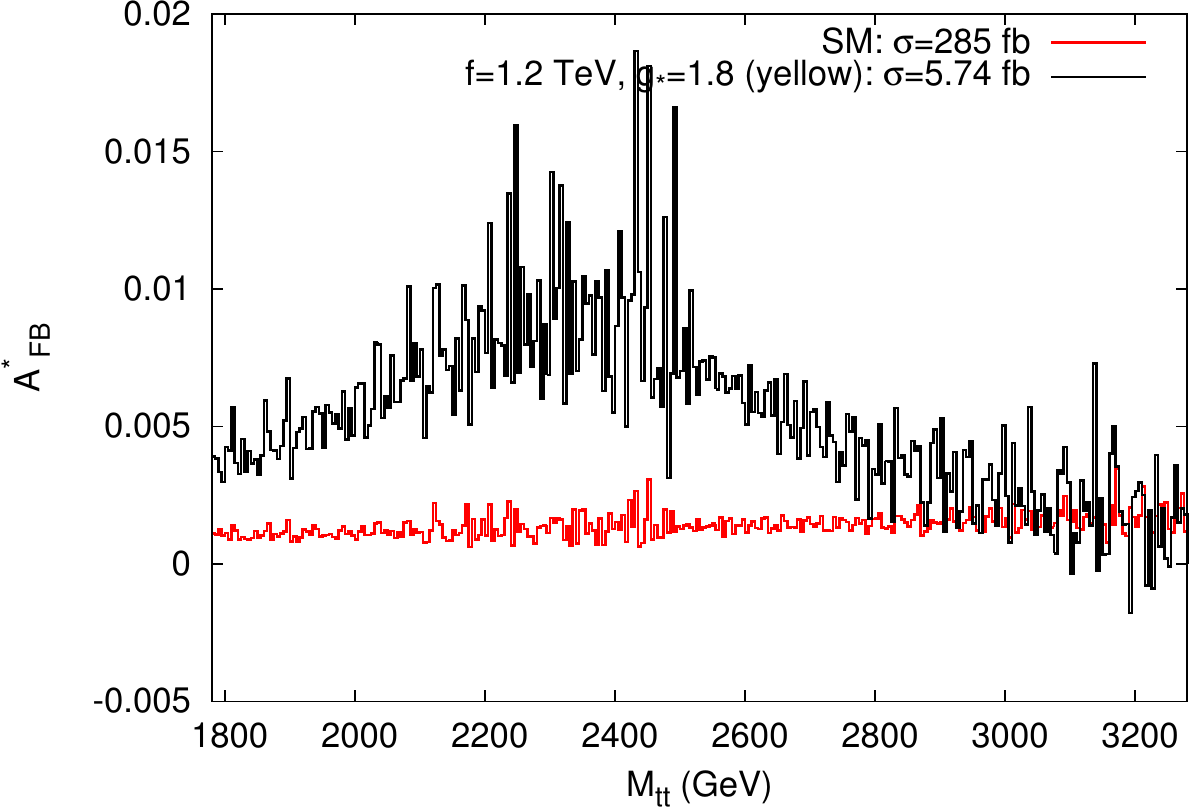}
\includegraphics[angle=0,width=0.45\linewidth ]{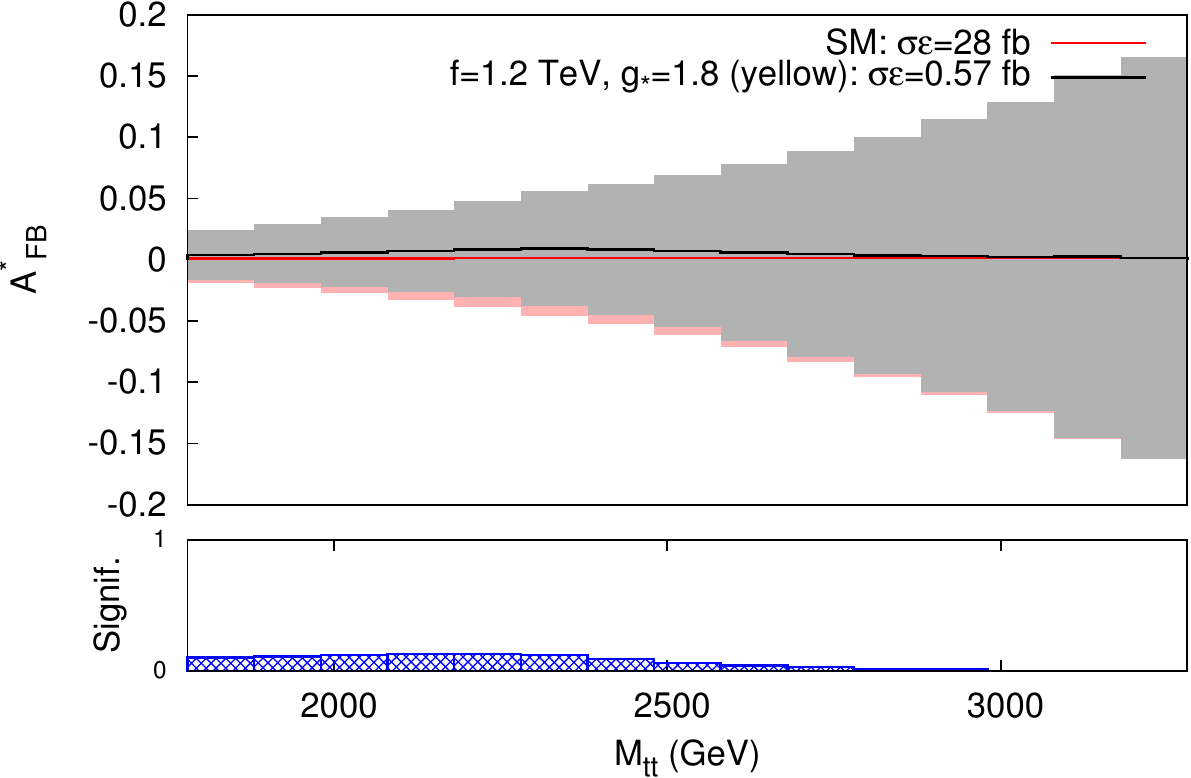}
\caption{\emph{(color online)} Cross section and asymmetries as a function of the $t\bar t$ invariant mass  for the $f=$1.2 TeV, $g_*=$1.8 (yellow) benchmark at the 14 TeV LHC with 300 fb$^{-1}$.
The left column shows the fully differential observable. 
Right plots (upper frames) include estimates of statistical uncertainty assuming a realistic 
100 GeV mass resolution and also display (lower frames) the theoretical significance assuming a 10\% reconstruction efficiency. Grey(Pink) shading represents the (statistical) error on the 4DCHM(SM) rates, in black(red) solid lines. Masses and widths of the gauge bosons are $M[\Gamma]_{Z_2,Z_3}=2249[1099]~{\rm GeV},2312[822]~{\rm GeV}$.}
\label{fig:yellow}
\end{figure}

\end{document}